\documentclass[12pt]{article}
\pdfoutput=1

\usepackage{putex}
\usepackage{autobreak}
\usepackage{graphicx}
\usepackage{caption}
\usepackage{amsmath}
\usepackage{array}
\usepackage{subcaption}
\usepackage{epstopdf}
\usepackage{enumerate}
\usepackage{cite}
\usepackage{youngtab}
\usepackage{tensor}
\usepackage{slashed}
\usepackage[aligntableaux=center]{ytableau}
\usepackage[utf8]{inputenc}
\usepackage{rotating}
\usepackage{multirow}
\usepackage[
      colorlinks=true,
      linkcolor=blue,
      urlcolor=blue,
      filecolor=black,
      citecolor=red,
      ]{hyperref}

\newcommand{\ec}{\,,}

\newcommand{\abs}[1]{\left\lvert #1 \right\rvert}

\newcommand {\be} {\begin {equation}}
\newcommand {\ee} {\end {equation}}

\newcommand {\bes} {\begin {equation*}}
\newcommand {\ees} {\end {equation*}}

\newcommand{\es}[2] {\begin{equation} \label{#1} \begin{split} #2 \end{split} \end{equation}}

\newcommand{\R}{\mathbb{R}}

\newcommand{\cD}{{\mathcal D}}

\newcommand{\cN}{{\mathcal N}}
\newcommand{\cO}{{\mathcal O}}
\newcommand{\cP}{{\mathcal P}}
\newcommand{\cQ}{{\mathcal Q}}
\newcommand{\cR}{{\mathcal R}}
\newcommand{\cS}{{\mathcal S}}

\newcommand{\cM}{{\mathcal M}}

\newcommand{\cZ}{{\mathcal Z}}

\newcommand{\beq}{\begin{equation}}
\newcommand{\eeq}{\end{equation}}

\def\ie{\begin{equation}\begin{aligned}}
\def\fe{\end{aligned}\end{equation}}

\numberwithin{equation}{section}

\def\<{\langle}
\def\>{\rangle}

\newcommand{\CR}{\color{red}}

\newcommand{\dk}{\delta}        \newcommand{\Dk}{\Delta}

\newcommand{\lk}{\lambda}       
          
\newcommand{\sk}{\sigma}        
          
          \newcommand{\Fk}{\Phi}
          
          \newcommand{\Yk}{\Psi}

\newcommand{\bra}[1]{\langle #1|} \newcommand{\ket}[1]{\left|#1\right\rangle}

\newcommand{\fcy}[1]{\mathcal{#1}}

\newcommand{\nb}{\partial} 

\begin{document}

\preprint{PUPT-2622 \\ LCTP-20-26}

\institution{PU}{Joseph Henry Laboratories, Princeton University, Princeton, NJ 08544, USA}
\institution{UMich}{Leinweber Center for Theoretical Physics, Randall Laboratory of Physics, \cr Department of Physics, University of Michigan, Ann Arbor, MI 48109, USA}
\institution{Exile}{Department of Particle Physics and Astrophysics, Weizmann Institute of Science, \cr Rehovot, Israel}

\title{
The 3d $\mathcal{N}=6$ Bootstrap: From Higher Spins to Strings to Membranes
}

\authors{Damon J.~Binder,\worksat{\PU} Shai M.~Chester,\worksat{\Exile} Max Jerdee,\worksat{\PU, \UMich}  and Silviu S.~Pufu\worksat{\PU}}

\abstract{
We study the space of 3d ${\cal N} = 6$ SCFTs by combining numerical bootstrap techniques with exact results derived using supersymmetric localization.  First we derive the superconformal block decomposition of the four-point function of the stress tensor multiplet superconformal primary. We then use supersymmetric localization results for the ${\cal N} = 6$ $U(N)_k\times U(N+M)_{-k}$ Chern-Simons-matter theories to determine two protected OPE coefficients for many values of $N,M,k$.  These two exact inputs are combined with the numerical bootstrap to compute precise rigorous islands for a wide range of $N,k$ at $M=0$, so that we can non-perturbatively interpolate between SCFTs with M-theory duals at small $k$ and string theory duals at large $k$. We also present evidence that the localization results for the $U(1)_{2M}\times U(1+M)_{-2M}$ theory, which has a vector-like large-$M$ limit dual to higher spin theory, saturates the bootstrap bounds for certain protected CFT data.  The extremal functional allows us to then conjecturally reconstruct low-lying CFT data for this theory.}
\date{}

\maketitle

\tableofcontents

\section{Introduction and Summary}
\label{intro}

The vast majority of studies so far that apply the numerical conformal bootstrap technique \cite{Rattazzi:2008pe}\footnote{For reviews, see \cite{Poland:2018epd,Simmons-Duffin:2016gjk,Chester:2019wfx}.} to superconformal field theories (SCFTs) with extended supersymmetry have been performed only for operators that belong to half-BPS supermultiplets \cite{Beem:2016wfs,Beem:2013qxa,Beem:2015aoa,Beem:2014zpa,Chester:2014mea,Agmon:2017xes,Chester:2014fya,Agmon:2019imm,Bobev:2015vsa,Bobev:2015jxa,Chester:2015qca,Chester:2015lej,Baggio:2017mas,Chang:2019dzt,Chang:2017cdx,Chang:2017xmr,Lemos:2016xke,Lemos:2015awa,Gimenez-Grau:2020jrx,Cornagliotto:2017snu,Bissi:2020jve}.\footnote{See, however, \cite{Cornagliotto:2017dup, Li:2018mdl}.}  Consequently, general constraints on the space of such SCFTs have been explored only when these SCFTs preserve the maximal amount of supersymmetry in their respective dimensions, because only then does the stress-energy tensor sit in a half-BPS multiplet. Our goal in this work is to perform a general study of SCFTs with ${\cal N} = 6$ supersymmetry in three spacetime dimensions. We will achieve this by studying the four-point function $\langle SSSS \rangle$ of the scalar superconformal primary $S$ of the stress tensor multiplet. Because $\cN = 6$ is less than the maximal $\cN = 8$ possible superconformal symmetry in three dimensions, the stress tensor multiplet is only $1/3$-BPS \cite{Dolan:2008vc,Cordova:2016emh}.\footnote{Half-BPS multiplets in 3d ${\cal N} = 6$ SCFTs have been studied in \cite{Liendo:2015cgi}.}

Three-dimensional ${\cal N} = 6$ SCFTs provide a unique window into theories of quantum gravity via the anti-de Sitter/Conformal Field Theory (AdS/CFT) correspondence \cite{Maldacena:1997re,Gubser:1998bc,Witten:1998qj}.\footnote{See, for instance, \cite{Aharony:1999ti} for a review.} Firstly, due to the large amount of supersymmetry these theories are amenable to exact computations of various protected quantities, as we will describe in more detail below.  Secondly, the less-than-maximal supersymmetry allows the existence of large families of such SCFTs \cite{Aharony:2008ug,Aharony:2008gk} that interpolate between weakly-coupled theories, theories with weakly-coupled supergravity duals in both ten and eleven dimensions, and theories with weakly-coupled higher-spin duals in $AdS_4$ \cite{Chang:2012kt}.  The known examples of ${\cal N} = 6$ SCFTs consist of the Aharony-Bergman-Jafferis-Maldacena (ABJM) and Aharony-Bergman-Jafferis (ABJ) gauge theories  with gauge groups $U(N)_k \times U(N+M)_{-k}$ \cite{Aharony:2008ug,Aharony:2008gk} or $SO(2)_{2k} \times USp(2 + 2M)_{-k}$ \cite{Aharony:2008gk, Hosomichi:2008jb}, with integer $N, M$ and Chern-Simons coefficient $k$, as well as version of these theories where the gauge group is quotiented out by a discrete subgroup,  where the $U$ groups are replaced by $SU$, or where extra $U(1)$ factors with certain Chern-Simons levels appear \cite{Schnabl:2008wj}.\footnote{We will find evidence from supersymmetric localization that all these more exotic versions have the same correlators we consider as certain $U(N)_k \times U(N+M)_{-k}$ or $SO(2)_{2k} \times USp(2 + 2M)_{-k}$ theories, so we can restrict to these standard theories for simplicity.}   This set of theories is very rich, as can be seen from various limits in parameter space. In the limit $k \gg N, M$, the SCFTs are weakly coupled and one can use perturbation theory. In the limit $M, k \gg N$ with $M/k$ fixed there is a weakly-interacting higher-spin dual description in $AdS_4$. In the limit $N \gg M, k$ there is a weakly-interacting M-theory dual description. Finally, in the limit $N, k \gg M$ with $N/k$ fixed there is a weakly-interacting type IIA string dual description.

Given a known CFT, the ideal outcome of a numerical conformal bootstrap study is to show that under a certain set of assumptions the CFT is unique, or that at least some of the CFT data can be uniquely determined.  This has so far been accomplished in one of two ways. The first is to find (small) islands in the space of CFT data surrounding the known CFT.  If one can argue that these islands shrink to a point as precision is increased, then one can determine part of the CFT data.  For instance, in the case of the 3d Ising model, a numerical bootstrap study of the four-point functions of the only two relevant operators gave allowed regions in the space of scaling dimensions of these operators that look like small islands surrounding the known values of these scaling dimensions for the Ising CFT \cite{Kos:2014bka,Kos:2016ysd,Simmons-Duffin:2015qma}.  Besides the 3d Ising model, there are very few other examples where similar small islands have been found with minimal physical assumptions---see, for instance, \cite{Kos:2015mba,Atanasov:2018kqw,Rong:2018okz,Chester:2019ifh}.  
One can also find islands in the space of OPE coefficients multiplying semishort superconformal blocks which, due to supersymmetry, cannot be deformed into long blocks.  Such islands were found for 3d ${\cal N} = 8$ SCFTs (which are, of course, a particular case of the 3d ${\cal N} = 6$ SCFTs we study here) in \cite{Chester:2014mea,Agmon:2019imm}, in which case exactly computable localization results for certain protected OPE coefficients were also inputted to further shrink the islands and identify them with known theories. The theories studied in \cite{Chester:2014mea,Agmon:2019imm} were strongly-interacting gauge theories which, in the limit where the rank of the gauge group is taken to infinity, are dual to weakly-coupled M-theory in eleven dimensions.  In the present work we will find similar islands for the much larger class of 3d ${\cal N} = 6$ SCFTs.

The other method used to ``solve'' for part of the CFT data of a known CFT applies to cases where one can argue that this CFT saturates certain bounds in the limit of infinite bootstrap precision.  In such a case, it is believed that there is a unique solution to the crossing equation and the CFT data can be extracted using the extremal functional method \cite{Poland:2010wg, ElShowk:2012hu, El-Showk:2014dwa}.\footnote{Ref.~\cite{Caron-Huot:2020adz} showed that it is sometimes possible that there could be several extremal functionals, but in all cases that were studied they produced the same CFT spectrum.} One application of this method has been to the 3d Ising model, which was argued to saturate the lower bound on the coefficient $c_T$ which appears in the stress-tensor two-point function  \cite{El-Showk:2014dwa,Simmons-Duffin:2016wlq}. In this paper we will find compelling evidence that the $U(1)_{2M} \times U(1+M)_M$ ABJ theory, which has a higher-spin limit at large $M$, also saturates certain bootstrap bounds.  This makes the theory amenable to a precision bootstrap study.  It is worth noting that the more restrictive bounds on ${\cal N} = 8$ SCFTs derived in \cite{Agmon:2017xes,Agmon:2019imm} were saturated by the $U(N)_2\times U(N+1)_{-2}$ ABJ theory with ${\cal N} = 8$ supersymmetry, which in the large $N$ limit has an M-theory dual description. So the present work shows that one can use the numerical bootstrap to study both SCFTs with supergravity and higher-spin dual descriptions.

As in the ${\cal N} = 8$ case studied in  \cite{Agmon:2017xes,Agmon:2019imm}, our bootstrap analyses are aided by supersymmetric localization, allowing us to analytically compute certain protected data that appears in $\langle SSSS\rangle$.  Exact computations using supersymmetric localization\footnote{For a review, see~\cite{Pestun:2016zxk}.} are possible in any ${\cal N} \geq 2$ SCFT in 3d \cite{Kapustin:2009kz,Jafferis:2010un}, but when $\mathcal{N}<4$ they can only give local CFT data related to conserved currents, such as $c_T$. For ${\cal N} \geq 4$ SCFTs, however, one can combine localization with the 1d topological sector discovered in \cite{Chester:2014mea,Beem:2016cbd}, whose explicit description for Lagrangian theories was determined in \cite{Dedushenko:2016jxl,Dedushenko:2017avn,Dedushenko:2018icp}, to compute all half-BPS (in $\mathcal{N}=4$ language) OPE coefficients. For instance, we can use localization in our $\mathcal{N}=6$ case to compute not only $c_T$ but also a certain $1/3$-BPS OPE coefficient (that would be half-BPS in $\mathcal{N}=4$) that appears in $\langle SSSS\rangle$. For the $U(N)_k \times U(N + M)_{-k}$ theory these computations take the form of $N$ dimensional integrals, which can be evaluated exactly for small $N$, and also to all orders in $1/N$ using the Fermi gas technique \cite{Marino:2011eh,Nosaka:2015iiw}.  For the $U(1)_k \times U(1+M)_{-k}$ and $SO(2)_{2k} \times USp(2 + 2M)_{-k}$ cases we derive one-dimensional integrals that can be computed exactly for all $M$ and $k$.

The rest of this paper is organized as follows. In Section~\ref{FOURPOINT} we review basic properties of $\langle SSSS\rangle$ and derive the superblock decomposition. In Section~\ref{EXACT}, we compute $c_T$ and the  squared $1/3$-BPS OPE coefficient $\lambda^2_{(B,2)_{2,0}^{022}}$ for various choices of $M,N,k$.  In Section~\ref{numerics}, we use the numerical bootstrap to compute non-perturbative bounds on CFT data, which we then combine with the localization results to derive precise islands and study the $U(1)_{2M}\times U(1+M)_{-2M}$ theory using the extremal functional. Finally, in Section~\ref{disc}, we end with a discussion of our results and of future directions. Various technical details are discussed in the appendices. We also include an attached \texttt{Mathematica} notebook with the explicit $\langle SSSS\rangle$ superconformal blocks.

\section{Superconformal block expansion of $\<SSSS\>$}
\label{FOURPOINT}

In this section we derive the superconformal block expansion for the $\<SSSS\>$ four-point correlator. We begin with a brief review of constraints that superconformal symmetry places on $\<SSSS\>$; a more detailed discussion can be found in Section~2 of~\cite{Binder:2019mpb}. Section~\ref{SSOPE} restricts the supermultiplets which can appear in the $S\times S$ OPE, and hence exchanged in $\<SSSS\>$, to a small number of possibilities. Section~\ref{casimir} applies the superconformal Casimir equation to each of the allowed supermultiplets to fix the superconformal blocks which contribute to $\<SSSS\>$. Finally, in Section~\ref{EXAMPLES} we determine the superblock decomposition for $\<SSSS\>$ in free fields theories.

\subsection{Constraints from conformal symmetry and R-symmetry}

In any 3d $\cN = 6$ SCFT, the stress tensor sits in a $1/3$-BPS multiplet \cite{Dolan:2008vc,Cordova:2016emh}.  The superconformal primary of this multiplet is a dimension $1$ scalar operator transforming in the ${\bf 15}$ of the $\mathfrak{so}(6)$ R-symmetry algebra.  Using the isomorphism $\mathfrak{so}(6) \cong \mathfrak{su}(4)$, we will write this operator as a $4 \times 4$ traceless hermitian matrix $S_{a}{}^b(\vec{x})$, where $a = 1, \ldots, 4$ and $b = 1, \ldots, 4$ are $\mathfrak{su}(4)$ fundamental and anti-fundamental indices, respectively.  To avoiding carrying around indices, we find it convenient to contract them with an auxiliary matrix $X$, thus defining
\es{SXDef}{
   S(\vec{x}, X) \equiv X_a{}^b S_b{}^a(\vec{x}) \,.
}
We normalize $S(\vec{x}, X)$ such that its two-point function is 
 \es{2pS}{
  \langle S(\vec x_1,X_1)S(\vec x_2,X_2)\rangle &= \frac{\text{tr}(X_1X_2)}{x_{12}^2} \,.
 }

We will not need many details about other operators in the stress tensor multiplet, but for completeness we list the conformal primaries of this multiplet in Table~\ref{stressTable}.    Apart from the superconformal primary, the only operators that will appear in the discussion below are the fermions $\chi_{\alpha}$, $F_{\alpha}$ and $\overline F_\alpha$, which all have dimension $3/2$ and transform in the ${\bf 6}$, ${\bf 10}$, and $\overline{{\bf 10}}$ of $\mathfrak{so}(6)_R$, as well as the (pseudo)scalar operator $P$ of dimension $2$ transforming in the ${\bf 15}$ of $\mathfrak{so}(6)_R$.\footnote{As can be seen from Table~\ref{stressTable}, all 3d $\mathcal{N}=6$ theories also have a $U(1)$ flavor symmetry whose conserved current $j$ also belongs to the stress tensor multiplet.  The superconformal primary $S$ is invariant under this flavor symmetry, so this symmetry will not play a role in this work.}

\begin{table}
\begin{center}
\renewcommand{\arraystretch}{1.2}
\begin{tabular}{c|c|c|c}
  Operator & $\Delta$ & Spin & $\mathfrak{so}(6)_R$ irrep \\\hline
  $S$    & 1   & $0$   & ${\bf 15} = [011]$              \\ \hline
  $\chi$ & 3/2 & 1/2 & ${\bf 6} = [100]$               \\
  $F$    & 3/2 & 1/2 & ${\bf 10} = [020]$               \\ 
  $\bar F$    & 3/2 & 1/2 & $\overline{\bf 10} = [002]$     \\ \hline
  $P$ & 2 & $0$ & ${\bf 15} = [011]$                   \\
  $J$ & 2 & $1$ & ${\bf 15} = [011]$                  \\
  $j$ & 2 & $1$ & ${\bf 1}=[000]$                        \\ \hline
  $\psi$ & 5/2 & 3/2 & ${\bf 6} = [100]$               \\ \hline
  $T$ & 3 & $2$ & ${\bf 1}=[000]$                     
  \end{tabular}
\caption{The conformal primary operators in the $\cN = 6$ stress tensor multiplet.  For each such operator, we list the scaling dimension, spin, and $\mathfrak{so}(6)_R$ representation.}
\label{stressTable}
\end{center}
\end{table}

As discussed in~\cite{Binder:2019mpb}, conformal and R-symmetry invariance imply that the four-point function of $S(\vec{x}, X)$ must take the form
 \es{SSSScor}{
   \langle S(\vec x_1,X_1)  S(\vec x_2,X_2) S(\vec x_3,X_3) S(\vec x_4,X_4)\rangle = 
 \frac{1}{\abs{\vec{x}_{12}}^2 \abs{\vec{x}_{34}}^2 } \sum_{i=1}^6 {\cal S}^i(U, V) {\cal B}_i \,,
 }
where we define the R-symmetry structures
 \es{BasisElems}{
  {\cal B}_1 &= \tr (X_1 X_2) \tr (X_3 X_4) \,, \\
  {\cal B}_2 &= \tr (X_1 X_3) \tr (X_2 X_4) \,, \\
  {\cal B}_3 &= \tr (X_1 X_4) \tr (X_2 X_3) \,, \\
  {\cal B}_4 &= \tr (X_1 X_4 X_2 X_3) + \tr (X_3 X_2 X_4 X_1) \,, \\
  {\cal B}_5 &= \tr (X_1 X_2 X_3 X_4) + \tr (X_4 X_3 X_2 X_1) \,, \\
  {\cal B}_6 &= \tr (X_1 X_3 X_4 X_2) + \tr (X_2 X_4 X_3 X_1) \,,
 }
and where $\cS^i$ are arbitrary functions of the conformally-invariant cross-ratios
\begin{equation} U \equiv \frac{x_{12}^2x_{34}^2}{x_{13}^2x_{24}^2} \,, \qquad  V \equiv \frac{x_{14}^2x_{23}^2}{x_{13}^2x_{24}^2} \,.
\end{equation}

While the form \eqref{SSSScor} has simple properties under crossing symmetry, namely
\es{CrossingP}{
\cS^1(U,V) &= \cS^1\left(U,\frac1V\right)\,, \quad \cS^2(U,V) = U\cS^1\left(\frac{1}{U},\frac{V}{U}\right)\,, \quad \cS^3(U,V) = \frac{U}{V}\cS^1(V,U)  \,, \\
\cS^4(U,V) &= \cS^4\left(U,\frac1V\right)\,, \quad \cS^5(U,V) = U\cS^4\left(\frac{1}{U},\frac{V}{U}\right)\,, \quad \cS^6(U,V) = \frac{U}{V}\cS^4(V,U) \,,
}
it is not the most convenient form to work with for writing a conformal block decomposition because each conformal block will contribute to several different $\cS^i$. To do better, we can take linear combinations of $\cS^i$ such that each such linear combination corresponds to a specific $\mathfrak{so}(6)$ irrep being exchanged in the $s$-channel OPE\@.  The possible such irreps are those appearing in the tensor product
 \es{irreps}{
{\bf15}\otimes{\bf15}&={\bf1}_s\oplus{\bf15}_a\oplus{\bf15}_s\oplus{\bf20'}_s\oplus({\bf45}_a\oplus{\bf\overline{45}}_a)\oplus{\bf84}_s\,.
 }
We define $\cS_{\bf{r}}$ to receive contributions only from operators in the $s$-channel OPE that belong to $\mathfrak{so}(6)_R$ irrep $\bf{r}$, so that \cite{Binder:2019mpb} 
\es{B}{
\vec \cS\cdot \bold{B}=&  \begin{pmatrix}\cS_{{\bf 1}_s} & \cS_{{\bf 15}_a} &\cS_{{\bf 15}_s} &\cS_{{\bf 20}'_s} &\cS_{{\bf 45}_a \oplus {\overline{\bf 45}}_a} &\cS_{{\bf 84}_s} \end{pmatrix}  \,,\\
\bold{B}=&\left(
\begin{array}{cccccc}
 1 & 0 & 0 & 0 & 0 & 0 \\
\frac{1}{15}  & -\frac{1}{8} & \frac{1}{6} & \frac{1}{24} & -\frac{1}{8}    &  \frac{1}{16}  \\
 \frac{1}{15} & \frac{1}{8}  & \frac{1}{6} & \frac{1}{24} & \frac{1}{8}   &  \frac{1}{16}  \\
-\frac{1}{30} & 0            & -\frac{1}{6}& -\frac{1}{12}& 0              &  \frac{1}{8}   \\
\frac{1}{2}   & \frac{1}{2}  & \frac{1}{2} & 0            & 0              &  0             \\
\frac{1}{2}   & -\frac{1}{2} & \frac{1}{2} & 0            & 0              &  0             \\
\end{array}
\right)\,.
}
Each of the functions $\cS_{\bf{r}}$ can then be expanded as a sum of conformal blocks
 \es{SDecomp}{
   \cS_{\bf r}(U,V) = \sum_{\text{conformal primaries ${\cal O}_{\Delta, \ell, {\bf r}}$}} a_{\Delta,\ell,\bf r}g_{\Delta,\ell}(U,V) \,,
 }
where the sum is taken over all the distinct conformal primary operators ${\cal O}_{\Delta, \ell, {\bf r}}$ transforming in the representation $\bf r$ which appear in the $S\times S$ OPE\@.  In \eqref{SDecomp}, $\Delta$ and $\ell$ are the scaling dimension and spin, respectively, of ${\cal O}_{\Delta, \ell, {\bf r}}$.

Invariance under the full $\mathfrak{osp}(6|4)$ superconformal algebra relates the various four-point functions of stress-tensor multiplet operators, and furthermore it imposes relations on the $\cS^i(U, V)$ defined above. As shown in~\cite{Binder:2019mpb}, there are two such relations obeyed by the $\cS^i$ and they take the form
\begin{equation}\begin{split}\label{SSSSward}
\partial_U \cS^6(U,V) &= \frac 1 {2U^2}\bigg[-(U^3\partial_U+U^2V\partial_V)\cS^1+(1-V+U(V-1)\partial_U+UV\partial_V)\cS^2\\
&\ \ \ +(1-U-V-U(1-2U+U^2-V)\partial_U+U(1-U)V\partial_V)\cS^3 \\
&\ \ \ +(2-U-2V+2U(U+V-1)\partial_U+2UV\partial_V)\cS^4\\
&\ \ \ -U(1+2U(U-1)\partial_U+2UV\partial_V)\cS^5 + U\cS^6\bigg] \,,\\
\partial_V \cS^6(U,V) &= \frac 1 {2U}\bigg[ U(U\partial_U+(V-1)\partial_V)\cS^1+(1-U\partial_U-U\partial_V)\cS^2\\
&\ \ \ +(1+U(U-1)\partial_U+UV\partial_V)\cS^3+(2-2U\partial_U)\cS^4\\
&\ \ \ +(2U^2\partial_U+2UV\partial_V)\cS^5\bigg] \,.
\end{split}\end{equation}

\subsection{The $S\times S$ OPE}
\label{SSOPE}

As a first step towards determining the superconformal block decomposition of the $\langle SSSS \rangle$ correlator, we turn to the task of determining which $\cN=6$ supermultiplets may appear in the $S\times S$ OPE\@. By using the R-symmetry selection rules and the fact that $S(\vec{x},X)$ is a $1/3$-BPS operator we can restrict our attention to only a handful of $\cN=6$ supermultiplets. In Section~\ref{casimir} we can then apply the superconformal Casimir equation in order to fully fix the superconformal blocks corresponding to each of these supermultiplets.

\subsubsection{$\cN = 6$ Supermultiplets}
\label{supermults}
The unitary multiplets of a 3d $\cN = 6$ theory are given in \cite{Dolan:2008vc,Cordova:2016emh}. Each multiplet can be labeled by the conformal dimension $\Delta$, spin $\ell$, and $\mathfrak{so}(6)$ R-symmetry irrep ${\bf r} = [a_1a_2a_3]$ of its superconformal primary.  These multiplets fall into three possible classes. Long multiplets have conformal dimension above the unitarity bound
 \es{DeltaLong}{
\Delta > \ell+ a_1 + \frac 12(a_2+a_3) + 1  
 }
and do not satisfy any shortening conditions. Semishort, or $A$-type, multiplets occur at the bottom of the continuum in \eqref{DeltaLong}
\begin{equation}
\Delta = \ell+ a_1 + \frac 12(a_2+a_3) + 1  
\end{equation}
and satisfy shortening conditions. Finally, if $\ell = 0$ we can also have short, or $B$-type, multiplets with dimension
\begin{equation}
\Delta = a_1 + \frac 12 (a_2+a_3)\,,
\end{equation}
 below the end of the lower continuum in \eqref{DeltaLong}, also obeying shortening conditions.   Multiplets can furthermore be distinguished by their BPSness. For generic representations $A$-type multiplets are $1/12$-BPS and $B$-type multiplets are $1/6$-BPS, but for specific R-symmetry representations the multiplets may be higher BPS\@. We list all possible multiplets in Table~\ref{n6mults}. Note that the stress-tensor multiplet discussed in the previous subsection is a $(B,2)$ multiplet in the notation of Table~\ref{n6mults}. 

\begin{table}
\begin{center}
\hspace{-.4in}
{\renewcommand{\arraystretch}{1.2}
\begin{tabular}{c|c|c||c|c|c}
Type & $\Delta$           & Spin   & Multiplet & $\mathfrak{so}(6)_R$ & BPS    \\\hline
Long & $>\Delta_B+\ell+1$ & $\ell$ & $\text{Long}$   & $[a_1a_2a_3]$        & 0      \\\hline
$A$  & $\Delta_B+\ell+1$  & $\ell$ & $(A,1)$   & $[a_1a_2a_3]$        & $1/12$ \\
     &                    &        & $(A,2)$   & $[0a_2a_3]$          & $1/6$  \\
     &                    &        & $(A,+)$   & $[0a_20]$            & $1/4$  \\
     &                    &        & $(A,-)$   & $[00a_3]$            & $1/4$  \\
     &                    &        & $(A, \text{cons.})$ & $[000]$              & $1/3$  \\\hline
$B$  & $\Delta_B$         & 0      & $(B,1)$   & $[a_1a_2a_3]$        & $1/6$  \\
     &                    &        & $(B,2)$   & $[0a_2a_3]$          & $1/3$  \\
     &                    &        & $(B,+)$   & $[0a_20]$            & $1/2$  \\
     &                    &        & $(B,-)$   & $[00a_3]$            & $1/2$  \\
     &                    &        & Trivial   & $[000]$              & $1$    
\end{tabular}}
\caption{Multiplets of $\mathfrak{osp}(6|4)$ and the quantum numbers of their superconformal primary, where $\Delta_B = a_1 + \frac 12 (a_2+a_3)$.}
\label{n6mults}
\end{center}
\end{table}

Of course, not all possible multiplets contain operators that can appear in the $S\times S$ OPE due to various selection rules. Note for instance that the operators in the $S\times S$ OPE must transform in the irreducible representations of $\mathfrak{so}(6)$ which appear in \eqref{irreps}.  Due to $1\leftrightarrow2$ crossing symmetry even spin operators must be in the ${\bf1}$, ${\bf15}$, ${\bf20'}$, or $\bf 84$ while odd spin operators must be in the $\bf 15$, $\bf 45$, or $\overline{\bf45}$. A large number of supermultiplets contain operators in at least one of these irreps, so by themselves these conditions are not very restrictive.

We can do better by using the fact that $S(\vec x, X)$ is a $1/3$-BPS operator, and as such is annihilated by certain Poincar\'e supercharges. If $\cQ$ is a Poincar\'e supercharge annihilating $S(\vec x, X)$ (for any $\vec{x}$ but a specific $X$), then it also annihilates $S(\vec x, X)S(\vec y, X)$.  We will explore the consequences of this fact in the next subsection.

\subsubsection{Operators in the $S \times S$ OPE}
\label{allowed84}

Let us begin by writing the generators of $\mathfrak{osp}(6|4)$ in terms of the $\mathfrak{so}(6)$ and $\mathfrak{sp}(4)$ Cartan subalgebras. The Lie algebra $\mathfrak{so}(6)$ has a three dimensional Cartan subalgebra, spanned by orthogonal operators\footnote{For instance, in the ${\bf 6}$ irrep of $\mathfrak{so}(6)_R$, we can represent the Cartan generators by the matrices $$
H_1 = 
\begin{pmatrix}
\sk_2 \\ & 0\\ & & 0
\end{pmatrix}\,,\qquad
H_2 = 
\begin{pmatrix}
0 \\ & \sk_2\\ & & 0
\end{pmatrix}\,,\qquad
H_3 = 
\begin{pmatrix}
0\\ & 0\\ & & \sk_2
\end{pmatrix}\,,
$$
where $\sigma_2$ is the second Pauli matrix.} $H_1$, $H_2$, and $H_3$. The other twelve R-symmetry generators take the form:
$$R_{\pm1,\pm1,0}\,,\quad R_{\pm1,0,\pm1}\,,\quad R_{0,\pm1,\pm1}\,,\quad R_{\pm1,\mp1,0}\,,\quad R_{\pm1,0,\mp1}\,,\quad R_{0,\pm1,\mp1}\,,$$
where for each $R$ the subscripts are correlated and label the weights of each of these generators under the Cartan subalgebra:
\begin{equation}
[H_i,R_{r_1,r_2,r_3}] = r_i R_{r_1,r_2,r_3}\,, \text{ for } i=1\,,2\,,3\,.
\end{equation}
We take the simple roots of $\mathfrak{so}(6)$ to be the raising operators
$$
\cR^+ = \big\{ R_{1,-1,0}\,, R_{0,1,1}\,, R_{0,1,-1}\big\}\,,
$$
while their corresponding lowering operators are
$$
\cR^- = \big\{ R_{-1,1,0}\,, R_{0,-1,-1}\,, R_{0,-1,1}\big\}\,.
$$
A highest weight state is one that is annihilated by each element of $\cR^+$; the highest weight state of the $\bf 15$ is then $R_{1,1,0}$. 

We perform a similar procedure with the conformal group $\mathfrak{sp}(4)$. We can take one Cartan element to be the dilatation operator $D$ and the other to be the rotation operator $J^0$. The other two rotation operators are the raising and lowering operators $J^\pm$. The $H_i$, $D$, and $J^0$ span a Cartan subalgebra of $\mathfrak{osp}(6|4)$.

We can now write the $\mathfrak{osp}(6|4)$ supercharges in terms of their charges under this subalgebra. The $Q$s and $S$s can be written as
$$
Q^\pm_{\pm1,0,0}\,,\quad Q^\pm_{0,\pm1,0}\,,\quad Q^\pm_{0,0,\pm1}\,, \quad \text{ and }S^\pm_{\pm1,0,0}\,,\quad S^\pm_{0,\pm1,0}\,,\quad S^\pm_{0,0,\pm1} \,,
$$
respectively, where the superscript is the $J^0$ charge and the subscripts are the $H_i$ charges. (The sign in the superscript is uncorrelated with the signs in the subscripts.) Note that the $Q$s have scaling dimension $+1/2$ and the $S$s have scaling dimension $-1/2$, so their charges under dilatation operator are also manifest in this notation.

Given an irreducible representation of $\mathfrak{osp}(6|4)$, the highest weight state $\ket{\Dk,\ell,r}$ is one which is annihilated by the raising operators of $\mathfrak{osp}(6|4)$:
\begin{equation}\label{raisingC}
K^\mu\ket{\Dk,\ell,r} = S^\pm_{a,b,c}\ket{\Dk,\ell,r} = J^+\ket{\Dk,\ell,r} = R^+\ket{\Dk,\ell,r} = 0\,,
\end{equation}
where $R^+ \in \cR^+$, and is an eigenstate of each of the Cartans:
\begin{equation}
H_i\ket{\Dk,\ell,r} = r_i\ket{\Dk,\ell,r}\,,\qquad D\ket{\Dk,\ell,r} = \Dk\ket{\Dk,\ell,r}\,,\qquad J^0\ket{\Dk,\ell,r} = 2\ell\ket{\Dk,\ell,r}\,.
\end{equation}
Here $\Dk$ and $\ell$ are the conformal dimension and spin of the superconformal primary, and the $r = (r_1,r_2,r_3)$'s are the highest weight states of the R-symmetry representation of the superconformal primary. These weights are related to the Dynkin label $[a_1a_2a_3]$ by the equation
\begin{equation}
r_1 = a_1+\frac{a_2+a_3}2\,,\qquad r_2 = \frac{a_2+a_3}2\,,\qquad r_3 = \frac{a_2-a_3}2
\end{equation}
and always satisfy $r_1\geq r_2\geq r_3$.

The highest weight state of the stress-tensor multiplet, $\ket{\fcy S^H},$ has conformal $\Delta = 1$, spin $\ell = 0$, and R-symmetry weights $(1,1,0)$. It can be created by acting with the operator\footnote{To avoid confusion between the superconformal generators $S^\pm_{r_1,r_2,r_3}$ and components of the stress-tensor superconformal primary $\hat S_{r_1,r_2,r_3}(x)$, in this section we adopt the convention that the latter operators are always hatted.} $\hat S_{1,1,0}(0)$ on the vacuum. The stress-tensor is a $1/3$-BPS multiplet, satisfying the shortening condition
\begin{equation}
Q \ket{\fcy S^H} = 0 \text{ for all } Q\in\cQ^+ = \left\{Q^\pm_{1,0,0}\,,\quad Q^\pm_{0,1,0}\right\}\,,
\end{equation}
which in turn implies that
\begin{equation}
Q\hat S_{1,1,0}(\vec x) = 0 \text{ for all } \vec x\in\mathbb R^3 \text{ and all } Q\in\cQ^+\,. 
\end{equation}
This is equivalent to imposing that $\hat S(\vec x,X)$ has no fermionic descendant in the $\bf 64$ of $\mathfrak{so}(6)$.  
We will find it useful to further define
\begin{equation}
\cQ^0 = \left\{Q^\pm_{0,0,\pm 1}\right\}\,,\qquad \cQ^- = \left\{Q^\pm_{-1,0,0}\,,\quad Q^\pm_{0,-1,0}\right\}\,,\qquad \cQ = \cQ^+\cup\cQ^0\cup \cQ^-
\end{equation}
along with analogous definitions for the $S$-supercharges.

Let $\Fk_{2,2,0}(\vec x)$ be any operator which appears in the OPE $\hat S_{1,1,0}\times \hat S_{1,1,0}$ and ${\ket\Fk = \Fk_{2,2,0}(0)\ket0}$ the associated state. This is the highest weight state of an $\bf 84$ multiplet which is annihilated by $\cR^+$ and $\cQ^+$. Without loss of generality we can take this operator to be a conformal primary which is annihilated by $J^+$; if it is not we can act with the raising operators $K^\mu$ and $J^+$ to construct such an operator. Because any operator $S^+\in\cS^+$ is of the form $[K,Q^+]$ for some $Q^+\in\cQ$, we find that $\cS^+$ also annihilates $\ket\Fk$. So in total, we have the conditions
\begin{equation}\label{FkConds}
Q^+\ket\Fk = J^+\ket\Fk = R^+\ket\Fk =  S^+\ket\Fk =  K^\mu\ket\Fk = 0 \text{ for any } R^+ \in\cR^+\,,\  Q^+\in\cQ^+\,,\  S^+\in\cS^+\,.
\end{equation}
Our task it to determine which supermultiplets $\ket\Fk$ may belong to.

By acting with operators in $\cS$ on $\ket\Fk$ we can construct states of lower conformal dimension. Consider first constructing a state $\ket{\cO'}$ by acting with all eight supercharges in $\cS^0\cup \cS^-$:
\begin{equation}
\ket{\cO'} = S^+_{0,+1,0}S^-_{0,+1,0}S^+_{0,-1,0}S^-_{0,-1,0}S^+_{0,0,+1}S^-_{0,0,+1}S^+_{0,0,-1}S^-_{0,0,-1}\ket\Fk\,.
\end{equation}
By assumption $\ket\Fk$ satisfies \eqref{FkConds}, and it is straightforward to see that $\ket{\cO'}$ then also satisfies \eqref{FkConds}. Because the $\cS$ operators anticommute with themselves, we furthermore find that any operator in $\cS^0\cup \cS^-$ annihilates $\ket{\cO'}$. The state $\ket{\cO'}$ is therefore annihilated by all of the ${\cal S}$ and by $J^+$ and $R^+$, and so either $\ket{\cO'}$ is the highest weight state of the superconformal primary of the supermultiplet, or $\ket{\cO'} = 0$. In either case we conclude that there exists some $0\leq k\leq8$ for which acting with any $k+1$ operators from $\cS^0\cup \cS^-$ annihilates $\ket{\Fk}$, but for which acting with just $k$ operators does not:
\begin{equation}\label{OfromS}
\ket{\cO} = S_1\cdots S_k\ket\Fk \neq 0
\end{equation}
for some string of $k$ operators $S_i \in \cS^0\cup \cS^-$. It is again easy to see that $\ket{\Fk}$ satisfies \eqref{FkConds} and is annihilated by the operators in $\cS^0\cup \cS^-$; we hence conclude that $\ket{\cO}$ is the highest weight state of the superconformal primary of the multiplet. Note that the different orderings of the operators $S_i$ in \eqref{OfromS} are equivalent, up to an overall minus sign.

Let us denote the $\mathfrak{so}(6)$ weights of $\ket{\cO}$ by
\begin{equation}
w = (2,2,0) + \sum_i v_i\,,
\end{equation} 
where $v_i = (v_{i1},v_{i2},v_{i3})$ are the $\mathfrak{so}(6)$ Cartans of the $S_i$ we act with in \eqref{OfromS}. Because $\ket{\cO}$ is a highest weight state we must have 
\begin{equation}\label{HWcond}
w_1\geq w_2\geq |w_3|\,,
\end{equation}
which provides a useful additional constraint on \eqref{OfromS}. 

As discussed in the previous section, $\ket{\cO}$ belongs to one of the three types of unitary representations of $\mathfrak{osp}(6|4)$. If $\ket{\cO}$ is part of a long multiplet, it is annihilated by all of the raising operators \eqref{raisingC} but satisfies no other conditions. If instead it belongs to an $A$-type multiplet it satisfies shortening conditions \cite{Dolan:2008vc}
\begin{equation}
\left(Q^-_{q_1,q_2,q_3}-\frac1{2\ell}Q^+_{q_1,q_2,q_3}J^-\right)\ket{\cO} = 0
\end{equation}
with the specific weights $q_i$ depending on the $\mathfrak{so}(6)$ weights of $\ket{\cO}$. Finally, if it is part of a $B$-type multiplet, it is annihilated by both $Q^+_{q_1,q_2,q_3}$ and $Q^-_{q_1,q_2,q_3}$ for specific weights $q_i$. Furthermore for $B$-type multiplets $\ket{\cO}$ is always a scalar.

With this information out of the way, we now simply enumerate all possibilities for \eqref{OfromS}, subject to the constraint \eqref{HWcond}. The simplest case is where $\ket\Fk$ is itself the highest weight primary. Then we have a $(B,2)$ multiplet in the $\bf 84$.

Next let us extend this reasoning to the case
\begin{equation}\label{case1}
\ket{\cO} = S_1\cdots S_n\ket{\Fk} \text{ where }S_i\in \cS^0\,.
\end{equation}
Because $\{\cQ^+,\cS^0\}$ consists only of positive R-symmetry generators, we see that $\ket{\cO}$ is annihilated by $\cQ^+$ and hence we still have a $(B,2)$ multiplet. The possible R-symmetry representations are the $[022]$ (which is the $\bf 84$), the $[031]$ and its conjugate $[013]$, and the $[040]$ and its conjugate $[004]$.  We can however eliminate the $[031]$ possibility, as in this case one needs to act with an odd number of supercharges to construct an operator in the $\bf 84$ from the superconformal primary.

Let us next consider the cases
\begin{subequations}
	\begin{equation}\label{caseB}
		\ket{\cO} = S_1\cdots S_n S_{-1,0,0}^\pm\ket{\Fk} \text{ where }S_i \in\cS^0\,,
    \end{equation}
	\begin{equation}\label{case2}
		\ket{\cO} = S_1\cdots S_n S_{0,-1,0}^\pm\ket{\Fk} \text{ where }S_i \in\cS^0\,,
    \end{equation}
    \begin{equation}\label{caseB2}
		\ket{\cO} = S_1\cdots S_n S_{-1,0,0}^+S_{-1,0,0}^-\ket{\Fk} \text{ where }S_i \in\cS^0\,.
	\end{equation}
    \begin{equation}\label{case3}
		\ket{\cO} = S_1\cdots S_n S_{0,-1,0}^+S_{0,-1,0}^-\ket{\Fk} \text{ where }S_i \in\cS^0\,,
	\end{equation}
\end{subequations}
Cases \eqref{caseB} and \eqref{caseB2} violate \eqref{HWcond} and so are not possible. For the other two possibilities we find that $\ket{\cO}$ is annihilated by $S_{1,0,0}^\pm$ and so $\ket{\cO}$ must be a $B$-type multiplet. Using \eqref{HWcond} we find that the possible multiplets for \eqref{case2} are
$$ (B,1) \text{ in the } [120]\,,\ [102]\,,\ \text{or a } (B,2) \text{ in the }[111]\,,$$
while for \eqref{case3} we can only have a $(B,1)$ in the $[200]$. We can furthermore eliminate the $(B,2)$ in the $[111]$ as an option because in this multiplet only fermions transform in the $\bf 84$.

The next cases to consider are
\begin{subequations}
\begin{equation}\label{case4}
\ket{\cO} = S_1\cdots S_n S_{-1,0,0}^\pm S_{0,-1,0}^\pm\ket{\Fk} \text{ where }S_i \in\cS^0\,,
\end{equation}
\begin{equation}\label{case5}
\ket{\cO} = S_1\cdots S_n S_{-1,0,0}^\pm S_{0,-1,0}^+S_{0,-1,0}^-\ket{\Fk} \text{ where }S_i \in\cS^0\,.
\end{equation}
\begin{equation}\label{case5b}
\ket{\cO} = S_1\cdots S_n S_{0,-1,0}^\pm S_{-1,0,0}^+S_{-1,0,0}^-\ket{\Fk} \text{ where }S_i \in\cS^0\,.
\end{equation}
\end{subequations}
Case \eqref{case5b} violates \eqref{HWcond} and so is forbidden. For other two cases we find some combination of $Q^-_{1,0,0}$ and $Q^+_{1,0,0}$ annihilate $\ket{\cO}$, so $\ket{\cO}$ must be either an $A$-type or $B$-type multiplet. For \eqref{case4} we find that \eqref{HWcond} restricts us to an $(A,+)$ or $(B,+)$ in the $[020]$, an $(A,-)$ or $(B,-)$ in the $[002]$, or an $(A,2)$ or $(B,2)$ in the $[011]$. For \eqref{case5} we instead find that $\ket{\cO}$ is an $(A,1)$ or $(B,1)$ multiplet in the $[100]$. However, we can rule out all $B$-type multiplets; the $(B,\pm)$ and $(B,1)$ only contains fermionic operators in the $\bf 84$, while due to its shortening conditions the $(B,2)$ does not contain any operator in the $\bf 84$. Thus only the $A$-type multiplets are possible.

Finally, we have the case
\begin{equation}\label{case6}
\ket{\cO} = S_1\cdots S_n S_{-1,0,0}^+S_{0,-1,0}^-S_{0,-1,0}^+S_{0,-1,0}^-\ket{\Fk} \text{ where }S_i \in\cS^0\,.
\end{equation}
Now $\ket{\cO}$ need not be annihilated by an supercharges, so it can be a long multiplet. The condition \eqref{HWcond} however forces it to be an $\mathfrak{so}(6)$ singlet. If $\ket{\cO}$ satisfies any shortening conditions it must be either a conserved current multiplet or the trivial (vacuum) multiplet, but neither of these contain an operator in the $\bf 84$ so these are both ruled out.

We summarize our results in the first 11 lines of Table~\ref{allowedBlocks}, where we give the full list of all possible superconformal blocks which contain an operator in the $\bf 84$. 

Our next task is to extend our arguments to operators $\Yk$, $\bar\Yk$ and $\Xi$ in the $\bf 45$, $\overline{\bf45}$ and $\bf 20'$ of $\mathfrak{so}(6)$ respectively. The highest weight state under $\mathfrak{so}(6)$ for each of these operators is
$$\Yk_{2,1,1}\,,\qquad \bar\Yk_{2,1,-1}\,,\quad \text{ and } \ \Xi_{2,0,0}$$
respectively, and so if these operators appear in the OPE $\hat S\times \hat S$ they must appear in
$$\Yk_{2,1,1}\in \hat S_{1,1,0}\times \hat S_{1,0,1}\,,\qquad \bar\Yk_{2,1,-1}\in \hat S_{1,1,0}\times \hat S_{1,0,-1}\,,\quad \text{ and } \quad \Xi_{2,0,0}\in \hat S_{1,1,0}\times \hat S_{1,-1,0}\,.$$
The shortening conditions on $S$ imply that $Q_{1,0,0}^\pm$ annihilates $\hat S_{1,\pm1,0}$ and $\hat S_{1,0,\pm1}$, and so must annihilate $\Yk_{2,1,1}$, $\bar \Yk_{2,1,-1}$ and $\Xi_{2,0,0}$. We can then repeat the analysis previously performed for $\Fk_{2,2,0}$, and recover the same list of multiplets that we found by analyzing the conditions for the operators in the $\bf 84$. We thus conclude that any supermultiplet appearing in $S\times S$ not listed in the first 11 lines of Table~\ref{allowedBlocks} can contain non-zero contributions only from operators in the $\bf 15$ and $\bf 1$.

Restricting the supermultiplets for which only operators in the ${\bf 15}$ and ${\bf 1}$ appear in the $S\times S$ OPE is more subtle and requires the use of superconformal Ward identities.  While we include the details of this analysis in Appendix~\ref{allowedHS}, the result is very simple.  There are only 3 such supermultiplets:  the identity supermultiplet (containing just the identity operator), the stress tensor multiplet itself, as well as a conserved multiplet $(A, \text{cons.})$ whose superconformal primary is an $\mathfrak{so}(6)$ singlet scalar with scaling dimension $1$.

\begin{table}
\begin{center}
{\renewcommand{\arraystretch}{1.2}
\begin{tabular}{ l | l | c | c | c}
Multiplet & \quad$\mathfrak{so}(6)_R$     & $\Delta$   &  $\ell$  & Case     \\ \hline
$(B,2)$   & $[022] = \bf 84$              & $2$        &  $0$     &   \eqref{case1}   \\
$(B,1)$   & $[200] = \bf 20'$             & $2$        &  $0$     &  \eqref{case3}    \\  \hline
$(A,+)$   & $[020] = \bf 10$              & $\ell+2$   &  half-integer & \eqref{case4} \\
$(A,-)$   & $[002] = \overline{\bf 10}$   & $\ell+2$   &  half-integer & \eqref{case4} \\
$(A,2)$   & $[011] = \bf 15$              & $\ell+2$   &  integer  &   \eqref{case4}  \\ 
$(A,1)$   & $[100] = \bf 6$               & $\ell+2$   &  half-integer & \eqref{case5} \\  \hline
$\text{Long}$      & $[000] = \bf 1$               & $>\ell+1$  &  integer  &  \eqref{case6}   \\ \hline
\CR$(B,+)$&\CR$[040] = \bf 35$            &\CR$2$      &\CR$0$  &   \CR\eqref{case1}     \\
\CR$(B,-)$&\CR$[004] = \overline{\bf 35}$ &\CR$2$      &\CR$0$  &  \CR\eqref{case1}      \\
\CR$(B,1)$&\CR$[120] = \bf 45$            &\CR$2$      &\CR$0$      &  \CR\eqref{case2}  \\
\CR$(B,1)$&\CR$[102] = \overline{\bf 45}$ &\CR$2$      &\CR$0$   &   \CR\eqref{case2}   \\  \hline
$(A, \text{cons.})$ & $[000] = \bf 1$               & $\ell+1$   &  integer  &   Appendix~\ref{allowedHS}  \\  
$(B,2)$   & $[011] = \bf 15$              & $1$        &  $0$    & Appendix~\ref{allowedHS}    \\ 
Trivial   & $[000] = \bf 1$               & $0$        &  $0$   &    Appendix~\ref{allowedHS}    \\ %\hline
\end{tabular}}
\caption{Table of superconformal blocks not eliminated by our analysis. The $\mathfrak{so}(6)_R$, $\Delta$ and $\ell$ given the R-symmetry, conformal dimension and spin of the superconformal primary of the exchanged multiplet. The rows in red are for multiplets which we do not eliminate, but for which the superconformal Casimir equation cannot be solved and so no superconformal block exists.}
\label{allowedBlocks}
\end{center}
\end{table}

Table~\ref{allowedBlocks} shows a summary of our analyses containing all the possible supermultiplets which can appear in the $S\times S$ OPE\@. By using the superconformal Casimir equation we shall find that most of these supermultiplets can in fact be exchanged; we mark those that cannot in red.

\subsection{Superconformal Casimir equation}
\label{casimir}

Just as the $s$-channel conformal blocks are eigenfunctions of the quadratic conformal Casimir when the Casimir acts only on the first two operators in a four-point function, superconformal blocks are eigenfunctions of the quadratic superconformal Casimir (see for instance \cite{Bobev:2015jxa,Bobev:2017jhk} for similar discussions with less supersymmetry). In the conformal case, this fact implies that the conformal blocks obey a second order differential equation.  In the superconformal case, the equation obeyed is more complicated because it mixes together four-point functions of operators with different spins.  In the case we are interested in, namely for the four-point function of the stress tensor multiplet superconformal primary, the superconformal Casimir equation involves both the $\langle SSSS\rangle$ four-point function as well as four-point functions of two scalar and two fermionic operators.

To fix conventions, let us denote by $M_\alpha{}^\beta$, $P_{\alpha \beta}$, $K^{\alpha \beta}$, and $D$ the Lorentz generators, the momentum generators, the special conformal generators, and the dilatation generator.  Here $\alpha, \beta = 1, 2$ are spinor indices raised and lowered with the epsilon symbol.  The precise normalization of these operators is fixed by our convention for the conformal algebra, which we give in Appendix~\ref{SUPERCONFORMAL}\@.  It is straightforward to check that in these conventions, the quadratic conformal Casimir 
 \es{ConfCasimir}{
 C_C 
   = \frac 12 M_\alpha{}^{\beta} M_\beta{}^{\alpha}  + D (D-3) -\frac 12  P_{\alpha \beta} K^{\alpha \beta}
 }
commutes with all conformal generators.  The normalization of $M_\alpha{}^\beta$ is such that when acting on an operator of spin $\ell$ placed at $\vec{x} = 0$, the first term in this expression evaluates to $\frac 12 M_\alpha{}^{\beta} M_\beta{}^{\alpha} = \ell(\ell+1)$.  Similarly, when acting on an operator of scaling dimension $\Delta$ also placed at $\vec{x} = 0$, the dilatation operator evaluates to $D = \Delta$.  Since a conformal primary ${\cal O}_{\Delta, \ell}$ of dimension $\Delta$ and spin $\ell$, placed at $\vec{x} = 0$, is annihilated by all  special conformal generators $K^{\alpha \beta}$, it follows that ${\cal O}_{\Delta, \ell}(0)$ is an eigenstate of $C_C$ with eigenvalue $\ell(\ell + 1) + \Delta(\Delta - 3)$.  By conformal symmetry this implies that for any operator ${\cal O}(\vec{x})$ that belongs to a conformal multiplet whose conformal primary has dimension $\Delta$ and spin $\ell$, we have
 \es{ConfEigenvalue}{
  C_C {\cal O} = \lambda_C(\Delta, \ell) {\cal O} \,, \qquad 
   \lambda_C(\Delta, \ell) \equiv \ell(\ell + 1) + \Delta(\Delta - 3) \,.
 }

The discussion in the previous paragraph can be generalized to the superconformal case for a theory with ${\cal N}$-extended superconformal symmetry.   (We will of course set ${\cal N} = 6$ shortly, but let us keep ${\cal N}$ arbitrary for now.) The superconformal algebra is generated by the conformal generators  $M_\alpha{}^\beta$, $P_{\alpha \beta}$, $K^{\alpha \beta}$, and $D$ described above, as well as the Poincar\'e supercharges $Q_{\alpha I}$, the superconformal charges $S^{\alpha I}$, and the R-symmetry generators $R_{IJ}$.  Here, $I = 1, \ldots, {\cal N}$ is an $\mathfrak{so}({\cal N})$ vector index, and $R_{IJ}$ is anti-symmetric.   The normalizations of these operators is fixed by the commutation and anti-commutation relations in Appendix~\ref{SUPERCONFORMAL}\@.  Using these commutation relations, one can check that the quadratic superconformal Casimir
 \es{SuperconfCasimir}{
   C_S = C_C + {\cal N} D  -  \frac 12 C_R      + \frac i2 Q_{\alpha I} S^\alpha_I \,, \qquad C_R \equiv  \frac{1}{2} R_{IJ} R_{IJ}
 }
commutes with all the conformal generators.  Here, the R-symmetry generators are such that when acting on an operator in a representation ${\bf r}$ of $\mathfrak{so}({\cal N})$, we have $C_R = \lambda_R({\bf r})$, where $\lambda_R({\bf r})$ is the eigenvalue of the quadratic Casimir of $\mathfrak{so}({\cal N})$ normalized so that $\lambda_R({\cal N}) = {\cal N} - 1$.  For the case of $\mathfrak{so}(6)$ and the various representations we will encounter, we have the quadratic Casimir eigenvalues in Table~\ref{SO6Casimir}.
\begin{table}[htp]
\begin{center}
\begin{tabular}{c|c}
 irrep ${\bf r}$ of $\mathfrak{so}(6)$ & $\lambda_R({\bf r})$  \\
   \hline
 ${\bf 1}$ & $0$ \\
${\bf 6}$ & $5$ \\
${\bf 15}$ & $8$ \\
${\bf 20}'$ & $12$ \\
${\bf 45}$, $\overline{\bf 45}$ & $16$ \\
${\bf 84}$ & $20$ 
\end{tabular}
\end{center}
\caption{Eigenvalues of $C_R$ in the ${\cal N} = 6$ case where the R-symmetry algebra is $\mathfrak{so}(6)_R$.}
\label{SO6Casimir}
\end{table}%
Eq.~\eqref{SuperconfCasimir} implies that when acting on the superconformal primary operator ${\cal O}_{\Delta, \ell, {\bf r}}$ of spin $\ell$, dimension $\Delta$, and R-symmetry representation ${\bf r}$, placed at $\vec{x} = 0$, the superconformal Casimir gives $\Delta (\Delta + \cN - 3) +  \ell(\ell+1) - \frac{1}{2} \lambda_R ({\bf r}) $; this follows because any such an operator is annihilated by $S^\alpha_I$.  Superconformal symmetry then implies that if ${\cal O}$ is any operator in a superconformal multiplet whose superconformal primary has dimension $\Delta$, spin $\ell$ and R-symmetry irrep ${\bf r}$, we have
 \es{lambdaSuperconf}{
   C_S {\cal O} = \lambda(\Delta, \ell, {\bf r}) {\cal O}\,, \qquad
    \lambda_S( \Delta, \ell, {\bf r}) \equiv \lambda_C(\Delta, \ell) + {\cal N} \Delta - \frac{1}{2} \lambda_R ({\bf r})  \,.
 }

Let us now use the Casimirs above to obtain an equation for the superconformal blocks.  Suppose we have four superconformal primary scalar operators $\phi_i$, $i = 1, \ldots, 4$, of dimension $\Delta_\phi$ and R-symmetry representation ${\bf r}_\phi$.  The four-point function has the conformal block decomposition
 \es{phiFourConf}{
  \langle \phi_1(\vec{x}_1) \phi_2(\vec{x}_2) \phi_3(\vec{x}_3) \phi_4(\vec{x}_4)  \rangle 
   = \frac{1}{\abs{\vec{x}_{12}}^{2 \Delta_\phi}\abs{\vec{x}_{34}}^{2 \Delta_\phi} }
   \sum_{\substack{\text{conf primaries } \\ \text{${\cal O}_{\Delta, \ell, {\bf r}} $ }}}  c_{\Delta, \ell, {\bf r}} \,  g_{\Delta, \ell}(U, V) \,.
 }
A superconformal block corresponding to the supermultiplet ${\cal M}_{\Delta_0, \ell_0}^{{\bf r}_0}$ whose superconformal primary has quantum numbers $(\Delta_0, \ell_0, {\bf r}_0)$ consists of the conformal primary operators in the sum on the RHS of \eqref{phiFourConf} that belong to the same supermultiplet as ${\cal O}_{\Delta_0, \ell_0, {\bf r}_0}$:
 \es{phiFourConfSuper}{
  \langle \phi_1(\vec{x}_1) \phi_2(\vec{x}_2) \phi_3(\vec{x}_3) \phi_4(\vec{x}_4)  \rangle \bigg|_{{\cal M}_{\Delta_0, \ell_0}^{{\bf r}_0}}
   = \frac{1}{\abs{\vec{x}_{12}}^{2 \Delta_\phi}\abs{\vec{x}_{34}}^{2 \Delta_\phi} }
    \sum_{\substack{\text{conf primaries } \\ \text{${\cal O}_{\Delta, \ell, {\bf r}} \in {\cal M}_{\Delta_0, \ell_0}^{{\bf r}_0}$ }}}  c_{\Delta, \ell, {\bf r}} \,  g_{\Delta, \ell}(U, V) \,.
 }
Let us now applying the superconformal Casimir operator \eqref{SuperconfCasimir}, assuming to act only on the first two operators.  To specify which of the four $\phi$'s an operator is acting on, let us use a subscript ``$(12)$'' if the operator is acting on $\phi_1$ and $\phi_2$ and a superscript ``$(i)$'' if the operator acts only on $\phi_i$.   From \eqref{lambdaSuperconf}, we see that
 \es{C12}{
  C_S^{(12)} - C_C^{(12)} + \frac 12 C_R^{(12)} 
   - \sum_{i=1}^2 \left( C_S^{(i)} - C_C^{(i)} + \frac 12 C_R^{(i)}  \right) 
    = \frac {i}{2}  \left( Q_{\alpha I}^{(1)} S_I^{(2) \alpha} +  Q_{\alpha I}^{(2)} S_I^{(1) \alpha}  \right)   \,.
 }
When we apply this expression to \eqref{phiFourConfSuper}, we act with the Casimirs with upper index $(i)$ on the LHS of the equation, and with the ones with upper index $(12)$ on the RHS of the equation---for instance $C_C^{(1)}$ simply gives $\lambda_C(\Delta_\phi, 0)$, while $C_R^{(12)}$ gives $\lambda_R({\bf r})$.   Thus, we obtain the following relation:
 \es{Relation}{
 & \frac{i}{2} \left( \langle (Q_{\alpha I} \phi_1)(\vec{x}_1) (S_I^\alpha \phi_2)(\vec{x}_2) \phi_3(\vec{x}_3) \phi_4(\vec{x}_4)  \rangle
  - \langle (S_I^\alpha \phi_1)(\vec{x}_1) (Q_{I \alpha} \phi_2)(\vec{x}_2) \phi_3(\vec{x}_3) \phi_4(\vec{x}_4)  \rangle 
  \right)  \bigg|_{{\cal M}_{\Delta_0, \ell_0, {\bf r}_0}} 
  \\
   &= \frac{1}{\abs{\vec{x}_{12}}^{2 \Delta_\phi}\abs{\vec{x}_{34}}^{2 \Delta_\phi} }
    \sum_{\substack{\text{conf primaries } \\ \text{${\cal O}_{\Delta, \ell, {\bf r}} \in {\cal M}_{\Delta_0, \ell_0}^{{\bf r}_0}$ }}}  \alpha_{\Delta, \ell, {\bf r}} c_{\Delta, \ell, {\bf r}} \,  g_{\Delta, \ell}(U, V)
 } 
where
 \es{Gotalpha}{
   \alpha_{\Delta, \ell, {\bf r}}  \equiv
      \lambda_S(\Delta_0, \ell_0, {\bf r}_0) - \lambda_C(\Delta, \ell) + \frac 12 \lambda_R ({\bf r}) 
       - 2 {\cal N} \Delta_\phi   \,.
 }
The RHS of Eq.~\eqref{Relation} can be easily evaluated provided we know all the conformal primaries occurring in the multiplet ${\cal M}_{\Delta_0, \ell_0}^{{\bf r}_0}$.  To evaluate the LHS, note that
 \es{SAction}{
  (S^{\alpha}_{I} \phi)(\vec{x}) = x^\mu \gamma_\mu^{\alpha \beta} (Q_{\beta I}  \phi)(\vec{x}) \,,
 }
and so Eq.~\eqref{Relation} becomes
 \es{Relation2}{
 & \frac{i}{2} x_{21}^\mu \gamma_\mu^{\alpha \beta}  
   \langle (Q_{\alpha I} \phi_1)(\vec{x}_1) (Q_{\beta I} \phi_2)(\vec{x}_2) \phi_3(\vec{x}_3) \phi_4(\vec{x}_4)  \rangle  \bigg|_{{\cal M}_{\Delta_0, \ell_0, {\bf r}_0}} \\
  &\qquad \qquad \qquad= \frac{1}{\abs{\vec{x}_{12}}^{2 \Delta_\phi}\abs{\vec{x}_{34}}^{2 \Delta_\phi} }
    \sum_{\substack{\text{conf primaries } \\ \text{${\cal O}_{\Delta, \ell, {\bf r}} \in {\cal M}_{\Delta_0, \ell_0}^{{\bf r}_0}$ }}}  \alpha_{\Delta, \ell, {\bf r}} c_{\Delta, \ell, {\bf r}} \,  g_{\Delta, \ell}(U, V) \,.
 }  
In general, there are Ward identities relating the LHS of \eqref{Relation2} to $\langle \phi_1 \phi_2 \phi_3 \phi_4 \rangle$, but the relations may not be sufficient to determine the LHS of \eqref{Relation2} completely in terms of $\langle \phi_1 \phi_2 \phi_3 \phi_4 \rangle$.

This general discussion can be applied to the case of interest to us, namely the $\langle SSSS \rangle$ correlator in 3d ${\cal N} = 6$ SCFTs\@.  If we replace $\phi_i(\vec{x}_i)$ by $S(\vec{x}_i, X_i)$, then $\langle SSSS \rangle$ can be expanded in R-symmetry channels as in \eqref{SSSScor}, and so can all the equations above.  In particular, we replace $c_{\Delta, \ell, {\bf r}}  \to c^i_{\Delta, \ell, {\bf r}} {\cal B}_i$ in all these equations, with $c^i_{\Delta, \ell, {\bf r}}$, placed in a row vector, determined in terms of the coefficients $a_{\Delta, \ell, {\bf r}}$ defined in \eqref{SDecomp} via
 \es{Gotci}{
    c^i_{\Delta, \ell, {\bf r}} = \begin{pmatrix}a_{\Delta, \ell, {\bf 1}_s} & a_{\Delta, \ell, {\bf 15}_a} &a_{\Delta, \ell,{\bf 15}_s} &a_{\Delta, \ell,{\bf 20}'_s} &a_{\Delta, \ell,{\bf 45}_a \oplus {\overline{\bf 45}}_a} &a_{\Delta, \ell,{\bf 84}_s} \end{pmatrix}  {\bf B}^{-1}  \,,
 }
with ${\bf B}$ defined in \eqref{B}.   Thus \eqref{Relation2} becomes
 \es{Relation2Particular}{
 &\frac{i}{2} x_{21}^\mu \gamma_\mu^{\alpha \beta}  
   \langle Q_{\alpha I} S(\vec{x}_1, X_1) Q_{\beta I} S (\vec{x}_2, X_2) S(\vec{x}_3, X_3) S(\vec{x}_4, X_4)  \rangle  \bigg|_{{\cal M}_{\Delta_0, \ell_0, {\bf r}_0}} \\ 
   &\qquad \qquad \qquad= \frac{1}{\abs{\vec{x}_{12}}^{2}\abs{\vec{x}_{34}}^{2} }
    \sum_{\substack{\text{conf primaries } \\ \text{${\cal O}_{\Delta, \ell, {\bf r}} \in {\cal M}_{\Delta_0, \ell_0}^{{\bf r}_0}$ }}}  \alpha_{\Delta, \ell, {\bf r}} c^i_{\Delta, \ell, {\bf r}}  \,  g_{\Delta, \ell}(U, V) {\cal B}_i \,,
 }  
with $\alpha_{\Delta, \ell, {\bf r}}$ evaluated in this particular case to
\es{GotalphaParticular}{
   \alpha_{\Delta, \ell, {\bf r}}  \equiv
      \lambda_S(\Delta_0, \ell_0, {\bf r}_0) - \lambda_C(\Delta, \ell) + \frac 12 \lambda_R ({\bf r}) 
       - 12   \,.
}

The remaining challenge is to evaluate the LHS of \eqref{Relation2Particular}.  This can be done by noting that $Q_{\alpha I} S(\vec{x}, X)$ is a linear combination of the fermions $\chi$, $F$, and $\overline{F}$ in the stress tensor multiplet, as given in Appendix~D of~\cite{Binder:2019mpb}.  Consequently, the LHS of \eqref{Relation2Particular} can be written in terms of the functions of $(U, V)$ appearing in the correlators $\langle \chi \chi SS \rangle$, $\langle \chi F SS \rangle$, $\langle FFSS \rangle$, and $\langle \overline{F} F SS \rangle$.  These functions are denoted by ${\cal C}^{i, a}$, ${\cal E}^{i, a}$, ${\cal F}^{i, a}$, and ${\cal G}^{i, a}$, respectively, in Appendix~D of~\cite{Binder:2019mpb}.  Here, the index $i$ runs over the R-symmetry structures and the index $a = 1, 2$ runs over the two spacetime structures of a fermion-fermion-scalar-scalar correlator.  Denoting ${\cal X}^{n, a} = ( {\cal F}^{1, a} , {\cal F}^{1, a} , {\cal G}^{1, a}, {\cal G}^{2, a}, {\cal G}^{3, a}, {\cal G}^{4, a},  {\cal E}^{1, a}, {\cal E}^{2, a}, {\cal E}^{3, a},  {\cal C}^{1, a}, {\cal C}^{2, a}, {\cal C}^{3, a}  )$, where $n = 1, \ldots, 12$,  we find
 \es{QSQSSS}{
   \frac{i}{2} x_{21}^\mu \gamma_\mu^{\alpha \beta}  
   \langle Q_{\alpha I} S(\vec{x}_1, X_1) Q_{\beta I} S (\vec{x}_2, X_2) S(\vec{x}_3, X_3) S(\vec{x}_4, X_4)  \rangle 
    =  \frac{  \sum_{i, n} \beta_{i, n} ( {\cal X}^{n, 1}  - \frac{V - U - 1}{2 U} {\cal X}^{n, 2}) {\cal B}_i}{\abs{\vec{x}_{12}}^{2}\abs{\vec{x}_{34}}^{2} }
 }
with the coefficients $\beta_{i, n}$ given by
 \begin{equation}
  \beta_{i, n} =  \begin{pmatrix}
   4 & 4 & -32 & -4 & -4 & 0 & 16 & 16 & 0 & -16  & -128 & -128 \\
   20 & 4 & 0 & 0 & 0 & 4 & 0 & 0 & -16 & 0  & -128 & -128 \\ 
   4 & 20 & 0 & 0 & 0 & 4 & 0 & 0 & 16 & 0  & -128 & -128 \\ 
   -12 & -12 & 0 & 0 & 0 & 4 & 0 & 0 & 0 & 0  & 128 & 128 \\ 
   -4 & -8 & 0 & 2 & -10 & -2 & -8 & -24 & -8 & 0  & 128 & 0 \\ 
   -8 & -4 & 0 & -10 & 2 & -2 & -24 & -8 & 8 & 0  & 0 & 128 
  \end{pmatrix} \,.
  \end{equation}
Thus, Eq.~\eqref{Relation2} reduces to the $6$ equations (one for each $i$):
 \es{Relation2Final}{
 \sum_{n=1}^{12} \beta_{i, n} \left[ {\cal X}^{n, 1}(U, V)  - \frac{V - U - 1}{2 U} {\cal X}^{n, 2}(U, V)\right]    \bigg|_{{\cal M}_{\Delta_0, \ell_0}^{{\bf r}_0}} = 
    \sum_{\substack{\text{conf primaries } \\ \text{${\cal O}_{\Delta, \ell, {\bf r}} \in {\cal M}_{\Delta_0, \ell_0}^{{\bf r}_0}$ }}}  \alpha_{\Delta, \ell, {\bf r}} c^i_{\Delta, \ell, {\bf r}}  \,  g_{\Delta, \ell}(U, V)  \,.
 }  

To use this equation for finding the coefficients $c^i_{\Delta, \ell, {\bf r}} $ of a given superconformal block, one should also expand the fermion-fermion-scalar-scalar correlators on the LHS in conformal blocks corresponding to operators belonging to the supermultiplet ${\cal M}_{\Delta_0, \ell_0}^{{\bf r}_0}$.  Fortunately, we do not have to do this for all 24 functions ${\cal X}^{n, a}$ because, as explained in Appendix~D of~\cite{Binder:2019mpb}, ${\cal C}^{i, a}$, ${\cal E}^{i, a}$, ${\cal F}^{i, a}$, and ${\cal G}^{i, a}$ can be completely determined from ${\cal S}^i$ and ${\cal F}^{1, a}$.  Since we have already expanded the ${\cal S}^i$ in conformal blocks, 
 \es{SiDecomBlocks}{
  {\cal S}^i(U, V)  \Big|_{{\cal M}_{\Delta_0, \ell_0}^{{\bf r}_0}}
   =  \sum_{\substack{\text{conf primaries } \\ \text{${\cal O}_{\Delta, \ell, {\bf r}} \in {\cal M}_{\Delta_0, \ell_0}^{{\bf r}_0}$ }}} 
    c^i_{\Delta, \ell, {\bf r}}  \,  g_{\Delta, \ell}(U, V)  \,,
 }
all that is left to do is to also expand  ${\cal F}^{1, a}$.  

The $s$-channel conformal block decomposition of a fermion-fermion-scalar-scalar four-point function was derived in \cite{Iliesiu:2015qra}.  For each conformal primary being exchanged, there are two possible blocks appearing with independent coefficients.  For ${\cal F}^{1, a}$, if we denote the corresponding coefficients by $ d_{\Delta, \ell, {\bf r}}$ for the first block and $e_{\Delta, \ell, {\bf r}}$ for the second block, we can then write:
 \es{FBlock}{
  \begin{pmatrix}
    {\cal F}^{1, 1}  \\
    {\cal F}^{1, 2} 
    \end{pmatrix}   \bigg|_{{\cal M}_{\Delta_0, \ell_0}^{{\bf r}_0}}
   =  \sum_{\substack{\text{conf primaries } \\ \text{${\cal O}_{\Delta, \ell, {\bf r}} \in {\cal M}_{\Delta_0, \ell_0}^{{\bf r}_0}$ }}} 
    d_{\Delta, \ell, {\bf r}}
     \begin{pmatrix}
      g_{\Delta, \ell} \\
      0 
     \end{pmatrix}
      + e_{\Delta, \ell, {\bf r}}
       \begin{pmatrix}
        {\cal D}_1 g_{\Delta, \ell} \\
        {\cal D}_2 g_{\Delta, \ell} 
       \end{pmatrix}  \,,
 }
where $g_{\Delta, \ell}$ are the scalar conformal blocks appearing above and ${\cal D}_{1, 2}$ are differential operators:
 \es{GotDDiff}{
  {\cal D}_1 &= 2 + 2 U \left[ - 2\partial_V - 2 V \partial_V^2 - \partial_U + 2 U \partial_U^2 \right] \,, \\
  {\cal D}_2 &= 4 U \left[ (V-1)( \partial_V + V \partial_V^2) + U (\partial_U + 2 V \partial_U \partial_V + U \partial_U^2 \right] \,.
 }
(Each doublet of functions $({\cal X}^{n, 1} , {\cal X}^{n, 2} )$ appearing on the LHS of \eqref{Relation2Final} has a similar block decomposition, but as mentioned above, we only need this decomposition for $( {\cal F}^{1, 1} , {\cal F}^{1, 2} )$.)

Using the relations between ${\cal X}^{n, a}$ and ${\cal S}^i$ and ${\cal F}^{1, a}$ given in Appendix~D of~\cite{Binder:2019mpb} together with the decompositions \eqref{SiDecomBlocks} and \eqref{FBlock}, we obtain a system of linear equations for $c^i_{\Delta, \ell, {\bf r}}$, $d_{\Delta, \ell, {\bf r}}$, and $e_{\Delta, \ell, {\bf r}}$ that has to be obeyed for all values of $(U, V)$.  Expanding $g_{\Delta, \ell}$ to sufficiently high orders in $U$ is then enough to determine the linearly-independent solutions of this system of equations, and thus determine the coefficients $c^i_{\Delta, \ell, {\bf r}}$ of the superconformal block corresponding to the supermultiplet ${\cal M}_{\Delta_0, \ell_0}^{{\bf r}_0}$.  

We performed this analysis for all the multiplets described in Table~\ref{allowedBlocks}.  The coefficients $c^i_{\Delta, \ell, {\bf r}}$ for each multiplet are included in the attached \texttt{Mathematica} notebook.  The multiplets marked in red in Table~\ref{allowedBlocks} did not give solutions to the system of equations that determines the $c^i_{\Delta, \ell, {\bf r}}$. For each of the remaining multiplets we found between one and three solutions.  Since any linear combination of superconformal blocks is a superconformal block, we are free to choose a basis of blocks with specific normalizations.  In other words, for the coefficients $a_{\Delta, \ell, {\bf r}}$ in \eqref{SDecomp} can be written as
 \es{aCoeffExpansion}{
  a_{\Delta, \ell, {\bf r}} = \sum_{I}  \lambda_{I}^2 a^{I}_{\Delta, \ell, {\bf r}}\,,
 }
where $I$ ranges over all superconformal blocks, $\lambda_I^2$ are theory-dependent coefficients, and  $a^{I}_{\Delta, \ell, {\bf r}}$ represent the solution to the super-Casimir equation for superconformal block $I$, normalized according to our choosing.  In Table~\ref{SupermultipletTable}, we list all the superconformal blocks as well as enough values for $a^{I}_{\Delta, \ell, {\bf r}}$ in order to determine the normalization of the blocks.\footnote{The $(A, +)$ and $(A, -)$ multiplets are each other's complex conjugates and they must appear together in the $S \times S$ OPE\@.} A superconformal block $\mathfrak{G}_I$ is simply
 \es{SuperconfBlock}{
  \mathfrak{G}_I^{\bf r}(U, V)= \sum_{\substack{\text{conf primaries } \\ \text{${\cal O}_{\Delta, \ell, {\bf r}} \in {\cal M}_{\Delta_0, \ell_0}^{{\bf r}_0}$ }}} 
    a^I_{\Delta, \ell, {\bf r}}  \,  g_{\Delta, \ell}(U, V) \,, \qquad
     I \equiv {{\cal M}_{\Delta_0, \ell_0}^{{\bf r}_0, n} } \,,
 }
where the index $I = {{\cal M}_{\Delta_0, \ell_0}^{{\bf r}_0, n} }$ of the block encodes both the supermultiplet $ {{\cal M}_{\Delta_0, \ell_0}^{{\bf r}_0} }$ as well as an integer $n = 1, 2, \ldots$ denoting which block this is according to Table~\ref{SupermultipletTable}.  (In the cases where there is a single superconformal block per multiplet, we omit the index $n$.)

As discussed in Appendix~B.3 of~\cite{Binder:2019mpb}, the stress-tensor multiplet forms a representation not only of the superconformal group $OSp(6|4)$, but also of a larger group $(\mathbb Z_2\times\mathbb Z_2)\ltimes OSp(6|4)$ which includes both a parity transformation $\cP$ and discrete R-symmetry transformation $\cZ$. The parity transformation $\cP$ extends the spacetime symmetries from $Spin(3, 2) \cong Sp(4, \R)$ to $Pin(3,2)$, while $\cZ$ extends the R-symmetry group from $SO(6)$ to $O(6)$.  In any local CFT the scalar three-point function $\<SSS\>$ is non-zero, which implies that in a $\cZ$-invariant theory the operator $S$ transforms as a pseudotensor ${\bf 15}^-$, while the supercharges transform as $O(6)$ vectors. 

Table \ref{SupermultipletTable} includes the ${\cal P}$ and ${\cal Z}$ charges relative to that of the superconformal primary, which are relevant for $\mathcal{N}=6$ theories that are invariant under these discrete symmetries. We derive $\cP$ charges for each superblock by noting that any two primaries $\fcy O_{\Dk,\ell}$ and $\fcy O_{\Dk',\ell'}$ in a supermultiplet have the same parity if and only if $\Dk'-\Dk\equiv\ell'-\ell\mod 2$. To derive the $\cZ$ charges we use the $O(6)$ tensor product of two pseudo-tensors:
\begin{equation}
{\bf 15}^- \otimes {\bf 15}^- = {\bf 1}^+_s \oplus {\bf 15}^+_a \oplus {\bf 15}^-_s \oplus {{\bf 20}'}^+_s \oplus {\bf 90}_a \oplus {\bf 84}_s^+\,.
\end{equation}

\begin{table}
\begin{center}
{\renewcommand{\arraystretch}{1.4}
\begin{tabular}{ l | c |   l|l }
Superconformal block    & {normalization} & $\mathcal{P}$ & $\mathcal{Z}$   \\  \hline
\multirow{2}{*}{$\text{Long}_{\Delta, 0}^{[000],n}$}  & $n=1:\quad(a_{\Delta, 0, {\bf 1}}, a_{\Delta+1, 0, {\bf 20'}})  = (1, 0)$ &  $+$ &  $+$    \\ 
 &    $n=2:\quad (a_{\Delta, 0, {\bf 1}}, a_{\Delta+1, 0, {\bf 20'}})  = (0, 1)$   &  $-$ &  $+$  \\ 
 \hline
$\text{Long}_{\Delta, \ell}^{[000]}$, $\ell \geq 1$ odd    &   $a_{\Delta+1, \ell+1, {\bf 15}_s} = 1$  & $+$ & $+$ \\ 
\hline
\multirow{3}{*}{$\text{Long}_{\Delta, \ell}^{[000],n}$, $\ell \geq 2$ even}    & $n=1:\quad(a_{\Delta, \ell, {\bf 1}}, a_{\Delta+1, \ell, {\bf 1}}, a_{\Delta+1, \ell, {\bf 15}_s})  = (1, 0, 0)$ &  $+$ &  $+$  \\ 
  & $n=2:\quad(a_{\Delta, \ell, {\bf 1}}, a_{\Delta+1, \ell, {\bf 1}}, a_{\Delta+1, \ell, {\bf 15}_s})  = (0, 1, 0)$&  $-$ &  $+$ \\ 
&   $n=3:\quad(a_{\Delta, \ell, {\bf 1}}, a_{\Delta+1, \ell, {\bf 1}}, a_{\Delta+1, \ell, {\bf 15}_s})  = (0, 0, 1)$&  $-$ &  $-$  \\ 
 \hline
 \multirow{2}{*}{$(A, 1)_{\ell + 2, \ell }^{[100],n}$, $\ell - \frac 12 \geq 1$ odd}
 & $n=1:\quad(a_{\ell + \frac 52, \ell + \frac 12, {\bf 1}}, a_{\ell + \frac 52, \ell + \frac 12, {\bf 15}_s} ) = (1, 0)$  &$+$  & $+$  \\
  &  $n=2:\quad(a_{\ell + \frac 52, \ell + \frac 12, {\bf 1}}, a_{\ell + \frac 52, \ell + \frac 12, {\bf 15}_s} ) = (0, 1)$ &$+$&$-$ \\
 \hline
$(A, 2)_{\ell + 2, \ell}^{[011]}$, $\ell \geq 0$ even & $a_{\ell+2, \ell, {\bf 15}_s} = 1$ & $+$ &  $-$ \\
\hline
$(A, 2)_{\ell + 2, \ell}^{[011]}$, $\ell \geq 0$ odd  & $a_{\ell+2, \ell, {\bf 15}_a} = 1$ & $+$ &  $+$ \\
\hline 
$(A, +)_{\ell + 2, \ell }^{[020]}$, $\ell - \frac 12 \geq 0$ even  & $a_{\ell+\frac 52, \ell + \frac 12, {\bf 15}_a} = 1$ & $+$ &  \\
\hline 
$(A, -)_{\ell + 2, \ell }^{[002]}$, $\ell - \frac 12 \geq 0$ even  & $a_{\ell+\frac 52, \ell + \frac 12, {\bf 15}_a} = 1$ & $+$ &  \\
\hline
$(A, \text{cons})_{\ell + 1, \ell}^{[000]}$, $\ell \geq 0$ even   & $a_{\ell + 1, \ell, {\bf 1}} = 1$ & $+$ &  $+$ \\
\hline
$(A, \text{cons})_{\ell + 1, \ell}^{[000]}$, $\ell \geq 1$ odd   & $a_{\ell + 2, \ell+1, {\bf 15}_s} = 1$ & $+$ &  $-$ \\
\hline
$(B, 1)_{2, 0}^{[200]}$  & $a_{2, 0, {\bf 20}'} = 1$ & $+$ &  $+$\\
\hline
$(B, 2)_{2, 0}^{[022]}$   & $a_{2, 0, {\bf 84}} = 1$& $+$ &  $+$ \\
\hline 
$(B, 2)_{1, 0}^{[011]}$  & $a_{1, 0, {\bf 15}_s} = 1$ & $+$ &  $-$
\end{tabular}}
\caption{A summary of the superconformal blocks and their normalizations in terms of a few OPE coefficients.  The values $a_{\Delta, \ell, {\bf r}}$ in this table correspond to $a^I_{\Delta, \ell, {\bf r}}$ in Eq.~\eqref{aCoeffExpansion}---we omitted the index $I$ for clarity. Note that the $(A,\pm)$ are complex conjugates and do not by themselves have well defined $\mathcal Z$ parity, but together they can be combined into a $\mathcal Z$-even and a $\mathcal Z$-odd structure.}
\label{SupermultipletTable}
\end{center}
\end{table}

Reflection positivity implies that the coefficients $a_{\Delta, \ell, {\bf r}}$ in \eqref{SDecomp} are non-negative for all~${\bf r}$. Because for each superconformal block in Table~\ref{SupermultipletTable} there exists an operator that receives contributions only from that block, it follows that the coefficients $\lambda_{I}^2$ in \eqref{aCoeffExpansion} are non-negative.  This is the reason why we wrote these coefficients in \eqref{aCoeffExpansion} manifestly as perfect squares.  They are the squares of real OPE coefficients.\footnote{In other words, for each multiplet for which there are several superconformal blocks, the number of superconformal 3-point structures equals the number of superconformal blocks.  This is so because each superconformal 3-point structure contains different operators from the exchanged multiplet.  }

Let us end this section by describing the unitarity limits of the long blocks obtained by taking $\Delta \to \ell + 1$.  For the scalar blocks, we obtain (up to normalization) either a spin-$0$ conserved block for the parity-even structure or a $(B,1)_{2,0}^{[200]}$ block for the parity odd structure:
 \es{LimitSpin0}{
  \text{Long}_{\Delta, 0}^{[000],1} \to (A, \text{cons})_{1,  0}^{[000]} \,, \qquad
   \text{Long}_{\Delta, 0}^{[000],2} \to (B,1)_{2,0}^{[200]} \,.
 }
For odd $\ell \geq 1$ there is a single block and it approaches a spin-$\ell$ conserved block: 
\es{LimitOddSpin}{
 \ell \geq 1 \text{ odd:} \qquad \text{Long}_{\Delta, \ell}^{[000]} \to (A, \text{cons})_{\ell+1,  \ell}^{[000]} \,.
}
Lastly, for even $\ell \geq 2$ we have three superconformal blocks.  The parity even one approaches a spin-$\ell$ conserved block, while the parity odd ones approach the two superconformal blocks for the $(A,1)_{\ell+3/2,\ell-1/2}^{[100]}$ multiplet:
 \es{LimitSpinEven}{
  \ell \geq 2 \text{ even:} \qquad \text{Long}_{\Delta, \ell}^{[000],1} &\to (A, \text{cons})_{\ell+1,  \ell}^{[000]} \,,\\
   \text{Long}_{\Delta, \ell}^{[000],2} &\to (A,1)_{\ell+3/2,\ell-1/2}^{[100], 1} \,, \\ 
      \text{Long}_{\Delta, \ell}^{[000],3} &\to (A,1)_{\ell+3/2,\ell-1/2}^{[100], 2} \,.
}
Even though the blocks on the RHS of \eqref{LimitSpin0}--\eqref{LimitSpinEven} involve short or semishort superconformal multiplets, they sit at the bottom of the continuum of long superconformal blocks. All other short and semishort superconformal blocks are isolated, as they cannot recombine into a long superconformal block. In particular, if the correlator $\<SSSS\>$ contains one of these isolated superconformal blocks, any sufficiently small deformation of $\<SSSS\>$ also must, while the other blocks can instead disappear by recombining into a long block.  This distinction will be important when we consider the numerical bootstrap.

\subsection{Examples:  GFFT and free ${\cal N} = 6$ hypermultiplet}
\label{EXAMPLES}

There are two theories for which we can determine the superconformal block decomposition and all CFT data. The first is the generalized free field theory (GFFT), where correlators of $S$ are computed using Wick contractions with the propagator $\langle S(\vec{x}_1, X_1) S(\vec{x}_2, X_2) \rangle = \frac{\tr (X_1 X_2)}{\abs{\vec{x}_{12}}}$.  In terms of the functions $\mathcal{S}^i(U,V)$, the $\<SSSS\>$ correlator is:
 \es{SiGFFT}{
 \cS^i_\text{GFFT}(U, V) =  \begin{pmatrix} 1 & U & \frac UV &0&0&0\end{pmatrix} \,.
 }
This theory does not have a stress-energy tensor, and it is thus non-local and therefore not of primary interest here.  However, the GFFT four-point function does represent the leading term in the strong-coupling limit of correlators of the local SCFTs that are discussed in the next section.  If we think about the $S$ operators as single-trace, then in the superconformal block decomposition of $\langle SSSS \rangle$ only double-trace operators appear, with schematic form $S \partial_{\mu_1} \cdots \partial_{\mu_\ell} \square^p S$ of spin $\ell$ dimension $2 + \ell + 2p$ with positive integer $\ell$ and $p$.   The GFFT defined as above does not necessarily have ${\cal N} = 6$ supersymmetry, but it can be completed into an ${\cal N} = 6$ preserving theory by considering similar rules for calculating correlators of any four stress-tensor multiplet operators from Table~\ref{stressTable}. We can expand \eqref{SiGFFT} in superconformal blocks to read off the CFT data given in Table~\ref{freeTable}.

The second theory that is exactly computable is a free theory.  Let us consider four complex scalar $\phi_a$, $a = 1, \ldots,4$ and their complex conjugates $\bar \phi^a$, with the two-point function normalized as $\langle \phi_a(\vec{x}_1) \phi^b(\vec{x}_2) \rangle = \frac{\delta_a^b}{\abs{\vec{x}_{12}}}$.  In this theory, we can consider the operator $S(\vec{x}, X) = \phi_a(\vec{x}) \bar \phi^b(\vec{x}) X_b{}^a$.  The $\langle SSSS \rangle$ correlator can then be computed using Wick contractions of the $\phi$ and $\bar \phi$'s, and in terms of the $\cS^i$ it is given by 
  \es{FreeHypers}{
  \cS^i_\text{free}(U, V)  = \begin{pmatrix} 1 & U & \frac UV & \frac{U}{\sqrt{V}} & \frac{\sqrt{U}}{\sqrt{V}}
   &  \sqrt{U} \end{pmatrix} \,.
 }
As was the case with the GFFT, this correlator does not necessarily correspond to an ${\cal N} = 6$ SCFT, but it can be embedded in one by considering the $\phi_a$ as the components of an ${\cal N} = 6$ hypermultiplet that also contains $4$ complex fermions.  The four-point function \eqref{FreeHypers} can then be expanded into superconformal blocks to give the CFT data given in Table~\ref{freeTable}. Note that the free theory has the same spectrum as the GFFT, except that it also contains conserved current multiplets for each spin, has a stress tensor multiplet, and does not have a $(B, 1)_{2, 0}^{[200]}$ multiplet.

\begin{table}[htp]
\begin{center}
{\renewcommand{\arraystretch}{1.2}
\begin{tabular}{l||c|c}
 & $\cS^i_\text{GFFT}$ & $\cS^i_\text{free}$   \\ 
\hline\hline
$\lambda^{2}_{(B,2)^{[011]}_{1,0}}$ & $0$ & $4$   \\
\hline
$\lambda^{2}_{(B,2)^{[022]}_{2,0}}$ & $2$ & $4$   \\
\hline
$\lambda^{2}_{(B,1)^{[200]}_{2,0}}$ & $\frac43$ & $0$   \\
\hline
$\lambda^{2}_{(A,\text{cons.})^{[000]}_{\ell+1,\ell}}$ for $\ell = 0, 1, 2, \ldots$ & $0$ & $4$  \\
\hline
$\lambda^{2}_{(A,2)^{[011]}_{\ell+2,\ell}}$ for $\ell = 0, 1, 2, \ldots$  & 
   $\frac{16}{3}$, $\frac{256}{45}$, $\frac{4096}{525}$, $\frac{32768}{3675}$, \ldots
   & $\frac83$, $\frac{32}{5}$, $\frac{3712}{525}$, $\frac{34304}{3675}$, \ldots  \\
\hline$\lambda^{2}_{(A,+)^{[002]}_{\ell + 5/2,\ell + 1/2}}$ for $\ell = 0, 2, \ldots$ & $\frac{16}{9}$, $\frac{6144}{1225}$, \ldots  
 & $\frac83$, $\frac{7872}{1225}$, \ldots   \\
\hline$\lambda^{2}_{(A,1)_{\ell + 7/2,\ell + 3/2}^{[100], 1}}$ for $\ell = 0, 2, \ldots$ & $\frac{512}{315}$, \ldots & $\frac{64}{63}$, \ldots  \\
\hline$\lambda^{2}_{(A,1)_{\ell + 7/2,\ell + 3/2}^{[100], 2}}$ for $\ell = 0, 2, \ldots$ & $\frac{1024}{105}$, \ldots & $\frac{128}{21}$, \ldots   \\
\hline$\Delta_{(0,1)}$ &  $2$, $4$, \ldots & $2$, $4$, \ldots   \\
\hline$\Delta_{(0,2)}$ & $3$, $5$, \ldots & $3$, $5$, \ldots  \\
\hline$\Delta_{\ell\geq 1}$, $\ell$ odd  & $\ell+2$, $\ell + 4$, \ldots & $\ell+2$, $\ell + 4$, \ldots   \\
\hline$\Delta_{(\ell \geq 2,1)}$, $\ell$ even  & $\ell+2$, $\ell + 4$, \ldots & $\ell+2$, $\ell + 4$, \ldots \\
\hline$\Delta_{(\ell \geq 2, 2)}$, $\ell$ even & $\ell+3$, $\ell + 5$, \ldots & $\ell+3$, $\ell + 5$, \ldots \\
\hline$\Delta_{(\ell \geq 2,3)}$, $\ell$ even & $\ell+3$, $\ell + 5$, \ldots & $\ell+3$, $\ell + 5$, \ldots  \\
\end{tabular}}
\caption{Low-lying CFT data for the generalized free field theory (GFFT) $\cS^i_\text{GFFT}$ and the free theory $\cS^i_\text{free}$. We write $\Delta_{(\ell,n)}$ to denote the scaling dimension of the superblock corresponding to the structure $\text{Long}_{\Delta, \ell}^{[000], n}$.}
\label{freeTable}
\end{center}
\end{table}

For both the GFFT theory and the free theory of an ${\cal N} = 6$ hypermultiplet, one can alternatively obtain the CFT data listed in Table~\ref{freeTable} by performing a decomposition of the correlators in the analogous ${\cal N} = 8$ SCFTs, as described in Appendix~\ref{8to6}.  Indeed, the ${\cal N} = 6$ GFFT is a subsector of the ${\cal N} = 8$ GFFT, where the ${\cal N} = 6$ stress tensor multiplet is embedded into the ${\cal N} = 8$ stress tensor multiplet.  Similarly, the theory of an ${\cal N} = 6$ free hypermultiplet has ${\cal N} = 8$ supersymmetry because a free ${\cal N} = 6$ hypermultiplet is identical in field content with an ${\cal N} = 8$ hypermultiplet: they both consist of eight real scalars and eight Majorana fermions.

\section{Exact results in ${\cal N} = 6$ SCFTs}
\label{EXACT}

\subsection{Known ${\cal N} = 6$ SCFTs}

$\mathcal{N}=6$ CFTs with Lagrangians were classified in \cite{Schnabl:2008wj}, up to discrete quotients that do not affect correlators of $S$.\footnote{See \cite{Tachikawa:2019dvq} for a conjectured classification that takes into account discrete quotients.} In ${\cal N } = 3$ SUSY notation, they are Chern-Simons-matter theories with two matter hypermultiplets. There are two possible families of gauge groups and representations:\footnote{The case $SU(N)_k\times SU(N)_{-k}$ describes the BLG theories \cite{Bagger:2007vi,Bagger:2007jr,Gustavsson:2007vu}.}
\es{families1}{
& SU(N)_k\times SU(N+M)_{-k}\times U(1)_K^L\,,\qquad  K^{ab}q_aq_b=\frac{1}{k}\left(\frac{1}{M+N}-\frac1N\right)\,,
}
for $N,M\geq1$ where the hypermultiplets are in the bifundamental of $SU(M)\times SU(N)$, and 
\es{families2}{
& USp(2+2M)_k\times  U(1)_K^L\,,\qquad\qquad\qquad  K^{ab}q_aq_b=-\frac{1}{2k}\,,\\
}
for $M\geq0$ where the hypermultiplets are in the fundamental of $USp(2 + 2M)$. In both cases, the hypermultiplets have equal and opposite charges $q_i$ for $i=1,\dots, L$ under the $U(1)$'s.  The matrix $K^{ab}$ is the inverse of the matrix $K_{ab}$ of Chern-Simons levels for the $L$ $U(1)$ gauge groups, and must satisfy the relations given in \eqref{families1} and \eqref{families2}.  Note that when $N=1$ in \eqref{families1}, the hypermultiplets are just in the fundamental of $SU(1+M)$ with appropriate charges under the $U(1)$'s. 

As we shall show in Appendix~\ref{AddU1s}, the $S^3$ partition function for both families of theories is independent of $L$, as long as the conditions in \eqref{families1} and \eqref{families2} are obeyed, up to an overall normalization constant. This leads us to conjecture that all these theories have the same $S$ correlators, so for this sector we only need consider two families of theories. One is the ABJ(M) family\footnote{When $N=1,M=0$, the ABJM theory describes a free SCFT equivalent to the theory of eight massless real scalars and eight Majorana fermions described in Section~\ref{EXAMPLES}. For $M=0$ and $N>1$, ABJM flows to the product of a free SCFT and a strongly-coupled SCFT, while for all other parameters ABJM theory has a unique stress tensor.}
\es{families1a}{
& U(N)_k\times U(N+M)_{-k}\,,\\
}
with $M \leq \abs{k}$ \cite{Aharony:2008ug,Aharony:2008gk}, which is the special case of \eqref{families1} where $L = 2$ with $q_1=q_2 = 1$ and $K_{11} = k N$, $K_{22} = -k(N+M)$, and $K_{12} = 0$. The other family is
\es{families2a}{
&SO(2)_{2k}\times USp(2+2M)_{-k}\,,\\
}
with $M+1\leq\abs{k}$ \cite{Hosomichi:2008jb,Aharony:2008gk}, is the $L = 1$, $q = 1$ case of \eqref{families2}. Sending $k \to -k$ gives a parity-conjugate theory, so without loss of generality we can focus on $k>0$. Seiberg duality imposes additional equivalences between each family:
\es{seiberg}{
U(N)_k \times U(N+M)_{-k} \ \ \ &\longleftrightarrow \ \ \ U(N)_{-k} \times U(N+\abs{k}-M)_{k}\,,\\
SO(2)_{2k} \times USp(2+2M)_{-k} \ \ \ &\longleftrightarrow \ \ \ SO(2)_{-2k} \times USp(2(\abs{k}-M-1)+2)_{k} \,.\\
}
In particular, the $k=2M$ case of ABJM is parity invariant, as is the $k+1=2M$ case of the $SO(2)\times USp(2+2M)$ theory.

The $U(N)_k \times U(N+M)_{-k}$ ABJ theories can be interpreted as effective theories on $N$ coincident M2-branes placed at a $\mathbb{C}^4/\mathbb{Z}_k$ singularity in the transverse directions, together with a discrete flux due to $M$ fractional M2-branes localized at the singularity. The $N\gg k^5$ limit is described by weakly coupled M-theory on $AdS_4\times S^7/\mathbb{Z}_k$, while the large $N,k$ limit with $N\sim k$ and finite $M$ is described by weakly coupled Type IIA string theory on $AdS_4\times \mathbb{CP}^3$.   When $M,k$ are large and $N$ is finite, ABJ theory becomes a vector model with weakly-broken higher-spin symmetry, which is dual to $\mathcal{N}=6$ higher-spin Vasiliev theory on $AdS_4$ \cite{Chang:2012kt}. The finite coupling $\lambda\equiv M/k$ with $\lambda\leq1$ is related to the parity breaking $\theta_0$ parameter in Vasiliev theory, such that $\lambda=0,1$ correspond to the free theory and $\lambda=1/2$ corresponds to a parity-invariant theory. The $SO(2)_{2k}\times USp(2+2M)_{-k}$ case in the large $M,k$ limit with fixed $\lambda=M/k$ is also a vector model dual to $\mathcal{N}=6$ higher-spin Vasiliev theory on $AdS_5$ with similar properties \cite{Honda:2017nku}.

\subsection{Exactly calculable CFT data}
\label{S3part}

In the next section, we will derive numerical bounds on the CFT data of 3d $\mathcal{N}=6$ SCFTs parameterized in terms of $c_T$, which is defined as the coefficient appearing in the two-point function of the canonically-normalized stress-tensor, 
 \es{CanStress}{
  \langle T_{\mu\nu}(\vec{x}) T_{\rho \sigma}(0) \rangle = \frac{c_T}{64} \left(P_{\mu\rho} P_{\nu \sigma} + P_{\nu \rho} P_{\mu \sigma} - P_{\mu\nu} P_{\rho\sigma} \right) \frac{1}{16 \pi^2 \vec{x}^2} \,, \qquad P_{\mu\nu} \equiv \eta_{\mu\nu} \nabla^2 - \partial_\mu \partial_\nu \,,
 }
and, in the normalization \eqref{2pS} for the external operator $S(\vec{x}, X)$, it is inversely related to the square of the OPE coefficient of the stress tensor multiplet:
\es{cT}{
  c_T = \frac{64}{\lambda^2_{(B,2)_{1,0}^{[011]}}}\,.
}
Here $c_T$ is defined such that it equals $1$ for a (non-supersymmetric) free massless real scalar or a free massless Majorana fermion. Hence $c_T = 16$ for the free ${\cal N} = 6$ hypermultiplet described in Section~\ref{EXAMPLES}, which also has $\mathcal{N}=8$ and is equivalent to ABJM theory with $M=0$ and $N=1$.

This $c_T$ is a particularly useful parameterization of physical theories. It can be computed exactly using supersymmetric localization for any ${\cal N} \geq 2$ SCFT with a Lagrangian description by taking two derivatives of the squashed sphere partition function with respect to the squashing parameter \cite{Closset:2012ru,Imamura:2011wg}.  For theories with at least ${\cal N} = 4$ supersymmetry, the stress tensor multiplet contains R-symmetry currents that from an ${\cal N} = 2$ point of view are flavor currents, and one can argue that $c_T$ is proportional to the two-point functions of such flavor currents \cite{Chester:2014fya}.  Such two-point functions can be computed by taking two derivatives of the round sphere partition function with respect to a mass parameter \cite{Closset:2012vg}.  For the ${\cal N} = 6$ theories that we focus on here, we will define a mass parameter $m$ with a normalization such that 
\es{cTFormula}{
c_T = -\frac{64}{\pi^2 }\frac{\nb^2 \log Z}{\nb m^2}\bigg|_{m=0}\,.
}
In these theories, supersymmetric localization \cite{Kapustin:2009kz} implies that the quantity $Z(m)$ can be expressed as an $N$-dimensional integral for any $k,M$,\footnote{For the orthogonal group case, it is always a one-dimensional integral.} and can be evaluated exactly at small $N$ and to all orders in $1/N$ for $M\leq k\leq N$  using the Fermi gas method \cite{Marino:2011eh,Nosaka:2015iiw,Agmon:2017xes}. In particular, for $N=1$ we can exactly compute the one-dimensional integral for any $M,k$, such as the large $M\sim k$ limit that describes the vector like limit, which can also be computed in a large $M$ expansion to any order as described in \cite{Hirano:2015yha}. For the various quantum gravity theories we discuss the leading order expressions for $c_T$ are then \cite{Chester:2014fya,Honda:2015sxa}:
\es{cTlargeN}{
&\text{M-theory, Type IIA string theory}:\qquad  c_T \approx \frac{64}{3\pi}\sqrt{2k}N^{3/2}\,,\\
&\text{Higher-spin theory}:\qquad\qquad\qquad\qquad\;\,\, c_T \approx \frac{16Nk}{\pi}\sin(\lambda\pi)\,,\\
}
so that M-theory (at finite $k$) scales like $N^{3/2}$, string theory (where $k\sim N$) has the typical matrix-like scaling $N^2$, while the higher-spin theory (where $k\sim M$ for finite $N$) has the typical vector-like linear scaling. 

We can similarly compute the OPE coefficient squared $\lambda^2_{(B,2)_{2,0}^{[022]}}$ by taking four derivatives of the mass deformed free energy as described in \cite{Agmon:2017xes,Chester:2018aca,Binder:2018yvd,Binder:2019mpb}, to get analytical results in the same range of $M,N,k$ as described for $c_T$ (see Eq.~(3.27) of \cite{Binder:2019mpb}):
\es{locB2}{
 \lambda^2_{(B,2)_{2,0}^{[022]}} = 2 +  \frac{ \partial_m^4 \log Z }{(\partial_m^2 \log Z)^2} \bigg|_{m=0}\,.
}
This short OPE coefficient is the $\mathcal{N}=6$ analogue of the $\mathcal{N}=8$ short OPE coefficient computed in \cite{Agmon:2017xes}.

We should point out that in the limit in which $\log Z$ and its mass derivatives go to infinity, we have $c_T \to \infty$ and $\lambda^2_{(B,2)_{2,0}^{[022]}} \to 2$.  This is expected because in this limit the CFT correlators factorize, and the $\langle SSSS \rangle$ correlator is that of the GFFT theory described in Section~\ref{EXAMPLES}.  Indeed, as can be seen from Table~\ref{freeTable} GFFT has $\lambda^2_{(B, 2)^{[011]}_{1, 0}} = 0$, corresponding to $c_T = \infty$, and $\lambda^2_{(B, 2)^{[022]}_{2, 0}} = 2$.

In the rest of this section we shall focus on computing $c_T$ and $\lambda^2_{(B,2)_{2,0}^{[022]}}$ in both the $U(N)_k\times U(N+M)_{-k}$ and $SO(2)_{2k}\times USp(2+2M)_{-k}$ theories. We will compute these quantities exactly for finite $M$ and $N$, as well as in the large $M$ expansion at fixed $\lambda=\frac{M}{k}$ and $N=1$, which describes the higher-spin limit of these theories. For the $U(N)_k\times U(N+M)_{-k}$ theory, we will also review the all orders in $1/N$ results at finite $M\leq k\leq N$ in \cite{Agmon:2017xes}, which describes the M-theory limit for finite $k,M$ and Type IIA string theory limit for $k\sim N$ and finite $M$.  As noted previously and proven explicitly in appendix \ref{AddU1s}, the sphere partition function $Z(m)$ is left unchanged by the presence of additional $U(1)$ factors, and so our results also hold for the more general families of $\cN=6$ theories \eqref{families1} and \eqref{families2}.

\subsection{$U(N)_k\times U(N+M)_{-k}$ theory}
\label{ABJOPES}
Using supersymmetric localization, the mass-deformed $U(N)_k\times U(N+M)_{-k}$ partition function can be reduced to $M+2N$ integrals \cite{Kapustin:2009kz,Hama:2011ea}:
\begin{equation}\begin{split}\label{ABJv1}
Z_{M,N,k}&(m)  \\
&=\int d^{M+N}\mu\, d^N\nu\  \frac{e^{-i\pi k(\sum_i \mu_i^2 - \sum_a \nu_a^2)}\prod_{i<j}4\sinh^2\left[\pi(\mu_i-\mu_j)\right]\prod_{a<b}4\sinh^2\left[\pi(\nu_a-\nu_b)\right]}{\prod_{i,a}4\cosh\left[\pi(\mu_i-\nu_a)+\frac{\pi m}2\right]\cosh[\pi(\mu_i-\nu_a)]}\,,
\end{split}\end{equation}
up to an overall $m$-independent normalization factor. Our first task will be to write \eqref{ABJv1} as an $N$-dimensional integral that we can then evaluate more easily. For the massless case $m = 0$ such a reduction was achieved in \cite{Honda:2013pea}. In Appendix~\ref{simpABJ}, we extended their methods to the massive partition function \eqref{ABJv1} and show that
\begin{equation}\label{ABJv2}\begin{split}
Z_{M,N,k}(m)
= Z_0 e^{-\frac \pi2MN m}&\int d^Ny\prod_{a<b} \tanh^2 \frac{\pi(y_a-y_b)}{k}\\
&\times\prod_{a=1}^N\left[\frac{e^{i\pi y_am}}{2\cosh\left(\pi y_a\right)}\prod_{l=0}^{M-1}\tanh \frac{\pi\big(y_a+i(l+1/2)\big)}{k} \right]\,,
\end{split}\end{equation}
where $Z_0$ is again an overall factor which is independent of the mass parameter $m$. Since our interest is ultimately in computing derivatives of $\log Z$ with respect to $m$, the value of $Z_0$ is unimportant.  Let us now discuss various limits in which we can evaluate \eqref{ABJv2} exactly or approximately.

\subsubsection{Small $M,N,k$}
\label{SMALLMNK}

When $M,N,k$ are small integers, we can evaluate \eqref{ABJv2} as contour integrals. Let us begin with the case $N = 1$. We must compute
\begin{equation}\label{ABJN1}\begin{split}
\hat Z_{M,1,k}(m) &\equiv \frac{Z_{M,1,k}(m)}{Z_0}= e^{-\frac \pi2Mm}\int_{-\infty}^{\infty} dx\ e^{i\pi xm} F_{M,k}(x) \,,
\end{split}\end{equation}
where we define
\begin{equation}
F_{M,k}(x) = \frac1{2\cosh\left(\pi x\right)}\prod_{ l = 0 }^{M-1}\tanh \frac{\pi(x+i(l+1/2))}{k} \,.
\end{equation}
All poles of $F_{M,k}(x)$ are located at $x = \frac i2 + iK$ for $K \in\mathbb Z\,.$ Furthermore $F_{M,k}(x)$ is periodic in the complex plane, with 
\begin{equation}
F_{M,k}(x+ik) = (-1)^kF_{M,k}(x)\,.
\end{equation}
By closing the integral \eqref{ABJN1} in the upper-half of the complex, we may therefore reduce it to a finite sum of poles 
\begin{equation}\begin{split}
\hat Z_{M,1,k}(m)
&= \frac{2\pi i e^{-\frac \pi2Mm}}{1-(-1)^k e^{-k\pi m}}\sum_{K = 1}^{k-1} \,\underset{x = \frac i2 + iK}{\text{Res}}\left[e^{i\pi mx}F_{M,k}(x)\right]\,.
\end{split}\end{equation}
We can then evaluate the residues and derive analytic expressions for $\hat Z_{M,1,k}(m)$ for any $M$ and $k$. We can then compute $c_T$ and $\lambda^2_{(B,2)_{2,0}^{[022]}}$ using \eqref{cTFormula} and \eqref{locB2}. Table~\ref{N1OPEs} lists these quantities for various values of $M$ and $k$. Note that the analytic results become increasingly elaborate as $M$ and $k$ become larger, and so we include analytic expressions in Table~\ref{N1OPEs} only if a concise expression exists.
\begin{table}
\begin{center}
{\renewcommand{\arraystretch}{1.2}
\begin{tabular}{ c | c | c | c }
$M$ & $k$ & $\frac{16}{c_T}$    & $\lambda^2_{(B,2)_{2,0}^{[022]}}$ \\ \hline
1 & 2 & $\frac34=0.75$          & $\frac{16}5 = 3.2$ \\ 
  & 3 & $\frac{5+2\sqrt3}{13}\approx0.6511$ & $\frac{612-62\sqrt3}{169}\approx2.986$ \\
  & 4 & $\frac{3(\pi-2)}{4(3\pi-8)}\approx0.6009$ & $\frac{2(512-441\pi+90\pi^2)}{5(8-3\pi)^2}\approx 2.921$ \\ \hline
2 & 4 & $\frac{3(\pi-4)}{15\pi-52}\approx0.5281$  & $\frac{4(12544-6936\pi+945\pi^2)}{5(52-15\pi)^2}\approx2.715$ \\
 & 5 & $0.4667$ & $2.618$ \\
 & 6 & $0.4309$ & $2.582$ \\ \hline
3 & 6  & $0.4005$ & $2.498$ \\
4 & 8  & $0.3211$ & $2.381$ \\
5 & 10 & $0.2674$ & $2.307$ \\
6 & 12 & $0.2290$ & $2.258$ 
\end{tabular}}
\caption{OPE coefficients $\frac{16}{c_T}$ and $\lambda^2_{(B,2)_{2,0}^{[022]}}$ in various $U(1)_k\times U(1+M)_{-k}$ ABJ theories.}
\label{N1OPEs}
\end{center}
\end{table}

The above analysis can be generalized to the $N>1$ case by repeatedly integrating over $z_k$.  When $N=2$, for instance, we must evaluate
\begin{equation}\label{eq:N2Int}\begin{split}
Z_{M,2,k}(m) =  e^{-\frac \pi2MN m}\int dz_1\,dz_2\, e^{i\pi(z_1+z_2)m}\tanh^2 \frac{\pi(z_1-z_2)}{k} \prod_{a=1}^2F_{M,k}(z_a)\,.
\end{split}\end{equation}
We evaluate this by first integrating over $z_1$ while fixing $|\text{Im}(z_2)|<\frac k2$. We can perform this integral by closing the contour in the upper half complex plane and then summing over the poles, which occur at
$$ z_1 = \frac{i N}2\quad \text{ and }\quad z_1 = z_2 + ik(K+1/2)\,,$$
where $K$ is a positive integer. Because both $K(z)$ and $\tanh z$ are periodic in the complex plane, we need only sum the poles with imaginary part less than $k$; the rest can be resummed as a geometric series. Having integrated over $z_1$, we perform the $z_2$ integral in a similar fashion. For general $N$ we must repeat this process for each of the $N$ integration variables. We list results in Table~\ref{N2OPEs} for the $U(2)_k\times U(M+2)_{-k}$ theory, and in Table~\ref{ABJMOPEs} for the $U(N)_k\times U(N)_{-k}$ ABJM theory.

\begin{table}
\begin{center}
{\renewcommand{\arraystretch}{1.2}
\begin{tabular}{ c | c | c | c }
$M$ & $k$ & $\frac{16}{c_T}$    & $\lambda^2_{(B,2)_{2,0}^{[022]}}$ \\ \hline
1 & 2  & $0.3177$ & $2.479$ \\
  & 3  & $0.2697$ & $2.384$ \\
  & 4  & $0.2425$ & $2.339$ \\ \hline
2 & 4  & $0.2242$ & $2.302$ \\
  & 5  & $0.1986$ & $2.262$ \\
  & 6  & $0.1822$ & $2.239$ \\ \hline
3 & 6  & $0.1736$ & $2.221$ \\ 
4 & 8  & $0.1419$ & $2.175$ \\
5 & 10 & $0.1201$ & $2.144$ \\
6 & 12 & $0.1041$ & $2.122$ \\
\end{tabular}}
\caption{OPE coefficients $\frac{16}{c_T}$ and $\lambda^2_{(B,2)_{2,0}^{[022]}}$ in various $U(2)_k\times U(M+2)_{-k}$ ABJ theories.}
\label{N2OPEs}
\end{center}
\end{table}

\begin{table}
\begin{center}
{\renewcommand{\arraystretch}{1.2}
\begin{tabular}{ c | c | c | c }
 $N$ & $k$ & $\frac{16}{c_T}$ & $\lambda^2_{(B,2)_{2,0}^{[022]}}$ \\ \hline
 2 & 2 & $\frac38 = 0.375$  & $\frac{13}5=2.6$  \\ \hline
 3 & 2 & $\frac{3(\pi^2-10)}{45\pi^2-446}\approx0.2095$ & $2.309$ \\
   & 3 & $0.1838$ & $2.258$ \\ \hline
 4 & 2 & $0.1381$ & $2.195$ \\
   & 3 & $0.1191$ & $2.161$ \\
   & 4 & $0.1071$ & $2.143$ 
\end{tabular}}
\caption{OPE coefficients $\frac{16}{c_T}$ and $\lambda^2_{(B,2)_{2,0}^{[022]}}$ in various $U(N)_k\times U(N)_{-k}$ ABJM theories}
\label{ABJMOPEs}
\end{center}
\end{table}

\subsubsection{Higher-spin limit}
\label{HSLimU1}
We now compute $Z_{M,1,k}(m)$ at large $M$ and fixed $\lk=\frac Mk$, which is the vector model limit of the $U(1)_k\times U(1+M)_{-k}$ theory. The special case where $m = 0$ has already been considered in \cite{Hirano:2015yha}, so our task is to generalize their work to non-zero mass.   Starting with \eqref{ABJN1} and performing a change of variables $x \rightarrow x - \frac{iM}2$, we find that
\begin{equation}\begin{split}
\hat Z_{M,1,k}(m) = \int dx\, e^{ix\pi m}\exp\left[\sum_{l = -\frac{M-1}2}^{\frac{M-1}2}\log\tanh \frac{\pi(x+ il)}{k} - R(x)\right]\,,
\end{split}
\end{equation}
where $R(x) = \log(2\cosh(\pi x))$ for even $M$ and $R(x) = \log(2\sinh(\pi x))$ for odd $M$. We can now use the asymptotic expansion
\begin{equation}\label{tanhAsymp}\begin{split}
f(x,\lk,k)\equiv \sum_{l = -\frac{M-1}2}^{\frac{M-1}2}&\log\tanh \frac{\pi(x+ il)}{k} - R(x) \sim \frac{\cos\frac{2x\nb_\lk}{k}}{\sinh \frac{\nb_\lk}k }\log\tan\frac{\pi \lk}2
\end{split}\end{equation}
derived in \cite{Hirano:2015yha}. The right-hand expression should be understood as a formal series expansion, which can be written more verbosely as
\begin{equation}\begin{split}
f(x,\lk,k) &= \sum_{n = 0}^\infty \frac{(-1)^nf_{2n}(k,\lk)}{(2n)!}\frac{x^{2n}}{k^{2n-1}}\,,\\
\text{ where }f_{2n}(k,\lk) &= \sum_{p = 0}^\infty \frac{4^n(2-4^p)B_{2p}}{(2p)!k^{2p}} \nb_\lk^{2p+2n-1}\log\tan\frac{\pi\lk}2\,.
\end{split}\end{equation}
Rather than working with $x$, it is more convenient to perform a change of variables $x = k^{1/2}\xi$, so that
\begin{equation}\begin{split}
F(\xi,\lk,k) \equiv f(k^{1/2}\xi,\lk,k) = \sum_{n = 0}^\infty \frac{(-1)^nf_{2n}(k,\lk)}{(2n)!}\frac{\xi^{2n}}{k^{n-1}}\,.
\end{split}\end{equation}
The first few terms of $F(\xi,\lk,k)$ are
\begin{equation}
F(\xi,\lk,k) = \text{cons.} -2\pi\csc(\pi\lk)\xi^2 + \frac13 \pi^3(\cos(2\lk \pi)+3)\csc^3(\lk\pi)\frac{\xi^4}{k}+O(k^{-2})\,,
\end{equation}
while higher order terms look more complicated but can also be easily computed. We can now evaluate
\begin{equation}
\hat Z_{M,1,k}(m) = \int d\xi\ e^{i \xi\sqrt{k}\pi m + F(\xi,\lk,k)}
\end{equation}
at large $k$, and at each order in $k^{-1}$ we must merely evaluate a Gaussian integral. After a little work we find that 
 \es{cTlamU1First}{
c_T &= \frac{16k\sin(\pi \lambda)}{\pi} + 4(\cos(2\pi \lambda)+3) - \frac{\pi(3+13\cos(2\pi \lambda))\sin(\pi \lambda)}{3k} \\
&+\frac{\pi^2(5+7\cos(2\pi \lambda))\sin^2(\pi \lambda)}{k^2}+\cdots \,,\\
\lambda^2_{(B,2)_{2,0}^{[022]}}  
 &= 2 + \frac{\pi(3+\cos(2\pi \lambda))\csc(\lk \pi)}{2k} + \frac{\pi^2(\cos(2\pi \lambda)-3)\cot^2(\pi \lambda)}{k^2}\\
& + \frac{\pi^3(1074+785\cos(2\pi \lambda)-450\cos(4\pi \lambda)+127\cos(6\pi \lambda))\csc^3(\pi \lambda)}{768k^3}+\cdots  \,.
 }
Solving for $\lambda^2_{(B,2)_{2,0}^{[022]}}$ in terms of $c_T$ we find
 \es{lamcTU}{
 \lambda^2_{(B,2)_{2,0}^{[022]}} =  2 + \frac{8(3+\cos(2\pi \lambda))}{c_T}-\frac{64\sin^2(\pi\lk)(3+5\cos(2\pi\lk))}{c_T^2}+\cdots \,,
 } 
from which we see that in the limit of large $c_T$, the theory with the lowest value of  $\lambda^2_{(B,2)_{2,0}^{[022]}}$ at fixed $c_T$ is that with $\lambda = 1/2$.  For this reason, the $\lk = \frac12$ theory, namely the case of $U(1)_{2M}\times U(1+M)_{-2M}$, will prove of special interest to us when considering numeric bootstrap bounds. Specializing to this case and going to a higher order than as given in \eqref{cTlamU1First}, we can expand $\lambda^2_{(B,2)_{2,0}^{[022]}}$ in $1/c_T$:
\begin{equation}\label{ctLamN1}\begin{split}
\lambda^2_{(B,2)_{2,0}^{[022]}} &= 2 + \frac{16}{c_T} + \frac12\left(\frac{16}{c_T}\right)^2+ \frac1{12}\left(\frac{16}{c_T}\right)^3+ \frac23\left(\frac{16}{c_T}\right)^4-\frac{217}{240}\left(\frac{16}{c_T}\right)^5\\
&\qquad+ \frac{979}{480}\left(\frac{16}{c_T}\right)^6-\frac{71291}{15120}\left(\frac{16}{c_T}\right)^7+\cdots \,.
\end{split}\end{equation}
Comparing to the exact values computed in Table~\ref{N1OPEs}, we find that \eqref{ctLamN1} gives answers to within $1\%$ of the exact results already for $M = 4$.

\subsubsection{Supergravity limit}
\label{ABJMAiry}

We will also be interested in the large $N$ expansion.  Taking $N$ large while keeping $M$ and $k$ fixed (with $M\leq k$) corresponds to the M-theory limit, while taking $N$ large while keeping $M$ and $N/k$ fixed describes the Type IIA string theory limit.  Results for the M-theory limit were already computed in \cite{Agmon:2017xes} to all orders in $1/N$, which we now briefly review. Using the Fermi gas method \cite{Marino:2011eh}, the mass deformed partition function was computed to all orders in $1/N$ \cite{Nosaka:2015iiw,Agmon:2017xes}:
 \es{GotZABJM}{
 & Z \approx e^A C^{-\frac 13} \text{Ai}\left[C^{-\frac 13} (N-B) \right] \,, \qquad C = \frac{2}{\pi^2 k (1 + m^2)} \,, \\
   B& = \frac{\pi^2 C}{3} - \frac{1}{6k} \left(1 + \frac{1}{1 + m^2} \right) - \frac{k}{12}+\frac k2\left(\frac12-\frac Mk\right)^2 \,, \\
  A&= \frac{2 {\cal A}[k]+  {\cal A}[k(1 + i m)] + {\cal A}[k(1 - i m)] }{4}  \,,
 } 
 where the constant map function ${\cal A}$ is given by \cite{Hanada:2012si}.
\es{constantMap}{
{\cal A}(k)&=\frac{2\zeta(3)}{\pi^2k}\left(1-\frac{k^3}{16}\right)+\frac{k^2}{\pi^2}\int_0^\infty dx\frac{x}{e^{kx}-1}\log\left(1-e^{-2x}\right)\,.
}
We can now simply take derivatives of these exact functions as in \eqref{cTFormula} and \eqref{locB2} to compute $c_T$ and $\lambda^2_{(B,2)_{2,0}^{[022]}}$, where for each $k$ the derivatives of the constant map take the form of a one-dimensional integral that can be easily computed numerically (or analytically by summing poles). For low $k=1,2$ and $M=1,2$, some explicit examples were given in \cite{Agmon:2017xes}, where it was shown that the large $N$ expansion compares well even down to the exact $N=2$ result. We will use this expansion in the numerics section specifically for $N=10$ and a range of $M,k$, which we summarize in Table \ref{tab10}.
\begin{table}
\begin{center}
{\renewcommand{\arraystretch}{1.2}
\begin{tabular}{ c | c | c | c }
$k$ & $M$ & $\frac{16}{c_T}$    & $\lambda^2_{(B,2)_{2,0}^{[022]}}$ \\ \hline
2 & 0 & $0.0361459$          & $2.04699$ \\  \hline
3 & 0 & $0.0301815$  & $2.03842$ \\ \hline
4 & 0 & $0.0265295$  & $2.03342$ \\
  & 1 & $0.0250946$  & $2.03158$ \\ \hline
6 & 0 & $0.0221553$  & $2.02766$ \\
  & 1 & $0.0208109$  & $2.02595$ \\
  & 2 & $0.0200682$  & $2.02501$ \\ \hline
10 & 0  & $0.0178216$ & $2.02218$ \\
   & 1  & $0.0166285$ & $2.02067$ \\
   & 2 & $0.0157899$ & $2.01961$ \\
   & 3 & $0.0152331$ & $2.01891$ \\
   & 4 & $0.0149146$ & $2.01851$ 
\end{tabular}}
\caption{OPE coefficients $\frac{16}{c_T}$ and $\lambda^2_{(B,2)_{2,0}^{[022]}}$ in various $U(10)_k\times U(10+M)_{-k}$ ABJ theories, as computed from the all orders in $1/N$ formula \eqref{GotZABJM}.}
\label{tab10}
\end{center}
\end{table}

\subsection{$SO(2)_{2k}\times USp(2+2M)_{-k}$ theory}
\label{so2OPES}

We now discuss the mass-deformed sphere partition function for the $SO(2)_{2k}\times USp(2+2M)_{-k}$ theory. Using supersymmetric localization, this quantity can be written as an $(M+1)$-dimensional integral \cite{Kapustin:2009kz,Gulotta:2012yd}:
\begin{equation}\begin{split}\label{so2v1}
Z_{M,k}&(m) \propto \int d\mu\,d^M\nu\,e^{2\pi i k(\mu^2 - \sum_a \nu_a^2)}\\
&\times \frac{\prod_a\sinh^2\left[2\pi\nu_a\right] \prod_{a<b} \sinh^2\left[\pi(\nu_a+\nu_b)\right]\sinh^2\left[\pi(\nu_a-\nu_b)\right]}{\prod_{b}\cosh\left[\pi (\mu - \nu_b) + \frac{\pi m}{2} \right]\cosh\left[\pi (\mu + \nu_b) + \frac{\pi m}{2} \right]\cosh\left[\pi (\mu-\nu_b)\right]\cosh\left[\pi(\mu+\nu_b)\right]}\,,
\end{split}\end{equation}
up to an overall $m$-independent factor.   In \cite{Moriyama:2016kqi} it was shown that the massless partition function $m = 0$ could be further simplified to a single integral. Generalizing their results to non-zero $m$ is straightforward, as we outline in Appendix~\ref{simpso2}. We show that
\begin{equation}\begin{split}\label{so2v2}
Z_{M,k}(m) = Z_0 \int dx\,e^{i\pi  m x}\frac{\cosh^2\frac{\pi x}{2k}}{\sinh \pi x \cosh\frac{\pi x}k}\prod_{l = -M}^M\tanh\frac{\pi( x+il)}{2k}
\end{split}\end{equation}
where $Z_0$ is an overall constant which is independent of $m$. We will now compute mass derivatives of this quantity, first at finite $M,k$, and then in the large $M$ and fixed $\lambda=M/k$ expansion.

\subsubsection{Finite $M,k$}
For computing $c_T$ and $\lambda^2_{(B,2)_{2,0}^{[022]}}$ let us use the first expression \eqref{so2v2}. We are thus led to evaluate
\begin{equation}\label{GInt}\begin{split}
\hat Z_{M,k}(m) = \int_{-\infty}^\infty dx\,  e^{i\pi m x}G_{M,k}(x)
\end{split}\end{equation}
where we define
\begin{equation}
G_{M,k}(x) = \frac{\cosh^2\frac{\pi x}{2k}}{\sinh \pi x \cosh\frac{\pi x}k}\prod_{l = -M}^M\tanh\frac{\pi( x+il)}{2k}\,.
\end{equation}
Similarly to $F_{M,k}(x)$ in Section~\ref{SMALLMNK}, all poles of $G_{M,k}(x)$ are located at $x = \frac {iK}2 $ for $K \in\mathbb Z\,,$ and furthermore $G_{M,k}(x)$ is periodic in the complex plane, with 
\begin{equation}
G_{M,k}(x+2ik) = G_{M,k}(x)\,.
\end{equation}
By closing the integral \eqref{GInt} in the upper-half of the complex, we may therefore reduce it to a finite sum of poles 
\begin{equation}\begin{split}
\hat Z_{M,k}(m)
&= \frac{2\pi i}{1-e^{-2k\pi m}}\sum_{K = 1}^{4k-1} \,\underset{x = \frac{iK}2}{\text{Res}}\left[e^{i\pi mx}G_{M,k}(x)\right]\,.
\end{split}\end{equation}

For small values of $M$ and $k$ we can easily sum over poles, and then compute $c_T$ and $\lambda^2_{(B,2)_{2,0}^{[022]}}$ using \eqref{cTFormula} and \eqref{locB2}. We list results for various $M$ and $k$ in Table~\ref{TSO2OPEs}.
\begin{table}
\begin{center}
{\renewcommand{\arraystretch}{1.2}
\begin{tabular}{ c | c | c | c }
$M$ & $k$ & $\frac{16}{c_T}$    & $\lambda^2_{(B,2)_{2,0}^{[022]}}$ \\ \hline
0 & 1  & $1$ & $4$ \\
  & 2  & $\frac34=0.75$ & $\frac {16}5 = 3.2$ \\
  & 3  & $0.6511$ & $2.986$ \\
  & 4  & $0.6009$ & $2.921$ \\\hline
1 & 2  & $\frac34=0.75$ & $\frac {16}5 = 3.2$   \\
  & 3  & $0.4778$ & $2.648$  \\
  & 4  & $0.3879$ & $2.503$  \\ \hline
2 & 4  & $0.3879$ & $2.503$  \\ 
  & 5  & $0.3021$ & $2.366$  \\ 
  & 6  & $0.2603$ & $2.312$  \\ \hline
3 & 6  & $0.2603$ & $2.312$  \\ 
4 & 8  & $0.1957$ & $2.225$  \\ 
5 & 10 & $0.1568$ & $2.175$  \\ 
6 & 12 & $0.1307$ & $2.144$  
\end{tabular}}
\caption{OPE coefficients $\frac{16}{c_T}$ and $\lambda^2_{(B,2)_{2,0}^{[022]}}$ in various $SO(2)_{2k}\times USp(2+2M)_{-k}$ theories}
\label{TSO2OPEs}
\end{center}
\end{table}

\subsubsection{Higher-spin limit}
Finally, we study the large $M$ limit of the $SO(2)_{2k}\times USp(2+2M)_{-k}$ theory, keeping $\lk = \frac{2M+1}{2k}$ fixed. We can write:
\begin{equation}\begin{split}\label{so2LargeM}
\hat Z_{M,k}&(m) = \int dx\,e^{i\pi  m x}\\
&\times\exp\left(2\log\cosh\frac{\pi x}{2k}-\cosh\frac{\pi x}k-\log \sinh \pi x+\sum_{l = -M}^M\log\tanh\frac{\pi( x+il)}{2k}\right)
\end{split}\end{equation}
Using \eqref{tanhAsymp} and defining the variable $\xi = k^{-1/2}x$, we find that 
\begin{equation}\begin{split}
\sum_{l = -M}^M&\log\tanh\frac{\pi( x+il)}{2k} - \log \sinh \pi x \\
& = \text{cons.} -\pi\csc(\pi\lk)\xi^2 + \frac1{24} \pi^3(\cos(2\lk \pi)+3)\csc^3(\lk\pi)\frac{\xi^4}{k}+O(k^{-2})\,,
\end{split}\end{equation}
while we can simply use a Taylor series expansion to compute
\begin{equation}
2\log\cosh\frac{\pi x}{2k}-\cosh\frac{\pi x}k = -\frac{\pi \xi^2}{4k} + \frac{7\pi^4\xi^4}{96k^2} + O(k^{-3}) \,.
\end{equation}
We can now hence evaluate
\begin{equation}
\hat Z_{M,1,k}(m) \propto \int d\xi\ e^{i \xi\sqrt{k}\pi m -\pi\csc(\pi\lk)\xi^2 + \dots}
\end{equation}
at large $k$, and at each order in $k^{-1}$ we must merely evaluate a Gaussian integral. After a little work we find that 
\begin{equation}\label{cTLamSO}\begin{split}
c_T &= \frac{32k \sin(\pi \lk)}{\pi} + 16\cos^2(\pi\lk)-\frac{\pi\sin(\pi \lk) [ 15+29\cos(2\pi \lk)]}{3k}+ O(k^{-2}) \\
\lambda^2_{(B,2)_{2,0}^{[022]}} &= 2 + \frac{\pi[3+\cos(2\pi \lk)]\csc(\pi \lk)}{4k} \\
&- \frac{\pi^2[39+44\cos(2\pi \lk)-19\cos(4\pi \lk)]\csc^2(\lk\pi)}{128k^2} + O(k^{-3}) \,.
\end{split}\end{equation}
Comparing to the exact results in Table~\ref{TSO2OPEs}, we see that already for $k = 2$ the approximations \eqref{cTLamSO} are within a couple percent of the exact answers.

Solving for $\lambda^2_{(B,2)_{2,0}^{[022]}}$ in terms of $c_T$, we find
 \es{lamOSpcT}{
  \lambda^2_{(B,2)_{2,0}^{[022]}}  &= 2 + \frac{8(\cos(2\lk\pi)+3)}{c_T}-\frac{32\sin^2(\pi\lk)(17+23\cos(2\pi\lk))}{c_T^2} + \cdots \,,
 }
from which we see that, at least at large $c_T$, the theory with $\lambda = 1/2$ has the smallest value of $\lambda^2_{(B,2)_{2,0}^{[022]}}$.  This fact will be useful in the next section.  Plugging in $\lambda = 1/2$ in \eqref{lamOSpcT} and expanding to higher orders in $1/c_T$, we obtain
\begin{equation}\label{lamHalfOSpcT}\begin{split}
  \lambda^2_{(B,2)_{2,0}^{[022]}}  = 2 &+ \frac{16}{c_T}+\frac{3}{4} \left( \frac{16 }{c_T} \right)^2 -\frac{25}{24}\left( \frac{16 }{c_T} \right)^3+\frac{437}{64}\left( \frac{16 }{c_T} \right)^4-\frac{20997}{640}\left( \frac{16 }{c_T} \right)^5\\
  &+\frac{259523}{1536}\left( \frac{16 }{c_T} \right)^6-\frac{897994667}{967680}\left( \frac{16 }{c_T} \right)^7+ \cdots \,.
\end{split}\end{equation}
Together with \eqref{ctLamN1}, this expression will be useful in the next section.

\section{Numerical bootstrap}
\label{numerics}

We will now use the results of the previous sections to derive the crossing equations for the superblock expansion of 3d $\mathcal{N}=6$ SCFTs suitable for numerical bootstrap study. These crossing equations allow us to numerically bootstrap upper/lower bounds on CFT data of general 3d $\mathcal{N}=6$ SCFTs. By imposing the specific values of $c_T$ and $\lambda^2_{(B,2)^{[022]}_{2,0}}$ as computed analytically for various theories in Section~\ref{EXACT}, we will find islands in the space of semishort OPE coefficients for each $N,k,M$. We will also argue that the $U(1)_{2M}\times U(1+M)_{-2M}$ theory saturates the numerical lower bound on $\lambda^2_{(B,2)^{[022]}_{2,0}}$ for a given value of $c_T$.

\subsection{Crossing equations}
\label{crossSec}

The position space crossing equations were written in \eqref{CrossingP}. For the $s$-channel superblock expansion the nontrivial constraint is the one given by $(x_1,X_1)\leftrightarrow(x_3,X_3)$. In terms of the ${\cal S}_{\bf r}(U,V)$ basis in \eqref{B}, the crossing equations \eqref{CrossingP} can be written in using a 6-component vector
\es{V}{
d^i(U,V)=\begin{pmatrix}
15 F_{-,{\bf 1}_s} + 80 F_{-,{\bf 45}_a \oplus {\overline{\bf 45}}_a} + 64 F_{-,{\bf 84}_s}\\
 F_{-,{\bf 15}_a} + F_{-,{\bf 45}_a \oplus {\overline{\bf 45}}_a} - 4 F_{-,{\bf 84}_s}\\
3 F_{-,{\bf 15}_s} - 12 F_{-,{\bf 45}_a \oplus {\overline{\bf 45}}_a} + 16 F_{-,{\bf 84}_s}\\
 3 F_{-,{\bf 20}'_s} - 2 F_{-,{\bf 45}_a \oplus {\overline{\bf 45}}_a} + 2 F_{-,{\bf 84}_s}\\ 
15 F_{+,{\bf 1}_s} - 15 F_{+,{\bf 15}_s} - 60 F_{+,{\bf 20}'_s} - 60 F_{+,{\bf 45}_a \oplus {\overline{\bf 45}}_a} - 56 F_{+,{\bf 84}_s}\\ 
3 F_{+, {\bf 15}_a} - 3 F_{+,{\bf 15}_s} - 9 F_{+,{\bf 20}'_s} - 3 F_{+,{\bf 45}_a \oplus {\overline{\bf 45}}_a} + 14 F_{+,{\bf 84}_s}\\
\end{pmatrix}\,,
}
where we define
\es{Fs}{
F_{\pm,{\bf r} }(U,V)\equiv V^2 {\cal S}_{\bf r} (U,V)\pm U^{2} {\cal S}_{\bf r} (V,U)\,.
}
Combining the crossing equations with the superconformal block decomposition, we can then define a $d^i_I$ for each superconformal block $I$ listed in Table \ref{n6mults} by replacing each ${\cal S}_{\bf r}$ in $d^i$ by $\mathfrak{G}_I^{\bf r}$ defined in \eqref{SuperconfBlock}.  The crossing equations in terms of these $d^i_I$ can then be written as
\es{crossing2}{
0=d^i_\text{Id}+\frac{64}{c_T}d^i_{(B,2)_{1,0}^{[011]}}+\sum_{I \neq \text{Id}, (B,2)_{1,0}^{[011]}}\lambda^2_{I}\, d^i_I\,,
}
where we normalized the squared OPE coefficient of the identity multiplet to $\lambda_\text{Id}^2=1$, and parameterized our theories by the value of $\lambda^2_{(B,2)_{1,0}^{[011]}} = \frac{64}{c_T}$ (see \eqref{cT}). The sum in \eqref{crossing2} should then be understood as running over all other superconformal blocks for multiplets appearing in the $S\times S$ OPE\@.

These six crossing equations are in fact redundant due to $\mathcal{N}=6$ superconformal symmetry, similar to the $\mathcal{N}=8$ case in \cite{Chester:2014fya,Agmon:2019imm}. It is important to remove these redundancies, since otherwise they cause numerical instabilities in the bootstrap algorithm. As in \cite{Agmon:2019imm}, we can do this using the explicit expressions for the crossing equations in \eqref{crossing2} in terms of superblocks, where each ${\cal S}_{\bf r} (U,V)$ is a linear combination of conformal blocks for each supermultiplet. We then expand in $z,\bar z$ derivatives as
\es{crossDer}{
F_{+,{\bf r} }(U,V)&=\sum_{\substack{p+q=\text{even}\\\text{s.t. }p\leq q}}\frac{2}{p!q!}\left(z-\frac12\right)^p\left(\bar z-\frac12\right)^q\partial^p_z\partial_{\bar z}^q F_{+,{\bf r}}(U,V)\vert_{z=\bar z=\frac12}\,,\\
F_{-,{\bf r} }(U,V)&=\sum_{\substack{p+q=\text{odd}\\\text{s.t. }p<q}}\frac{2}{p!q!}\left(z-\frac12\right)^p\left(\bar z-\frac12\right)^q\partial^p_z\partial_{\bar z}^q F_{-, {\bf r}}(U,V)\vert_{z=\bar z=\frac12}\,,
}
where $z,\bar z$ are written in terms of $U,V$ as
 \es{UVtozzbar}{
 U=z \bar z\,,\qquad V=(1-z)(1-\bar z)\,.
 }
In the sums in Eqs.~\eqref{crossDer} we only consider terms that are nonzero and independent according to the definition \eqref{Fs}. We then truncate these sums to a finite number of terms by imposing that
\es{nmax}{
p+q\leq\Lambda\,,
}
and then consider the finite dimensional matrix $\widetilde d_i^{(p,q)}$ whose rows as labeled by $i=1,\dots6$ are those of $d^i$, and whose columns as labeled by ${(p,q)}$ are the coefficients of the $\partial^p_z\partial^q_{\bar z} {\cal S}_{\bf r} (U,V)\vert_{z=\bar z=\frac12}$ that appear in each entry of $d^i$ after expanding like \eqref{crossDer} using the definition \eqref{Fs} of $F_{\pm,{\bf r} }(U,V)$ in terms of ${\cal S}_{\bf r} (U,V)$.  Finally, we check numerically to see which crossing equations are linearly independent for each value of $\Lambda$, and find that a linearly independent subspace for any $\Lambda$ is given by
\es{indies2}{
\{
d^{3}\,,
d^{4}\,,
d^{5}\,,
d^{6}\}\,,
}
where we include all nonzero $z,\bar z$ derivatives for the crossing equations listed.\footnote{In the analogous $\mathcal{N}=8$ case studied in \cite{Chester:2014fya}, the linearly independent set consisted of just one crossing equation with all of its derivatives, as well as a second crossing equation with only derivatives in $z$.} 

We now have all the ingredients to perform the numerical bootstrap using the crossing equations \eqref{crossing2}, where we restrict to the linearly independent set of crossing equations \eqref{indies2}.  We can now derive numerical bounds on both OPE coefficients and caling dimensions using numerical algorithms that are by now standard (see for instance \cite{Agmon:2019imm,Kos:2014bka}) and can be implemented using {\tt SDPB} \cite{Simmons-Duffin:2015qma,Landry:2019qug}.   In each case, the numerical algorithms involve finding functionals $\alpha$ that act on the vector of functions $d^i(U, V)$ and return a linear combination of derivatives of these functions evaluated at the crossing-symmetric point $U = V = 1/4$.  In all the numerical studies presented below, we will restrict the total derivative order $\Lambda$ (see \eqref{nmax}) to be $\Lambda = 39$, and we will only consider acting with $\alpha$ on blocks that have spin up to $\ell_\text{max} = 50$.

\subsection{Bounds on short OPE coefficients and the extremal functional conjecture}
\label{genbound}

We begin by deriving numerical bootstrap bounds on the squared OPE coefficients ${\lambda_{(B, 2)_{1, 0}^{[011]}}^2 = \frac{64}{c_T}}$ and $\lambda_{(B,2)_{2,0}^{[022]}}^2$ that were computed using supersymmetric localization in specific ${\cal N} = 6$ SCFTs from the ABJ(M) family in the previous section. 

First, we can derive a lower bound on $c_T$ that applies to all ${\cal N} = 6$ SCFTs.  To do so, we consider linear functionals $\alpha$ satisfying 
 \es{CondcT}{
  &\alpha(d^i_{(B,2)_{1,0}^{[011]}}) = 1 \,,  \\
  &\alpha(d^i_I) \geq 0 
    \,, \qquad \qquad \qquad\;   \text{for all superconf.~blocks $I \notin \{ \text{Id}, (B,2)_{1,0}^{[011]} \}$} \,.
 }
From \eqref{crossing2}, the existence of such an $\alpha$ implies
 \es{cTBoundAbstract}{
   \frac{64}{c_T}  \leq - \alpha(d^i_\text{Id}) \,.
 }
We performed such a numerical study, and we found
\es{cTbound}{
c_T\geq15.5\,,
}
where recall that $c_T=16$ corresponds to the theory of a free ${\cal N} = 6$ massless hypermultiplet, which also has ${\cal N} = 8$ SUSY\@. The bound \eqref{cTbound} can be compared to the analogous $\mathcal{N}=8$ bound $c_T\geq15.9$ computed in \cite{Agmon:2017xes} with $\Lambda=43$. In both cases, we expect the numerics should be converging to $c_T \geq 16$ in the infinite $\Lambda$ limit, simply because there are no known ${\cal N} = 6$ SCFTs with $c_T$ smaller than $16$.  The fact that the ${\cal N} = 6$ bound \eqref{cTbound} is weaker than the ${\cal N} = 8$ one suggests that the $\mathcal{N}=6$ numerics are slightly less converged than the $\mathcal{N}=8$ numerics. In the remainder of this paper, we will only show results for $c_T\geq16$.

Let us now compute bounds on the squared OPE coefficient $\lambda_{(B,2)_{2,0}^{[022]}}^2$ as a function of $c_T$. In general, to find upper/lower bounds on the OPE coefficient of an isolated superblock $I^*$ that appears in \eqref{crossing2}, we consider linear functionals $\alpha$ satisfying 
 \es{CondOPE}{
  &\alpha(d^i_{I^*}) = s \,, \qquad\qquad \qquad\text{$s=1$ for upper bounds, $s=-1$ for lower bounds} \,,  \\
  &\alpha(d^i_{I}) \geq 0 
    \,, \qquad \qquad \qquad\;   \text{for all short and semi-short ${I} \notin \{ \text{Id}, (B,2)_{1,0}^{[011]}, I^* \}$} \,, \\
  &\alpha(d^i_{I}) \geq 0  \,, \qquad\qquad\qquad\; \text{for all long $I$ with $\Delta_{I}\geq \ell + 1$}\,.
 }
The existence of such an $\alpha$ implies that
 \es{UpperOPE}{
  &\text{if $s=1$, then}\qquad\;\;\;\lambda_{{I}^*}^2 \leq - \alpha (d^i_\text{Id})  - \frac{64}{c_T}\alpha( d^i_{(B,2)_{1,0}^{[011]}} ) \,,\\
    &\text{if $s=-1$, then}\qquad \lambda_{{I}^*}^2 \geq  \alpha (d^i_\text{Id})  + \frac{64}{c_T}\alpha( d^i_{(B,2)_{1,0}^{[011]}} ) \,, 
 }
thus giving us both an upper and a lower bound on $\lambda_{{I}^*}^2$.
\begin{figure}[]
\begin{center}
        \includegraphics[width=.85\textwidth]{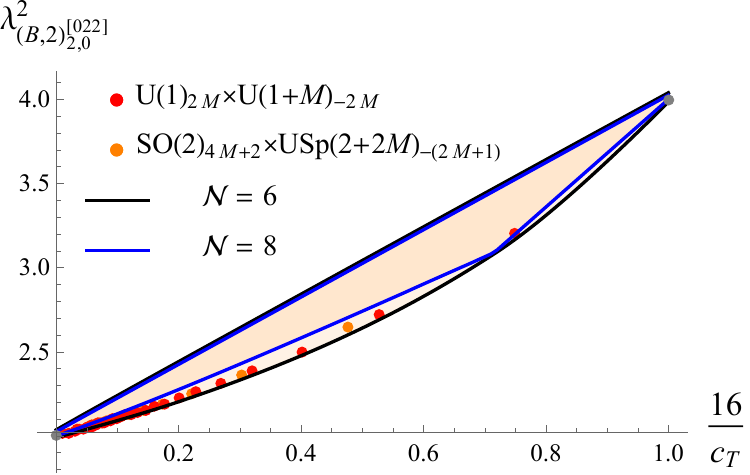}
\caption{Upper and lower bounds on the $\lambda_{(B,2)_{2,0}^{[022]}}^2$ OPE coefficient in terms of the stress-tensor coefficient $c_T$, where the orange shaded region is allowed, and the plot ranges from the generalized free field theory (GFFT) limit $c_T\to\infty$ to the free theory $c_T=16$. The black lines denote the $\mathcal{N}=6$ upper/lower bounds computed in this work with $\Lambda=39$, the blue lines denotes the $\mathcal{N}=8$ upper/lower bounds computed in \cite{Agmon:2017xes} with $\Lambda=43$. The red and orange dots denote the exact values in Tables \ref{N1OPEs} and \ref{TSO2OPEs} for the $U(1)_{2M}\times U(1+M)_{-2M}$ and $SO(2)_{4M+2}\times USp(2+2M)_{-(2M+1)}$ theories, respectively, for $M=1,2,\dots$, while the gray dots denote the GFFT and free theory values from Table \ref{freeTable}.
}
\label{B2fig}
\end{center}
\end{figure} 
Using this procedure, our numerical study gives the upper and lower bounds shown in black in Figure~\ref{B2fig}.  On the same plot, we indicated in blue the bounds obtained with $\Lambda = 43$ in the $\mathcal{N}=8$ case, as derived in \cite{Agmon:2017xes}.  While the upper bounds for the ${\cal N} = 6$ and ${\cal N} = 8$ cases are very similar and likely differ only because of the different value of $\Lambda$ that was used, the lower bounds are qualitatively different.  Indeed, the $\mathcal{N}=6$ and $\mathcal{N}=8$ lower bounds meet at $\frac{16}{c_T} = 0$, $1$, and at around $.71$, where the $\mathcal{N}=8$ bound has a kink.\footnote{This $\mathcal{N}=8$ kink was previously observed in \cite{Chester:2014fya,Chester:2014mea}.}   At other values of $\frac{16}{c_T}$, the ${\cal N} = 6$ lower bound is significantly weaker than the ${\cal N} = 8$ one, which suggests that it may be saturated by ${\cal N} = 6$ SCFTs that do not have ${\cal N}= 8$ SUSY\@.  

Indeed, in Figure~\ref{B2fig} we also mark in red and orange the values of the OPE coefficients computed analytically for the $U(1)_{2M} \times U(1+M)_{-2M}$ and $SO(2)_{4M+2} \times USp(2M+ 2)_{-(2M+1)}$ in Eqs.~\eqref{ctLamN1} and \eqref{lamHalfOSpcT}, respectively, and we see that these values do lie outside the ${\cal N} = 8$ region  and come close to saturating the ${\cal N} = 6$ bounds.  We chose to plot the exact results for these particular theories because, of all the exact results that we computed, these ones are the SCFTs in their respective families that come closest to saturating the lower bounds in Figure~\ref{B2fig}.  The red dots are slightly closer to the lower bound than the orange ones, as can also be seen analytically at large $c_T$ by comparing the $1/c_T^2$ terms in \eqref{ctLamN1} and \eqref{lamHalfOSpcT}. We hence conjecture that, in the infinite $\Lambda$ limit, it is the $U(1)_{2M} \times U(1+M)_{-2M}$ theory that saturates the numerical lower bound.  This is reminiscent of the $\mathcal{N}=8$ case in \cite{Agmon:2017xes}, where the $U(N)_{2}\times U(N+1)_{-2}$ theory was found to saturate the corresponding lower bound. 

To orient the reader about the spectrum of the superconformal block decomposition of the $\langle SSSS \rangle$ correlator in the higher-spin limit, we list the low-lying CFT data for a parity-preserving theory such as $U(1)_{2M} \times U(1+M)_{-2M}$ in Table~\ref{HSSpectrum}.\footnote{We only discuss single and double trace operators here, since higher trace operators have squared OPE coefficients that are suppressed as $c_T^{-n}$ for $n>2$ traces, and so would be very difficult to see numerically. }  The spectrum is similar to that in Table~\ref{freeTable}, except that the conserved multiplets combine with double trace multiplets according to \eqref{LimitSpin0}--\eqref{LimitSpinEven} to form single trace long multiplets whose scaling dimensions are close to unitarity.  Since these multiplets are single trace, their OPE coefficients are $O(c_T^{-1})$, as opposed to all the other double trace operators whose OPE coefficients are $O(c_T^{0})$. For ${\cal N} = 6$ SCFTs that do not preserve parity, all the scaling dimensions for any given spin should appear in all structures.
\begin{table}[htp]
\begin{center}
{\renewcommand{\arraystretch}{1.4}
\begin{tabular}{l||c}
 & parity-preserving higher-spin limit   \\ 
\hline\hline
$\lambda^{2}_{(B,2)^{[011]}_{1,0}}$ & $\frac{64}{c_T} + O(c_T^{-2}) $   \\
\hline
$\lambda^{2}_{(B,2)^{[022]}_{2,0}}$ & $2 + O(c_T^{-1}) $    \\
\hline
$\lambda^{2}_{(A,2)^{[011]}_{\ell+2,\ell}}$ for $\ell = 0, 1, 2, \ldots$  & 
   $\frac{16}{3} + O(c_T^{-1}) $, $\frac{256}{45} + O(c_T^{-1}) $, $\frac{4096}{525}+ O(c_T^{-1}) $, \ldots
     \\
\hline$\lambda^{2}_{(A,+)^{[002]}_{\ell + 5/2,\ell + 1/2}}$ for $\ell = 0, 2, \ldots$ & $\frac{16}{9} + O(c_T^{-1}) $, $\frac{6144}{1225} + O(c_T^{-1}) $, \ldots  
    \\
\hline$\lambda^{2}_{(A,1)_{\ell + 7/2,\ell + 3/2}^{[100], 1}}$ for $\ell = 0, 2, \ldots$ & $\frac{512}{315}+O(c_T^{-1})$   \\
\hline$\lambda^{2}_{(A,1)_{\ell + 7/2,\ell + 3/2}^{[100], 2}}$ for $\ell = 0, 2, \ldots$ & $\frac{1024}{105}+O(c_T^{-1})$   \\
\hline$\Delta_{(0,1)}$ &  $2 + O(c_T^{-1})$, $4 + O(c_T^{-1})$, \ldots   \\
\hline$\lambda^2_{(0,1)}$ &  $ \frac{32}{15}+O(c_T^{-1})$, $\frac{256}{315} + O(c_T^{-1})$, \ldots   \\
\hline$\Delta_{(0,2)}$ & $1 + O(c_T^{-1})$, $3 + O(c_T^{-1})$, \ldots \\
\hline$\lambda^2_{(0,2)}$ & $\frac{4}{9} + O(c_T^{-1})$, $\frac{64}{315} + O(c_T^{-1})$, \ldots \\
\hline$\Delta_{\ell\geq 1}$, $\ell$ odd  & $\ell+1 + O(c_T^{-1}) $, $\ell + 2+ O(c_T^{-1}) $, $\ell + 4+ O(c_T^{-1}) $, \ldots    \\
\hline$\lambda^2_{\ell\geq 1}$, $\ell$ odd  & $ O(c_T^{-1}) $, $ O(c_T^{0}) $, $ O(c_T^{0}) $, \ldots    \\
\hline$\Delta_{(\ell \geq 2,1)}$, $\ell$ even  &  $\ell + 2+ O(c_T^{-1}) $, $\ell + 4+ O(c_T^{-1}) $, \ldots  \\
\hline$\lambda^2_{(\ell \geq 2,1)}$, $\ell$ even  &  $ O(c_T^{0}) $, $ O(c_T^{0}) $, \ldots    \\
\hline$\Delta_{(\ell \geq 2, 2)}$, $\ell$ even & $\ell+1 + O(c_T^{-1}) $, $\ell+3 + O(c_T^{-1}) $, $\ell+5 + O(c_T^{-1}) $, \ldots  \\
\hline$\lambda^2_{(\ell \geq 2, 2)}$, $\ell$ even & $\ O(c_T^{0}) $, $O(c_T^{0}) $, $O(c_T^{0}) $, \ldots  \\
\hline$\Delta_{(\ell \geq 2,3)}$, $\ell$ even & $\ell+3 + O(c_T^{-1}) $, $\ell+5 + O(c_T^{-1}) $, \ldots   \\
\hline$\lambda^2_{(\ell \geq 2,3)}$, $\ell$ even & $O(c_T^{0}) $, $ O(c_T^{0}) $, \ldots   \\
\end{tabular}}
\caption{Low-lying single and double trace CFT data for a parity-preserving ${\cal N} = 6$ SCFT such as the $U(1)_{2M} \times U(1+M)_{-2M}$ theory in the higher-spin limit.  We write $\Delta_{(\ell,n)}$ and $\lambda^2_{(\ell,n)}$ to denote the scaling dimensions and OPE coefficients squared, respectively, of the superblock corresponding to the structure $\text{Long}_{\Delta, \ell}^{[000], n}$.}
\label{HSSpectrum}
\end{center}
\end{table}

\subsection{Bounds on semishort OPE coefficients}

\begin{figure}[]
\begin{center}
        \includegraphics[width=.49\textwidth]{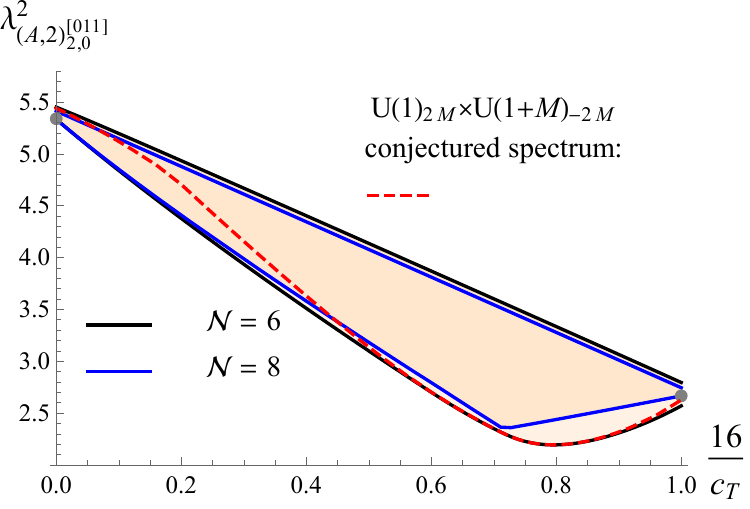} \includegraphics[width=.49\textwidth]{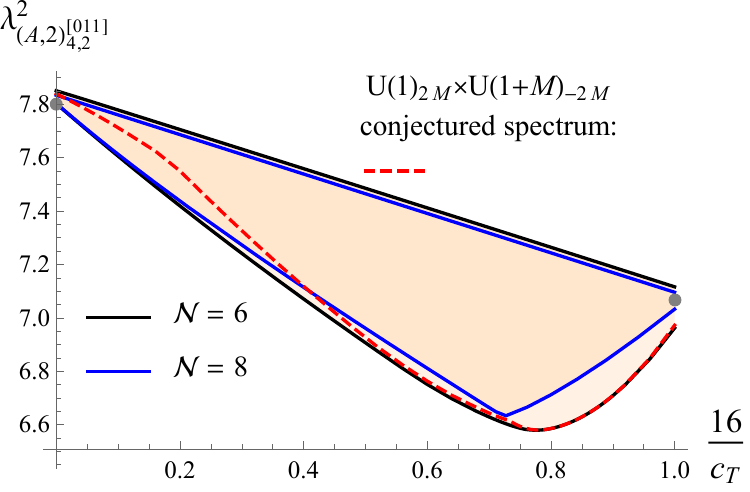}
           \includegraphics[width=.49\textwidth]{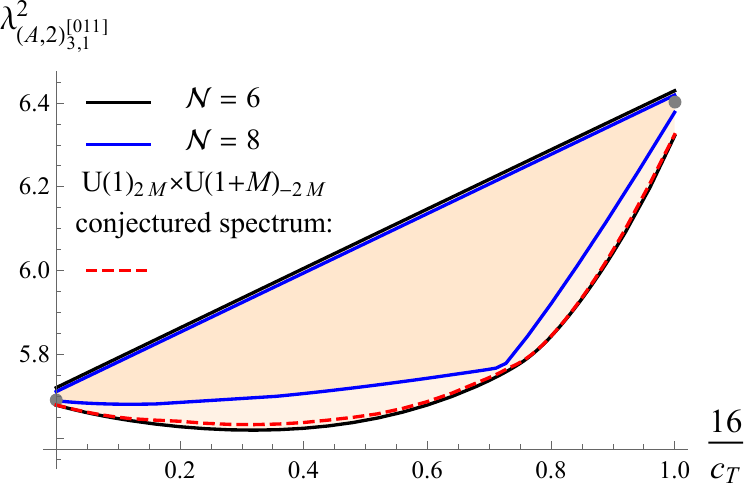} \includegraphics[width=.49\textwidth]{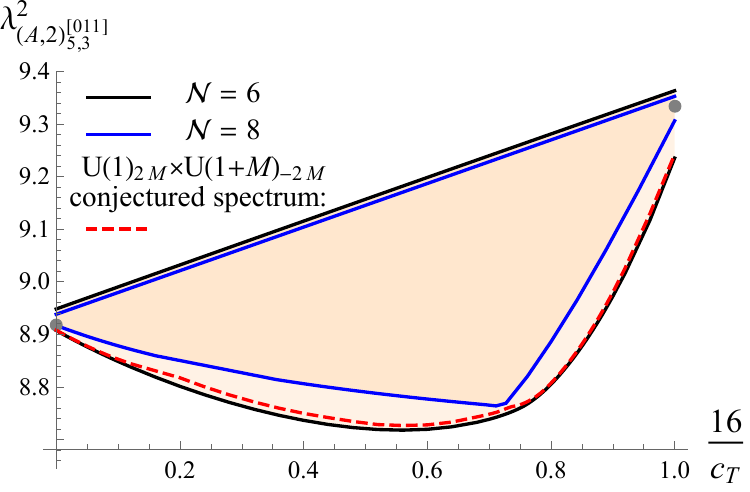}
              \includegraphics[width=.49\textwidth]{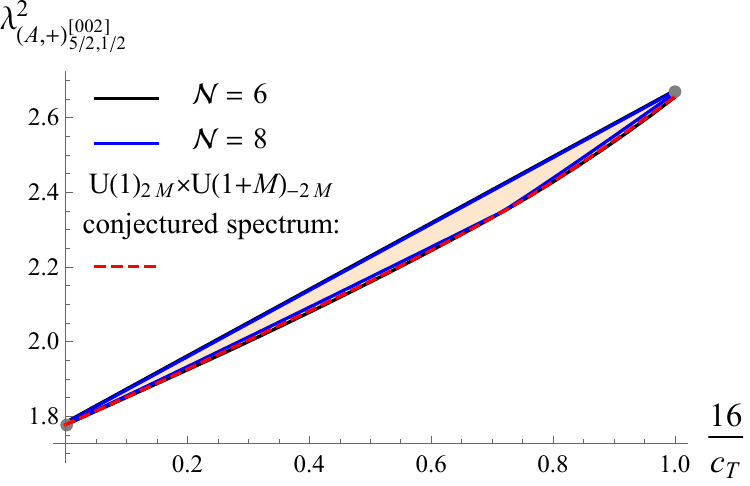} \includegraphics[width=.49\textwidth]{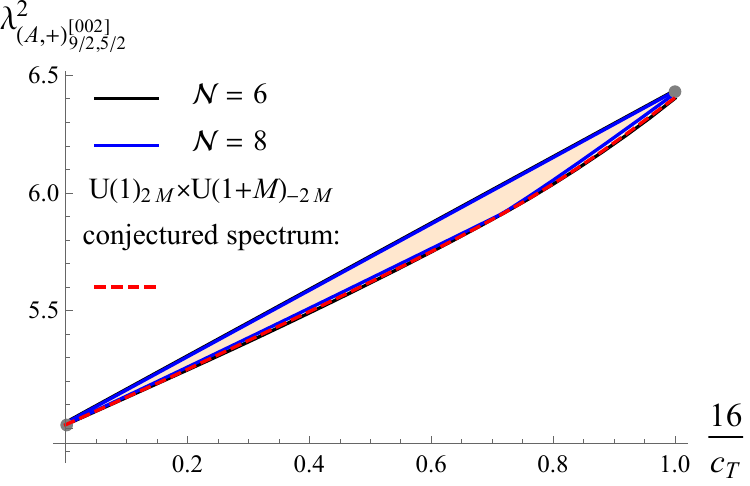}
\caption{Upper and lower bounds on various semishort OPE coefficients squared in terms of $c_T$, where the orange shaded regions are allowed, and the plots ranges from the GFFT limit $c_T\to\infty$ to the free theory $c_T=16$. The black lines denote the $\mathcal{N}=6$ upper/lower bounds computed in this work with $\Lambda=39$, the blue lines denote the $\mathcal{N}=8$ upper/lower bounds computed in \cite{Agmon:2017xes} with $\Lambda=43$. The red dashed lines denotes the spectrum read off from the functional saturating the lower bound in Figure \ref{B2fig}, which we identify with the $U(1)_{2M}\times U(1+M)_{-2M}$ theory. The gray dots denote the GFFT and free theory values from Table \ref{freeTable}.}
\label{A20fig}
\end{center}
\end{figure}

Let us now discuss upper and lower bounds on OPE coefficients for isolated superconformal blocks that appear in the $\<SSSS\>$.   (Recall that the isolated superconformal blocks are the short and semishort superblocks in Table~\ref{SupermultipletTable} which do not appear on the RHS of \eqref{LimitSpin0}--\eqref{LimitSpinEven}.)  Using the algorithm presented in \eqref{CondOPE}--\eqref{UpperOPE}, we determined such bounds as shows in Figure~\ref{A20fig}.  In these plots, our $\Lambda=39$ $\mathcal{N}=6$ upper/lower bounds are shown in black, and they can be compared to the $\Lambda=43$ $\mathcal{N}=8$ bounds computed in \cite{Agmon:2017xes}, which in these figures are shown in blue.  As in Figure~\ref{B2fig} discussed above, in all these plots, the $\mathcal{N}=6$ and $\mathcal{N}=8$ lower bounds meet at around $\frac{16}{c_T}\sim .71$.  Note that the $\mathcal{N}=6$ upper/lower bounds do not converge at the GFFT and free theory points yet, whose exactly known values where listed in Table~\ref{freeTable} and are denoted by gray dots, which is evidence that they are not fully converged. The exception is the bound on the OPE coefficient for $(A,+)_{\ell+2,\ell}^{[002]}$, which is our most constraining plot.  

In addition to the upper and lower bounds, in Figure~\ref{A20fig} we also plotted in dashed red the values of the OPE coefficients as extracted from the extremal functional under the assumption that the lower bound of Figure~\ref{B2fig} is saturated.  As we can see, the extremal functional values for the OPE coefficients come close to saturating several of the bounds in this figure, but not all.

\subsection{Bounds on long scaling dimensions}

Lastly, let us show bounds on the scaling dimensions of the long multiplets. To find upper bounds on the scaling dimension $\Delta^*$ of the lowest dimension operator in a long supermultiplet with spin $\ell^*$ that appears in \eqref{crossing2}, we consider linear functionals $\alpha$ satisfying
  \es{CondLong}{
  &\alpha(d^i_\text{Id})+ \frac{64}{c_T}\alpha( d^i_{(B,2)_{1,0}^{[011]}} ) = 1 \,,   \\
  &\alpha(d^i_{I}) \geq 0 
    \,, \;\;\quad\qquad \text{for all short and semi-short ${I} \notin \{ \text{Id}, (B,2)_{1,0}^{[011]} \}$} \,, \\
  &\alpha(d^i_{I}) \geq 0  \,, \quad\;\; \qquad\text{for all long $I$ with $\Delta_{I}\geq \Delta'_{I}$}\,,  
 }
 where we set all $\Delta'_{I}$ to their unitarity values except for $\Delta'_{I*}$. If such a functional $\alpha$ exists, then this $\alpha$ applied to \eqref{crossing2} along with the reality of $\lambda_{I}$ would lead to a contradiction. By running this algorithm for many values of $(c_T,\Delta'_{I*})$ we can find an upper bound on $\Delta'_{I*}$ in this plane. 
 
Since for the long multiplets $\text{Long}_{\Delta, \ell}^{[000]}$ of even spin $\ell$ there are several superconformal blocks (two for $\ell = 0$ and three for $\ell \geq 2$), we can ask what the upper bound on $\Delta$ is independently for each superconformal structure $\text{Long}_{\Delta, \ell}^{[000], n}$.  To be explicit, we denote by $\Delta_{(\ell, n)}$ the bound obtained from the structure $\text{Long}_{\Delta, \ell}^{[000], n}$.  (For odd $\ell$, we simply denote the bound by $\Delta_\ell$.). 

For general ${\cal N} = 6$ SCFTs, the bounds for different $n$ need not be the same, but we do expect that a long multiplet $\text{Long}_{\Delta, \ell}^{[000]}$ in a generic ${\cal N} = 6$ SCFTs will contribute to all superconformal structures and, if this is the case, the lowest dimension long multiplet must obey all the bounds obtained separately from each superconformal structure.  Since the superconformal structures are distinguished by they parity $\cP$ and $\cZ$ charges (see Table~\ref{SupermultipletTable}), in an SCFT that preserves these symmetries, $\Delta_{(\ell, n)}$ represents the upper bound on the lowest long multiplet with the $\cP$ and $\cZ$ charges that correspond to the structure  $\text{Long}_{\Delta, \ell}^{[000], n}$ as given in Table~\ref{SupermultipletTable}.

\begin{figure}[]
\begin{center}
        \includegraphics[width=.85\textwidth]{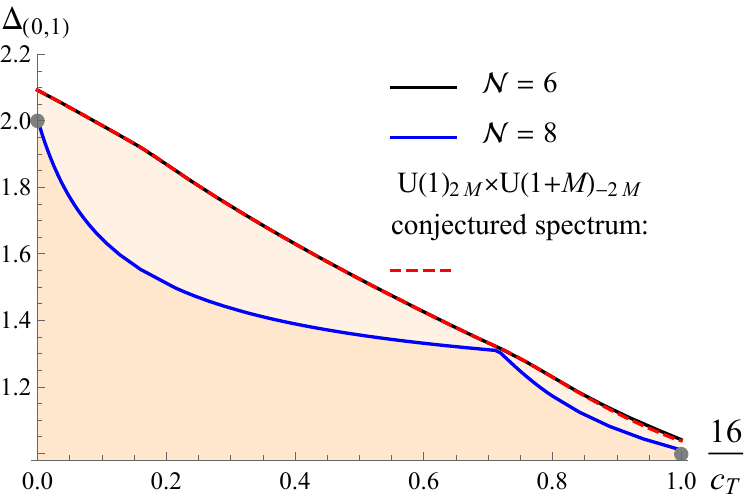}
\caption{Upper bounds on the scaling dimension of the lowest dimension $\ell=0$ long multiplet in terms of $c_T$ for the $\text{Long}_{\Delta, 0}^{[000], 1}$ superconformal structure, which for parity preserving theories has the same parity as the superprimary. The orange shaded region is allowed, and the plot ranges from the GFFT limit $c_T\to\infty$ to the free theory $c_T=16$. The black line denotes the $\mathcal{N}=6$ upper bound computed in this work with $\Lambda=39$, the blue line denotes the $\mathcal{N}=8$ upper bound computed in \cite{Agmon:2017xes} with $\Lambda=43$. The red dashed line denotes the spectrum read off from the functional saturating the lower bound in Figure \ref{B2fig}, which we identify with the $U(1)_{2M}\times U(1+M)_{-2M}$ theory. The gray dots denote the GFFT and free theory values from Table \ref{freeTable}.}
\label{scal0fig}
\end{center}
\end{figure}

\begin{figure}[]
\begin{center}
        \includegraphics[width=.85\textwidth]{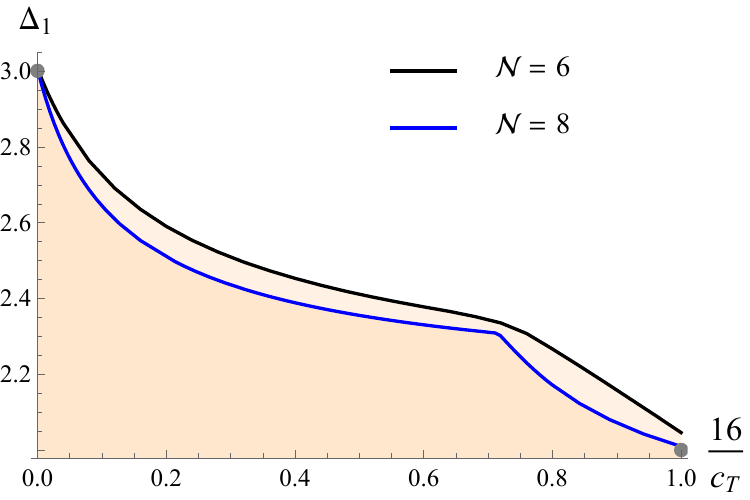}
\caption{Upper bounds on the scaling dimension of the lowest dimension $\ell=1$ long multiplet (which has a unique superconformal structure) in terms of $c_T$, where the orange shaded region is allowed, and the plot ranges from the GFFT limit $c_T\to\infty$ to the free theory $c_T=16$. The black line denotes the $\mathcal{N}=6$ upper bound computed in this work with $\Lambda=39$, the blue line denote the $\mathcal{N}=8$ upper bound computed in \cite{Agmon:2017xes} with $\Lambda=43$.  The gray dots denote the GFFT and free theory values from Table \ref{freeTable}.}
\label{scal1fig}
\end{center}
\end{figure}

\begin{figure}[]
\begin{center}
        \includegraphics[width=.85\textwidth]{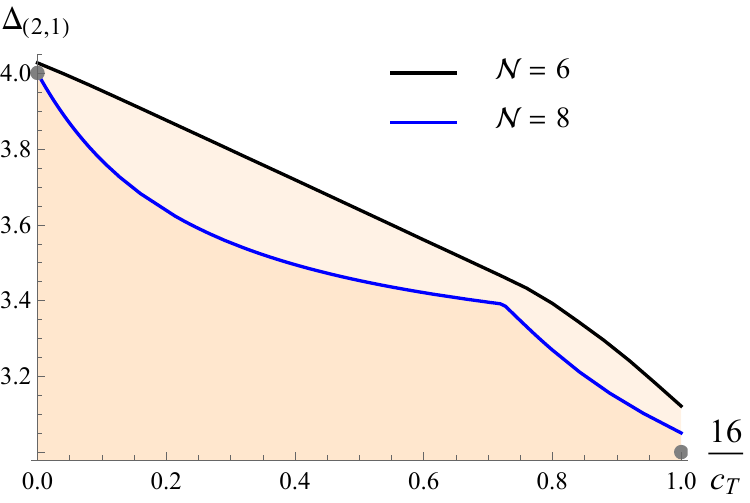}
\caption{Upper bound on the scaling dimension of the lowest dimension $\ell=2$ long multiplet in terms of $c_T$ for the $\text{Long}_{\Delta, 2}^{[000], 1}$ superconformal structure, which for parity preserving theories has the same parity as the superprimary. The orange shaded region is allowed, and the plot ranges from the GFFT limit $c_T\to\infty$ to the free theory $c_T=16$. The black line denotes the $\mathcal{N}=6$ upper bound computed in this work with $\Lambda=39$, the blue line denote the $\mathcal{N}=8$ upper bound computed in \cite{Agmon:2017xes} with $\Lambda=43$.  The gray dots denote the GFFT and free theory values from Table \ref{freeTable}.}
\label{scal2fig}
\end{center}
\end{figure}

\begin{figure}[]
\begin{center}
        \includegraphics[width=.85\textwidth]{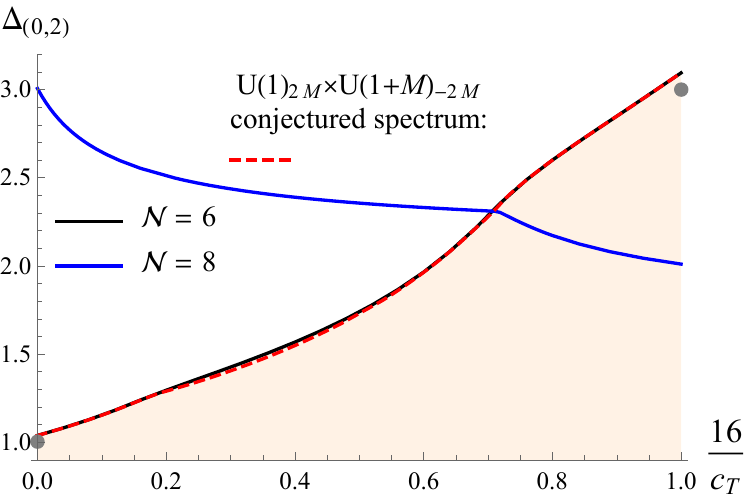}
\caption{Upper bounds on the scaling dimension of the lowest dimension $\ell=0$ long multiplet in terms of $c_T$ for the second superconformal structure $\text{Long}_{\Delta, 0}^{[000], 2}$, which for parity preserving theories has the opposite parity as the superprimary. The orange shaded region is allowed, and the plot ranges from the GFFT limit $c_T\to\infty$ to the free theory $c_T=16$. The black line denotes the $\mathcal{N}=6$ upper bound computed in this work with $\Lambda=39$, the blue lines denote the $\mathcal{N}=8$ upper bounds computed in \cite{Agmon:2017xes} with $\Lambda=43$. At $c_T=16$ the $\mathcal{N}=8$ upper bound does not apply, as the $\mathcal{N}=8$ superblock becomes a conserved current that does not decompose to $\text{Long}_{\Delta, 0}^{[000], 2}$. The red dashed line denotes the spectrum read off from the functional saturating the lower bound in Figure \ref{B2fig}, which we identify with the $U(1)_{2M}\times U(1+M)_{-2M}$ theory. The gray dots denote the GFFT and free theory values from Table \ref{freeTable}.}
\label{scal0pfig}
\end{center}
\end{figure}

\begin{figure}[]
\begin{center}
        \includegraphics[width=.49\textwidth]{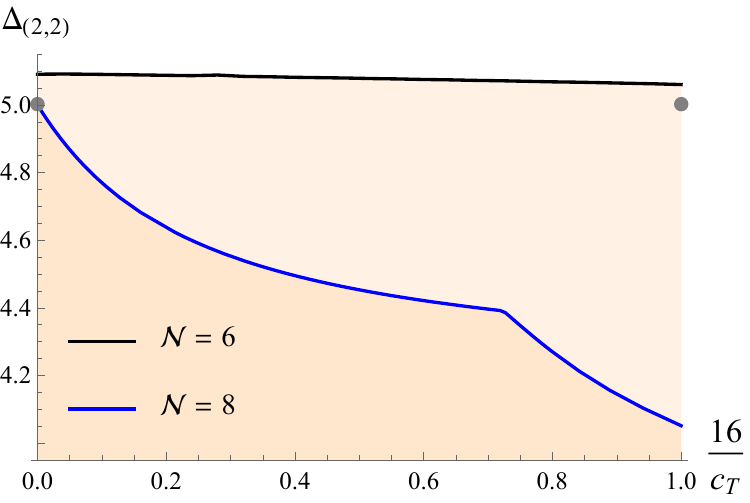} \includegraphics[width=.49\textwidth]{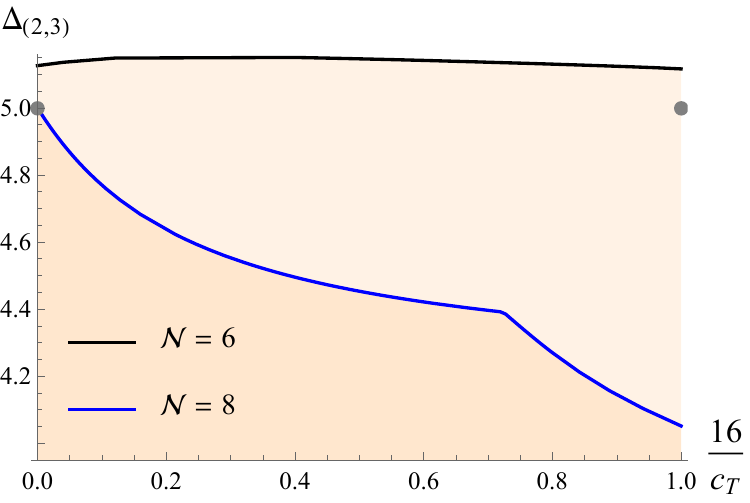}
\caption{Upper bounds on the scaling dimension of the lowest dimension $\ell=2$ long multiplet in terms of $c_T$ for the $\text{Long}_{\Delta, 2}^{[000], 2}$ ({\bf left}) and $\text{Long}_{\Delta, 2}^{[000], 3}$ ({\bf right}) superconformal structures, which for parity preserving theories have the opposite parity as the superprimary, and for $\mathcal{Z}$ preserving theories has the same charge for $\text{Long}_{\Delta, 2}^{[000], 2}$ and the opposite charge for $\text{Long}_{\Delta, 2}^{[000], 3}$. The orange shaded region is allowed, and the plot ranges from the GFFT limit $c_T\to\infty$ to the free theory $c_T=16$. The black lines denotes the $\mathcal{N}=6$ upper bounds computed in this work with $\Lambda=39$, the blue lines denote the $\mathcal{N}=8$ upper bounds computed in \cite{Agmon:2017xes} with $\Lambda=43$. At $c_T=16$ the $\mathcal{N}=8$ upper bound does not apply, as the $\mathcal{N}=8$ superblock becomes a conserved current that does not decompose to $\text{Long}_{\Delta, 2}^{[000], 2}$ or $\text{Long}_{\Delta, 2}^{[000], 3}$. The gray dots denote the GFFT and free theory values from Table \ref{freeTable}.}
\label{scal2pfig}
\end{center}
\end{figure}

Our bounds on the scaling dimensions of long multiplets of spin $\ell = 0, 1, 2$ are presented in Figures~\ref{scal0fig}--\ref{scal2pfig}.  The bounds on the first superconformal structure, namely $\Delta_{(0, 1)}$ for $\ell=0$, $\Delta_1$ for $\ell=1$, and $\Delta_{(2, 1)}$ for $\ell = 2$, shown in Figures~\ref{scal0fig},~\ref{scal1fig}, and~\ref{scal2fig}, respectively, are relatively straightforward to understand:  they interpolate smoothly between the values of the corresponding scaling dimensions at the free ${\cal N} = 6$ hypermultiplet theory at $\frac{16}{c_T} = 1$ and the GFFT at $\frac{16}{c_T} = 0$.  This is unlike in the ${\cal N} = 8$ case where the upper bounds on the scaling dimension exhibit kinks at $\frac{16}{c_T}\sim .71$.

The bounds on the other structures, namely $\Delta_{(0, 2)}$ for $\ell=0$ and $\Delta_{(2, 2)}$ and $\Delta_{(2, 3)}$ for $\ell=2$, are more subtle.  Let us start discussing $\Delta_{(0, 2)}$ first.  Recall that, as per \eqref{LimitSpin0}, the unitarity limit of the $\text{Long}_{\Delta, 0}^{[000], 2}$ superconformal block is precisely given by the $(B,1)_{2,0}^{[200]}$ superconformal block, so what bound $\Delta_{(0, 2)}$ we find depends on what assumptions we make about the possibility of having a $(B,1)_{2,0}^{[200]}$ multiplet appearing in the $S \times S$ OPE\@. If we assume that there are no $(B,1)_{2,0}^{[200]}$ operators that appear in the $S \times S$ OPE, then we obtain the bound in Figure~\ref{scal0pfig}.  As we can see from this figure, the bound $\Delta_{0,2}$ smoothly goes from the GFFT value 1 at $\frac{16}{c_T} = 0$ to the free theory value 3 at $\frac{16}{c_T} = 1$. This suggests that it is possible for ${\cal N} = 6$ SCFTs to not contain $(B,1)_{2,0}^{[200]}$ multiplets, and indeed the $U(1)_{2M} \times U(1+M)_{-2M}$ theory is an example of an ${\cal N} = 6$ SCFT with this property.  

For $\Delta_{(0, 2)}$, the extremal functional result that we identify with the $U(1)_{2M} \times U(1+M)_{-2M}$ theory, shown in red, is close to the unitarity value for large $c_T$. This is suggestive of an approximately broken higher spin current, as one generically expects for such vector-like theories. For $\Delta_1$ and $\Delta_{(2, 1)}$ we do not yet have sufficient precision to accurately show the extremal functional results. We would also expect to see approximately conserved currents in  $\Delta_1$, but since it is single trace its OPE coefficient start at $O(c_T^{-1})$, which make it especially difficult to see from numerics.

In \cite{Chester:2014mea}, it was shown that all $\mathcal{N}=8$ SCFTs with $\frac{16}{c_T}<.71$ contain a short multiplet (namely $(B, 2)_{2, 0}^{[0200]}$) that upon reduction to ${\cal N} = 6$ includes a $(B,1)_{2,0}^{[200]}$ multiplet---see~\eqref{b8to6} for the reduction of ${\cal N} = 8$ superconformal blocks to ${\cal N} = 6$ ones.  Thus, the bound presented in Figure~\ref{scal0pfig} does not have to apply to ${\cal N} = 8$ SCFTs with $\frac{16}{c_T}<.71$.  Indeed, the upper bound determined in \cite{Chester:2014mea} on the lowest long multiplet $\text{Long}_{\Delta, 0}^{[0000]}$ of an ${\cal N} = 8$ SCFT, which upon reduction to ${\cal N} = 6$ contributes to the $\text{Long}_{\Delta, 0}^{[000], 2}$ superconformal block, is given by the blue line in Figure~\ref{scal0pfig}.  Because for $\frac{16}{c_T}<.71$ this line lies outside the ${\cal N} = 6$ bound in Figure~\ref{scal0pfig}, it follows that any ${\cal N} = 8$ SCFT that saturates the ${\cal N} = 8$ bound must contain a $(B,1)_{2,0}^{[200]}$ multiplet from the ${\cal N} = 6$ point of view.\footnote{It may seem curious that in Figure~\ref{scal0pfig}, at $c_T = 16$, where the free theory has in fact ${\cal N} = 8$ SUSY,  the lowest operator (marked by a gray dot) that contributes to the $\text{Long}_{\Delta, 0}^{[000], 2}$  block does not obey the ${\cal N} =8$ bound in blue.  This is because in that case the $\text{Long}_{\Delta, 0}^{[0000]}$ multiplet that gives the ${\cal N} = 8$ bound is replaced by an ${\cal N} = 8$ conserved current multiplet $(A, \text{cons.})_{1, 0}^{[0000]}$ which no longer contributes to the $\text{Long}_{\Delta, 0}^{[000], 2}$.  The gray dot in Figure~\ref{scal0pfig} instead comes from a $\text{Long}_{2, 0}^{[0000]}$ multiplet in ${\cal N} = 8$.}

For the $\Delta_{(2,2)}$ and $\Delta_{(2,3)}$ bounds, we should recall that, at unitarity, the $(A,0)_{\Delta,\ell+1}^{[000], 2}$ and $(A,0)_{\Delta,\ell+1}^{[000], 3}$ superblocks become $(A,1)_{\ell+5/2,\ell+1/2}^{[100],1}$ and $(A,1)_{\ell+5/2,\ell+1/2}^{[100],2}$, as per \eqref{LimitSpinEven}.  Thus, the bound $\Delta_{(2, n)}$, with $n=2, 3$ depends on what we assume about $(A,1)_{\ell+5/2,\ell+1/2}^{[100],n-1}$.  Our first result is that if we assume that $(A,1)_{7/2,3/2}^{[100],n-1}$ does not appear in the $S\times S$ OPE, we find that the long multiplet bound is at the unitarity value $\Delta_{(2, n)} = 3$, which implies that the assumption that  $(A,1)_{7/2,3/2}^{[100],n-1}$ does not appear in the $S\times S$ OPE was incorrect.  Thus, all ${\cal N} = 6$ SCFTs must contain $(A,1)_{7/2,3/2}^{[100]}$ multiplets!  This is consistent with the result in \cite{Chester:2014mea} that all $\mathcal{N}=8$ SCFTs must contain an $\mathcal{N}=8$ $(A, 2)_{3, 1}^{[0020]}$ multiplet, which reduces to the $(A,1)_{7/2,3/2}^{[100]}$ $\mathcal{N}=6$ multiplet as per \eqref{b8to6}.

Next, we can derive revised bounds $\Delta_{(2, n)}$, with $n=2, 3$, under the assumption that the $S\times S$ OPE contains the $(A,1)_{7/2,3/2}^{[100],n-1}$ superblocks.    As we can see from Figure~\ref{scal2pfig}, we found that the bounds $\Delta_{(2, n)}$ are slightly above $5$ for all $c_T$, with little dependence on $c_T$. This is consistent with the value at both GFFT and free theory. For comparison, we also showed the second lowest operator for $\mathcal{N}=8$ theories, which corresponds to the lowest long spin 2 $\mathcal{N}=8$ operator.\footnote{It may again seem curious that at $c_T = 16$, where the free theory has ${\cal N} = 8$ SUSY, the gray dot does not obey the ${\cal N} = 8$ bound.  At exactly the free theory point, this $\mathcal{N}=8$ operator becomes a conserved current which no longer decomposes to a parity odd $\mathcal{N}=6$ long multiplet, which is why the $\mathcal{N}=6$ free theory value of the second lowest operator, denoted by the second lowest gray dot, does not coincide with the $\frac{16}{c_T}\to1$ limit of the $\mathcal{N}=8$ upper bound.} We do not show any extremal functional results for these plots, because we do not yet have sufficient numerical precision.

\subsection{Islands for semishort OPE coefficients}
\label{islands}

\begin{figure}[]
\begin{center}
        \includegraphics[width=\textwidth]{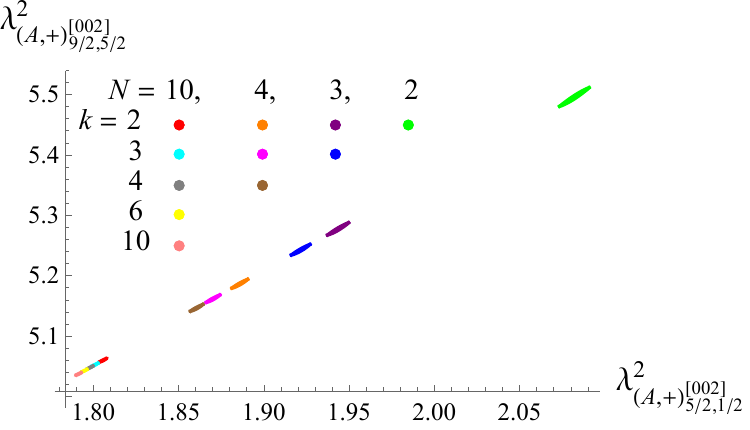}
\caption{Islands in the space of the semi-short OPE coefficients $\lambda_{(A, +)_{5/2,1/2}^{[002]}}^2$, $\lambda_{(A, +)_{9/2,5/2}^{[002]}}^2$ (to be defined precisely later) for $U(N)_k\times U(N)_{-k}$ ABJM theory for various $N,k$. These bounds are derived from the $\mathcal{N}=6$ bootstrap with $\Lambda=39$ derivatives, and with the short OPE coefficients (i.e. $c_T$ and $\lambda^2_{(B,2)_{2,0}^{[022]}}$) fixed to their values in each theory using the exact localization results of \cite{Chester:2014mea} for $N=2,3,4$ as shown in Table \ref{ABJMOPEs} and the all orders in $1/N$ formulae in \cite{Agmon:2017xes} for $N=10$ as shown in Table \ref{tab10}. }
\label{islandsfig}
\end{center}
\end{figure}

In the previous subsection we discussed numerical bounds that apply to all 3d $\mathcal{N}=6$ SCFTs. In particular, we noticed that the upper/lower bounds on $(A,+)_{\ell+2,\ell}^{[002]}$ for $\ell=1/2,5/2$ were extremely constraining. This implies that for a given value of $c_T$, we could find a small island in the space of $(\lambda^2_{(A,+)_{5/2,1/2}^{[002]}},\lambda^2_{(A,+)_{9/2,5/2}^{[002]}})$ using the OPE island algorithm described in the previous subsection. We can make this island even smaller, and correlate it to a physical theory, by imposing values of $c_T$ and $\lambda^2_{(B,2)_{{2,0}}^{[022]}}$ computed using localization in Section~\ref{EXACT}. We show the results of these plots for $U(N)_k\times U(N)_{-k}$ for a variety of $N,k$ in the Figure~\ref{islandsfig}. Note that the islands are small enough that we can distinguish each value of $N$ and $k$, which allows us to non-perturbatively interpolate between M-theory at small $k$ and Type IIA at large $k$.

\begin{figure}[]
\begin{center}
        \includegraphics[width=\textwidth]{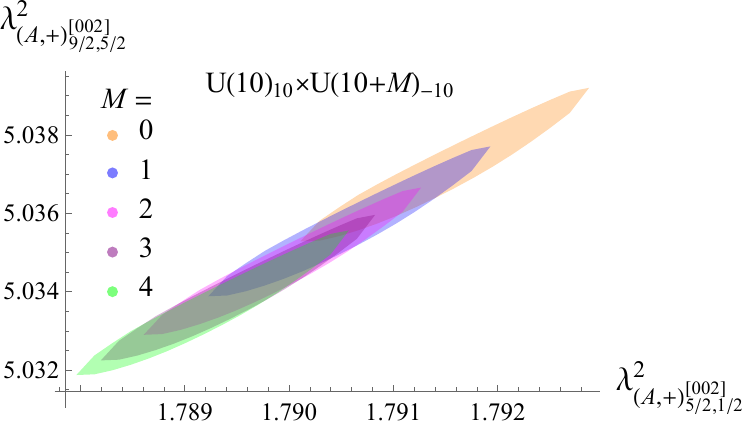}
\caption{Islands in the space of the semi-short OPE coefficients $\lambda_{(A, +)_{5/2,1/2}^{[002]}}^2$, $\lambda_{(A, +)_{9/2,5/2}^{[002]}}^2$ for $U(10)_{10}\times U(10+M)_{-10}$ ABJM theory for  $M<k/2=5$. These bounds are derived from the $\mathcal{N}=6$ bootstrap with $\Lambda=39$ derivatives, and with $c_T$ and $\lambda^2_{(B,2)_{2,0}^{[022]}}$ fixed to their values in each theory using the all orders in $1/N$ localization formulae in \cite{Agmon:2017xes} for $N=10$. Note that the axes describe a very narrow range in parameter space.}
\label{islands2fig}
\end{center}
\end{figure}

One difficulty with trying to fix a physical theory by imposing two exactly computed quantities, $c_T$ and $\lambda^2_{(B,2)_{{2,0}}^{[022]}}$, is that the most general $\mathcal{N}=6$ ABJ(M) theory has gauge group $U(N)_k\times U(N+M)_{-k}$ and so is described by 3 parameters $N,k,M$. While for physical theories these parameters should be integers, we expect that the numerical bootstrap should find theories with any real value of these parameters, so we are effectively trying to parameterize a 3-dimensional space of theories. Since we are only imposing two quantities, these islands are expected to have a finite area even at high numerical precision corresponding to the third direction in ``theory space''. Thankfully, this third direction appears to be very small. We can quantify this by fixing $N=k=10$ and computing islands for several different values of $M\leq k/2=5$. As shown in Figure \ref{islands2fig}, the island is not very sensitive to the value of $M<N$, which explains why we were able to get such small islands in a 3-dimensional conformal manifold by just imposing two values of the parameters.

\section{Conclusion}
\label{disc}

In this paper, we have three main results.  Firstly, we setup up the conformal bootstrap for the stress tensor multiplet scalar bottom component four-point function $\langle SSSS\rangle$, by computing the superblock decomposition and deriving the four independent crossing equations. Secondly, we computed both $c_T$ and the short OPE coefficient squared $\lambda^2_{(B,2)_{2, 0}^{[022]}}$ using the supersymmetric localization \cite{Kapustin:2009kz} formula for the mass deformed free energy in both the $U(N)_{k}\times U(N+M)_{-k}$ and $SO(2)_{2k}\times USp(2+2M)_{-k}$ ABJM theories for a wide range of $N,k,M$, including the exact results for the vector-like limit $N=1$ and arbitrary $k,M$, which extends the all orders in $1/N$ results for the $M\leq k\leq N$ case \cite{Agmon:2017xes}. 
Finally, we combined all these ingredients to non-perturbatively study the ABJM theories. In particular, by inputting the exact values of $c_T$ and $\lambda^2_{(B,2)_{2,0}^{[022]}}$ for a given $k,N$ and $M=0$, we found precise rigorous islands in the space of semishort OPE coefficients that interpolate between M-theory at small $k$ and type IIA string theory at $k\sim N$. We also conjectured that in the infinite precision limit, the numerical lower bound on $\lambda^2_{(B,2)_{2,0}^{[022]}}$ is saturated by the family of $U(1)_{2M}\times U(1+M)_{-2M}$ theories, which allowed us to non-rigorously read off all CFT data in $\langle SSSS\rangle$ using the extremal functional method.  Interestingly, in the regime of large $c_T$, we found a spin zero long multiplet whose scaling dimension approaches a zero spin conserved current multiplet at large $c_T$, as expected from weakly broken higher spin symmetry.   

There are several ways we can improve upon our 3d $\mathcal{N}=6$ bootstrap study. From the numerical perspective, it will be useful to improve the precision of our study. This is parameterized by the parameter $\Lambda$ defined in the main text. While we used $\Lambda=39$ in this work, which is close to the $\Lambda=43$ values used in the analogous $\mathcal{N}=8$ studies \cite{Agmon:2017xes,Agmon:2019imm}, for $\mathcal{N}=6$ this value has not led to complete convergence. For instance, we found the lower bound $c_T\geq 15.5$, compared to the $\mathcal{N}=8$ result $c_T\geq 15.9$; both are expected to converge to the free theory $c_T=16$. More physically, we expect that approximately conserved currents should appear in the extremal functional that conjecturally describes the $U(1)_{2M}\times U(1+M)_{-2M}$ theory. We found such an operator in the zero spin sector as shown in Figure \ref{scal0pfig}, but do not yet have sufficient precision to see them for higher spin.  The main obstacle to increasing $\Lambda$ at the moment is not SDPB, which due to the recent upgrade \cite{Landry:2019qug} can easily handle four crossing equations at very high $\Lambda$, but simply the difficulty in computing numerical approximations to the superblocks at large $\Lambda$. In particular it would be extremely useful to have an efficient code for approximations of linear combinations of conformal blocks with $\Delta$ dependent coefficients around the crossing symmetric point. Currently the code \texttt{scalar\_blocks} code, found on the bootstrap collaboration website,\footnote{This code can be found at \texttt{https://gitlab.com/bootstrapcollaboration/scalar\_blocks/blob/\\master/Install.md}.} is only able to efficiently compute single conformal blocks.

From the analytical perspective, it would be interesting to derive the $1/c_T$ corrections to $\langle SSSS \rangle$ for ABJ theories in the vector-like limit, such as for $U(N)_{k}\times U(N+M)_{-k}$ at finite $N$ and large $M,k$ with fixed $M/k$. This would complement the order $1/c_T$ correlator that was computed in the supergravity limit in \cite{Zhou:2017zaw} and was successfully matched to numerical bootstrap results in \cite{Chester:2018lbz}. We expect that the vector-like correlator can be computed by generalizing \cite{Turiaci:2018nua,Maldacena:2012sf} to $\mathcal{N}=6$, and hope to report on these results soon in a future publication.\footnote{The result can be found at \cite{Binder:2021cif}.}

We could also make further use of localization to improve our results. In this study we were only able to impose two analytic constraints, $c_T$ and $\lambda^2_{(B,2)_{2,0}^{[022]}}$, while ABJM is parameterized by three parameters $N,M,k$. For this reason there are not enough constraints to uniquely pick out a single ABJM theory and so we should not expect our islands to shrink indefinitely as we increase $\Lambda$. We think this is the reason why the islands shown in Figure~\ref{islandsfig}, while small, are still much bigger than the $\mathcal{N}=8$ islands computed in \cite{Agmon:2019imm}.   Localization conveniently provides us a third quantity given by four mixed mass derivatives of the mass deformed free energy, which as shown \cite{Binder:2019mpb} is related to an integral of $\langle SSSS\rangle$. While this integrated constraint can be imposed analytically in a large $N$ expansion as in \cite{Binder:2018yvd,Binder:2019mpb}, it is not yet known how to impose it on the numerical bootstrap in our case. Perhaps the method used in \cite{Lin:2015wcg}, where a similar integrated constraint was successfully imposed on the numerical bootstrap of a certain supersymmetric 2d theory, could be applied to our case.  Another option would be to look at a larger system of correlators involving fermions.  This would allow us to impose parity, which would restrict the set of known ${\cal N} = 6$ SCFTs to a few families such as $U(N)_k \times U(N)_{-k}$ parameterized by only two parameters each.

Once we can fully fix the three-parameters ABJM theory, it would be interesting to see if we can match  integrability results computed for the lowest dimension singlet 	scaling dimension in the leading large $N$ 't Hooft limit at fixed $\lambda_\text{'t Hooft}=N/k$ and $M=0$. On the integrability side some results are available, for instance, in  \cite{Bombardelli:2018bqz,LevkovichMaslyuk:2011ty}.  On the localization side, we would need to compute the derivatives of the mass deformed free energy in the $1/N$ expansion at finite $\lambda_\text{'t Hooft}$. In fact, the zero mass free energy has already been computed in this limit in \cite{Drukker:2010nc} by applying topological recursion to the Lens space $L(2,1)$ matrix model, so computing $c_T$ and $\lambda^2_{(B,2)_{2,0}^{[022]}}$ should correspond to just computing two- and four-body operators in this matrix model. This could potentially lead to the first precise comparison between integrability and the numerical conformal bootstrap.

Finally, it would be interesting to consider the superconformal block decomposition of other correlators involving operators which are less than half-BPS but still have scalar superprimaries, which makes them feasible to numerical bootstrap. For instance, in $\cN=4$ theories the stress tensor multiplet is only $1/4$-BPS but still has a scalar superprimary. Similarly, while conserved current multiplets are half-BPS for $\cN=4$ theories (and were studied using the numerical bootstrap in \cite{Chang:2019dzt}), for $\cN=3$ theories they are $1/3$-BPS and for $\cN = 2$ theories are only $1/4$-BPS\@. Many localization results exist which could be applied to these cases, including $\cN = 4$ results computed in \cite{Chester:2020jay}.

\section*{Acknowledgments} 

We thank Ofer Aharony, Oren Bergman, Simone Giombi, and Yuji Tachikawa for useful discussions and correspondence.   DJB, MJ, and SSP are supported in part by the Simons Foundation Grant No.~488653, and by the US NSF under Grant No.~1820651\@.  DJB is also supported in part by the General Sir John Monash Foundation.  SMC is supported by the Zuckerman STEM Leadership Fellowship.

\appendix

\section{Multiplets with only $\bf 15$s and $\bf1$s in the $S \times S$ OPE}
\label{allowedHS}

In the main text we were able to restrict the possible superconformal blocks where operators in the $\bf 84$, $\bf 45$, $\overline{\bf 45}$ or $\bf 20'$ are exchanged. This leaves us to consider superblocks where the only exchanged operators are in the $\bf 1$ or $\bf 15$. We will analyze this possibility using the superconformal Ward identities.

Let us fix some supermultiplet $\cM$ and define
\begin{equation}
\cS_{\bf r}^{(\cM)}(U,V) = \sum_{(\Dk,\ell,{\bf r})\in\cM} a_{\Dk,\ell,\bf r}g_{\Dk,\ell}(U,V)
\end{equation}
to be the contribution from $s$-channel $\cM$ exchange to $\cS_{\bf r}(U,V)$. The superconformal Ward identities apply to each superblock independently, and so $\cS_{\bf r}^{(\cM)}(U,V)$ must satisfy the Ward identities \eqref{SSSSward}. If we demand that no operators in the $\bf 84$, $\bf 45$, $\overline{\bf 45}$ or $\bf 20'$ are exchanged, we then find that
\begin{equation}\label{wardno15}\begin{split}
U^2\partial_{U}\left(\cS_{\bf 1}^{(\cM)} + \cS_{{\bf 15}_s}^{(\cM)}\right)+\left[2U(U+2V-2)\partial_U+4UV\partial_V-2(V-1)\right] \cS_{{\bf 15}_a}^{(\cM)} &= 0 \\
U\partial_V(\cS_{\bf 1}^{(\cM)} + \cS_{{\bf 15}_s}^{(\cM)})+\left[-4U\partial_U-2U\partial_V+2\right]\cS_{{\bf 15}_a}^{(\cM)} &= 0\,.
\end{split}\end{equation}

To make further progress we can consider the correlator
\begin{equation}\begin{aligned}
\langle S(\vec x_1,X_1)S(\vec x_2,X_2)&P(\vec x_3,X_3)P(\vec x_4,X_4)\rangle 
  = \frac 1{x_{12}^2 x_{34}^4} \sum_{i=1}^6 \cP^i(U,V) {\cal B}_i \,,
\end{aligned}\end{equation}
where ${\cal B}_{i}$ are defined as in \eqref{BasisElems}. The Ward identities relating $\cP^i(U,V)$ to $\cS^i(U,V)$ have been computed in \cite{Binder:2019mpb}. Note that because both $S$ and $P$ transform in the $\bf 15$ of $\mathfrak{so}(6)$, the correlators $\<SSSS\>$ and $\<SSPP\>$ have the same R-symmetry structures. Since we are interested in $s$-channel conformal block expansion, we are led to define $\cP_{\bf r}(U,V)$ in an analogous fashion to $\cS_{\bf r}(U,V)$ in \eqref{B}. 

If $s$-channel exchange of $\cM$ only contributes to the $\bf 1$ and $\bf 15$ channels in the $\<SSSS\>$ correlator, this must also be true of $\<SSPP\>$ and so
\begin{equation}
\cP_{\bf 20'}^{(\cM)}(U,V) = \cP_{{\bf 45}\oplus\overline{\bf45}}^{(\cM)}(U,V) = \cP_{\bf 84}^{(\cM)}(U,V) = 0\,.
\end{equation}
Combining this with \eqref{wardno15} and the Ward identities derived in \cite{Binder:2019mpb}, we find that
\begin{equation}\label{Srconds}
\cD \cS^{(\cM)}_{\bf 1}(U,V) = \cD\cS^{(\cM)}_{{\bf 15}_s}(U,V) = \cD\cS^{(\cM)}_{{\bf 15}_a}(U,V)= 0 \,,
\end{equation}
where $\cD$ is the differential operator
\begin{equation}
\cD=2U^2\partial_U^2 + 2(U+V-1)\partial_U\partial_V+2UV\partial_V^2+U\partial_U+(1+2U-V)\partial_V\,.
\end{equation}

Our next step is to rewrite the cross-ratios $(U,V)$ using radial coordinates $(r,\eta)$
\begin{equation}
U = \frac{16r^2}{(1+r^2+2r\eta)^2}\,,\qquad V = \frac{(1+r^2-2r\eta)^2}{(1+r^2+2r\eta)^2}\,.
\end{equation}
Conformal blocks have a relatively simple form in radial coordinates:
\begin{equation}
g_{\Delta,\ell}(r,\eta) = r^{\Delta}\sum_{k = 0}^{\infty}r^{2k}p_{\Dk,\ell,k}(\eta) \,,
\end{equation}
where each $p_{\Dk,\ell,k}(\eta)$ is polynomial in $\eta$ \cite{Hogervorst:2013sma}. In particular, the leading term is given by
\begin{equation}
p_{\Dk,\ell,0}(\eta) = P_\ell(\eta)\,,
\end{equation}
where $P_n(x)$ is the $n^{\text{th}}$ Legendre polynomial. Since $\cS_{\bf r}^{(\cM)}(U,V)$ is the sum of a finite number of conformal blocks, we expect that
\begin{equation}\label{srblock}
\cS_{\bf r}^{(\cM)}(U,V) = r^\Delta\left(q_{\bf r}(\eta) + O(r^2)\right)
\end{equation}
for some polynomial $q_{\bf r}(\eta)$.

Let us translate \eqref{Srconds} into radial coordinates:
\begin{equation}
\left[r^2(r^2-1)\partial_r^2 + 2r^3\partial_r-(r^2-1)(\eta^2-1)\partial_\eta^2-2(r^2-1)\eta\partial_\eta\right]\cS_{\bf r}^{(\cM)}(r,\eta) = 0\,.
\end{equation}
Substituting \eqref{srblock} into this equation we find that $q_{\bf r}(\eta)$ satisfies Legendre's equation
\begin{equation}\label{qr2}
(1-\eta^2)q_{\bf r}''(\eta)-2\eta q_{\bf r}'(\eta)+\Dk(\Dk-1)q_{\bf r}(\eta) = 0\,.
\end{equation}
Hence $q_{\bf r}(\eta)$ is a polynomial if and only if $\Delta\in\mathbb Z$, in which case ${q_{\bf r}(\eta) = a P_{\Delta+1}(\eta)}$ for some arbitrary constant $a$. Since unitarity implies that $\Delta\geq0$, we conclude that $\cS_{\bf r}^{(\cM)}(r,\eta)$ includes a contribution from either an operator with twist $\Delta - \ell = 1$, or else from the identity operator $\Delta = \ell = 0$.

Any operator in a superconformal multiplet has twist greater than or equal to the twist of the superconformal primary. Thus if $\cM$ is not the trivial supermultiplet then its superconformal primary must have twist one. Examining Table~\ref{n6mults}, we see that aside from the stress-tensor multiplet the only other such multiplets are conserved currents: A-type multiplets whose superprimary is an R-symmetry singlet with conformal dimension $\Delta = \ell+1$.  We conclude that any superblock in which the only exchanged operators transform in the $\bf 1$ or $\bf 15$ must correspond to the exchange of the trivial, stress-tensor, or a conserved current multiplet.

\section{Superconformal algebra}
\label{SUPERCONFORMAL}

We follow the conventions of Appendix~B of \cite{Chester:2014mea} for both the conformal and superconformal generators.  In particular, the commutation relations for the conformal generators are
 \es{Conformal}{
[M_{\alpha}^{\,\,\beta}, P_{\gamma\delta}] &= \delta_{\gamma}^{\,\,\beta}P_{\alpha\delta} + \delta_{\delta}^{\,\,\beta}P_{\alpha\gamma} - \delta_{\alpha}^{\,\,\beta}P_{\gamma\delta} \,, \\
[M_{\alpha}^{\,\,\beta}, K^{\gamma\delta}] &= - \delta_{\alpha}^{\,\,\gamma}K^{\beta\delta} - \delta_{\alpha}^{\,\,\delta} K^{\beta\gamma} + \delta_{\alpha}^{\,\,\beta} K^{\gamma\delta} \,,  \\
[M_{\alpha}^{\,\,\beta}, M_{\gamma}^{\,\,\delta}] &= -\delta_{\alpha}^{\,\,\delta}M_{\gamma}^{\,\,\beta} + \delta_{\gamma}^{\,\,\beta} M_{\alpha}^{\,\,\delta} \,, \\
 [D, P_{\alpha\beta}] &= P_{\alpha\beta} \,, \\
   [D,K^{\alpha\beta}] &= -K^{\alpha\beta} \,,  \\
[K^{\alpha\beta}, P_{\gamma\delta}] &= 4\delta_{(\gamma}^{\,\,(\alpha}M_{\delta)}^{\,\,\beta)} + 4\delta_{(\gamma}^{\,\,\alpha}\delta_{\delta)}^{\,\,\beta}D \,. 
 }
In radial quantization, the conjugation properties of the conformal generators are
 \es{ConjugConf2}{
  (P_{\alpha\beta})^\dagger &= K^{\alpha\beta} \,, \qquad (K_{\alpha\beta})^\dagger = P^{\alpha\beta} \,, \\
  (M_{\alpha}{}^\beta)^\dagger &= M_\beta{}^\alpha \,, \qquad D^\dagger = D \,.
 } 

The extension of the conformal algebra to the $\mathfrak{osp}(\cN|4)$ superconformal algebra is given by
\begin{alignat}{3}
\{Q_{\alpha r} , Q_{\beta s}\} &= 2\delta_{rs} P_{\alpha\beta} \,, \qquad & \{S^{\alpha}_{\,\,r}, S^{\beta}_{\,\,s}\} &= -2\delta_{rs} K^{\alpha\beta} \,, \\
[K^{\alpha\beta},Q_{\gamma r}] &= -i\left(\delta_{\gamma}^{\,\,\alpha} S^{\beta}_{\,\,r} + \delta_{\gamma}^{\,\,\beta} S^{\alpha}_{\,\, r} \right) \,, \qquad & [P_{\alpha\beta}, S^{\gamma}_{\,\,r}] &= -i \left( \delta_{\alpha}^{\,\,\gamma} Q_{\beta r} + \delta_{\beta}^{\,\,\gamma} Q_{\alpha r} \right) \,, \\
[M_{\alpha}^{\,\,\beta}, Q_{\gamma r} ] &= \delta_{\gamma}^{\,\,\beta} Q_{\alpha r} - \frac{1}{2} \delta_{\alpha}^{\,\,\beta} Q_{\gamma r} \,, & [M_{\alpha}^{\,\,\beta}, S^{\gamma}_{\,\,r}] &= - \delta_{\alpha}^{\,\,\gamma} S^{\beta}_{\,\,r} + \frac{1}{2}\delta_{\alpha}^{\,\,\beta} S^{\gamma}_{\,\, r} \,, \\
[D,Q_{\alpha r}] &= \frac{1}{2} Q_{\alpha r} \,, & [D, S^{\alpha}_{\,\, r}] &= -\frac{1}{2} S^{\alpha}_{\,\,r} \,, \\
[R_{rs}, Q_{\alpha t}] &= i\left( \delta_{rt} Q_{\alpha s} - \delta_{st} Q_{\alpha r} \right) \,, & [R_{rs}, S^{\alpha}_{\,\,t}] &= i \left( \delta_{rt} S^{\alpha}_{\,\,s} - \delta_{st} S^{\alpha}_{\,\,r} \right) \,, \\
[R_{rs},R_{tu}] &= i \left(\delta_{rt} R_{su} + \cdots\right) \,, \qquad & \{Q_{\alpha r}, S^{\beta}_{\,\,s}\} &= 2i\left( \delta_{rs}\left(M_{\alpha}^{\,\,\beta} + \delta_{\alpha}^{\,\,\beta} D \right) - i \delta_{\alpha}^{\,\,\beta} R_{rs} \right) \,,
\end{alignat}
where $R_{rs}$ are the anti-symmetric generators of the $\mathfrak{so}(\cN)$ R-symmetry.  In addition to \eqref{ConjugConf2}, we also have the following conjugation properties:
 \es{SuperconfConj}{
  (Q_{\alpha r})^\dagger &= -i S^\alpha_r \,, \qquad (S^{\alpha}_r)^\dagger = -i Q_{\alpha r} \,, \\
  (R_{rs})^\dagger &= R_{rs} \,.
 }
 
\section{Characters of $\mathfrak{osp}(6|4)$}
\label{characters}

In this section we will review the character formulas of $\mathfrak{osp}(6|4)$, which were computed in \cite{Dolan:2008vc}, as well as their decomposition under  $\mathfrak{osp}(6|4)\rightarrow \mathfrak{so}(3,2)\oplus\mathfrak{so}(6)$. This decomposition was used in Section~\ref{FOURPOINT} to determine which conformal primaries reside in each supermultiplet appearing in the $S\times S$ OPE.

The $\mathfrak{osp}(6|4)$ characters are defined in terms of the quantum numbers and generators given in Section \ref{SSOPE} as
\begin{align}
\chi_{(\Delta;j;\mathrm{r})}(s,x,y) \equiv \mathrm{Tr}_{\cR_{(\Delta;j;\mathrm{r})}} \left(s^{2D} x^{2J_3} y_1^{H_1}y_2^{H_2} y_3^{H_3}\right)\,.
\end{align}
Their explicit form for the multiplets we consider are
\begin{align}
\chi^{(A,\pm)}_{(\Delta;j;r,r,r)}(s,x,y) &= s^{2\Delta} P(s,x) \sum_{a_1\ec a_2\ec a_3=0}^2 \,  \sum_{\bar{a}_1\ec\bar{a}_2\ec\bar{a}_3=0}^{1}  s^{a_1+a_2+a_3+\bar{a}_1+\bar{a}_2+\bar{a}_3} \chi_{2j+\bar{a}_1+\bar{a}_2+\bar{a}_3}(x) \notag\\
&\times\left(\prod_{i=1}^3\chi_{j_{a_i}}(x)\right) \, \chi_{(r+\bar{a}_1-a_1\ec r+\bar{a}_2-a_2\ec \pm r\pm\bar{a}_3\mp a_3)}(y) \ec \label{Apchar}\\
\chi^{(B,+)}_{(\Delta;0;r,r,r)}(s,x,y) &= s^{2\Delta} P(s,x) \sum_{a_1\ec a_2\ec a_3=0}^2 s^{a_1+a_2+a_3} \left(\prod_{i=1}^3\chi_{j_{a_i}}(x)\right) \, \chi_{(r-a_1\ec r-a_2\ec r-a_3)}(y) \ec
\end{align}
\begin{align}
\chi^{(A,n)}_{(\Delta;j;r_1,\ldots,r_1,r_{n+1},\ldots,r_3)}(s,x,y) &= s^{2\Delta} P(s,x) \sum_{a_1\ec a_2\ec a_3=0}^2 \sum_{\bar{a}_{n+1},\ldots,\bar{a}_3=0}^{2} \sum_{\bar{a}_{1},\ldots,\bar{a}_n=0}^{1} s^{a_1+a_2+a_3+\bar{a}_1+\bar{a}_2+\bar{a}_3} \chi_{2j+\bar{a}_1+\cdots+\bar{a}_n}(x) \notag\\
&\times \left(\prod_{i=n+1}^3\chi_{j_{\bar{a}_i}}(x) \right) \left(\prod_{i=1}^3 \chi_{j_{a_i}}(x)\right)  \, \chi_{(r_1+\bar{a}_1-a_1\ec r_2+\bar{a}_2-a_2\ec  r_3+\bar{a}_3-a_3)}(y)\ec \\
\chi^{(B,n)}_{(\Delta;0;r_1,\ldots,r_1,r_{n+1},\ldots,r_3)}(s,x,y) &= s^{2\Delta}P(s,x) \sum_{a_1\ec a_2\ec a_3\ec\bar{a}_{n+1},\ldots,\bar{a}_3=0}^{2} s^{a_1 + a_2 + a_3 + \bar{a}_{n+1} + \cdots + \bar{a}_3} \left(\prod_{i=n+1}^3\chi_{j_{\bar{a}_i}}(x) \right) \notag\\
&\times \left(\prod_{i=1}^3 \chi_{j_{a_i}}(x)\right)  \, \chi_{(r_1-a_1,\ldots,r_1-a_n,r_{n+1}+\bar{a}_{n+1}-a_{n+1},\ldots,r_3+\bar{a}_3-a_3)}(y) \ec\label{Bnchar}
\end{align}
where the long multiplet corresponds to $(A,0)$, we define $j_a\equiv a\,\,(\!\!\!\!\mod{2})$, the $\mathfrak{su}(2)$ and $\mathfrak{so}(6)$ characters are 
\begin{align}
\chi_j(x) &= \frac{x^{j+1}-x^{-j-1}}{x-x^{-1}}\ec\\
\chi_{\mathrm{r}}(y) &=\frac{\det\left[y_i^{r_j+3-j}+y_i^{-r_j-3+j}\right] + \det\left[y_i^{r_j+3-j}-y_i^{-r_j-3+j}\right]}{2 \prod_{1\leq i<j\leq 3} (y_i+y_i^{-1}-y_j-y_j^{-1}) }\,,\label{Rchar}
\end{align}
and the function $P(s,x)$ is related to the $\mathfrak{so}(3,2)$ character and takes the form
\es{PP}{
P(s,x) &= \frac{1}{1-s^4}\sum_{n=0}^{\infty} s^{2n}\chi_{2n}(x) \,.
}

The products of the $\mathfrak{su}(2)$ characters in \eqref{Apchar}--\eqref{Bnchar} are easily transformed into sums of such characters by decomposing $\mathfrak{su}(2)$ tensor products. After doing so, we see that \eqref{Apchar}--\eqref{Bnchar} become sums over $\mathfrak{so}(3,2)\oplus\mathfrak{so}(6)$ characters, as desired.\footnote{Sometimes the $\mathfrak{so}(6)$ characters in \eqref{Apchar}--\eqref{Bnchar} appear with negative Dynkin labels. One can then try to use the identity
\begin{align}
\chi_{\mathrm{r}^{\omega}}(y) = (-)^{\ell(\omega)} \chi_{\mathrm{r}}(y) \ec\notag
\end{align} 
to obtain a character with non-negative Dynkin labels. In this identity $\omega$ is an element of the $\mathfrak{so}(6)$ Weyl group $S_4$, $\mathrm{r}^{\omega} = \omega(\mathrm{r}+\rho)-\rho$ is a Weyl reflection, $\rho=(2,1,0)$ is the Weyl vector, and $(-)^{\ell(\omega)}$ is the signature of the Weyl transformation. If there is no Weyl transformation such that $\mathrm{r}^{\omega}$ correspond to non-negative integer Dynkin labels, then $\chi_{\mathrm{r}}=0$.}

\section{Decomposing $\mathcal{N}=8$ superblocks to $\mathcal{N}=6$}
\label{8to6}

In this appendix we discuss how the superblocks that appeared in the four-point function of the $\mathcal{N}=8$ stress tensor superprimary $\overline{S}$ decompose into the $\mathcal{N}=6$ superblocks discussed for $\langle SSSS\rangle$ in the main text. This serves as both a consistency check of our $\mathcal{N}=6$ superblocks, and also allows us to translate the $\mathcal{N}=8$ numerical bootstrap results of \cite{Chester:2014fya,Agmon:2017xes,Chester:2014mea,Agmon:2019imm} into $\mathcal{N}=6$ language, which we will use to compare to the $\mathcal{N}=6$ results in Section \ref{numerics}. $\overline{S}$ transforms in the $\bf35_c$ of the $\mathcal{N}=8$ R-symmetry group $SO(8)$, which decomposes to $SO(6)\times U(1)$ as
\es{35to15}{
{\bf35}\to{\bf 15}_0\oplus{\bf10}_{2}\oplus\overline{\bf10}_{-2}\,,
}
so $\overline{ S}$ decomposes to $S$ as well as the superprimaries of the multiplets $(B,+)_{1,0}^{020}$ and $(B,+)_{1,0}^{002}$ that are charged under $U(1)$. Since we are only interested in correlators of $S$, we will always restrict to $U(1)$ singlets when decomposing from $\mathcal{N}=8$ to $\mathcal{N}=6$ in this appendix, which we will denote using an arrow instead of an equality.

\begin{table}
\centering
\begin{tabular}{|c|c|r|c|c|}
\hline
Type    & $(\Delta,\ell)$     & $\mathfrak{so}(8)_R$ irrep  &spin $\ell$ & BPS \\
\hline
$(B,+)$ &  $(2,0)$         & ${\bf 294}_c = [0040]$& $0$ & $1/2$ \\ 
$(B,2)$ &  $(2,0)$         & ${\bf 300} = [0200]$& $0$ & $1/4$ \\
$(B,+)$ &  $(1,0)$         & ${\bf 35}_c = [0020]$ & $0$ & 1/2 \\
$(A,+)$ &  $(\ell+2,\ell)$       & ${\bf 35}_c = [0020]$ &even & $1/4$ \\
$(A,2)$ &  $(\ell+2,\ell)$       & ${\bf 28} = [0100]$ & odd & $1/8$ \\
$\text{Long}$ &  $\Delta\ge \ell+1$ & ${\bf 1} = [0000]$ & even & $0$\\
$(A,\text{cons.})$ &  $(\ell+1, \ell)$ & ${\bf 1} = [0000]$ & even & $5/16$\\
\hline
\end{tabular}
\caption{The possible superconformal multiplets in the $\overline{S}\times  \overline{S}$ OPE\@.  The $\mathfrak{so}(3, 2) \oplus \mathfrak{so}(8)_R$ quantum numbers are those of the superconformal primary in each multiplet.}
\label{opemult}
\end{table}

The $\mathcal{N}=8$ multiplets that appear in $\overline{S}\times \overline{S}$ are listed in Table \ref{opemult}.\footnote{In \cite{Chester:2014fya}, the long multiplet was denoted as $(A,0)$.} We can decompose the characters for these superblocks, as computed in \cite{Chester:2014fya}, into the characters of the $\mathcal{N}=6$ superblocks as computed from the previous Appendix to get the following decomposition of multiplets:
\es{8to6mult}{
(B,+)^{[0020]}_{1,0}\qquad&\to \qquad(B,2)^{[011]}_{1,0}\,,\\
(B,+)^{[0040]}_{2,0}\qquad&\to \qquad(B,2)^{[022]}_{2,0}\,,\\
(B,2)^{[0200]}_{2,0}\qquad&\to \qquad (B,2)^{[022]}_{2,0}\oplus (B,1)^{[200]}_{2,0} \oplus  (A,2)^{[011]}_{2,0} \oplus  (A,0)^{[000]}_{2,0}\,, \\
(A,+)^{[0020]}_{\ell+2,\ell}\qquad&\to \qquad (A,+)^{[020]}_{\ell+5/2,\ell+1/2} \oplus (A,-)^{[002]}_{\ell+5/2,\ell+1/2} \oplus   (A,2)^{[011]}_{\ell+2,\ell} \oplus  (A,2)^{[011]}_{\ell+3,\ell+1}\,,\\
(A,2)^{[0100]}_{\ell+2,\ell}\qquad&\to \qquad    (A,2)^{[011]}_{\ell+2,\ell} \oplus  (A,2)^{[011]}_{\ell+3,\ell+1}  \oplus  2\times(A,1)^{[100]}_{\ell+5/2,\ell+1/2}  \oplus  (A,0)^{[000]}_{\ell+3,\ell+1}  \oplus  (A,0)^{[000]}_{\ell+2,\ell}\,,\\
\text{Long}^{[0000]}_{\Delta,\ell}\qquad&\to \qquad \text{Long}^{[000]}_{\Delta,\ell} \oplus  \text{Long}^{[000]}_{\Delta+1,\ell-1}  \oplus  2\times\text{Long}^{[000]}_{\Delta+1,\ell}  \oplus  \text{Long}^{[000]}_{\Delta+1,\ell+1}  \oplus  \text{Long}^{[000]}_{\Delta+2,\ell}\,,\\
(A,\text{cons.})^{[0000]}_{\ell+1,\ell}\qquad&\to \qquad (A,\text{cons.})^{[000]}_{\ell+1,\ell}\oplus (A,\text{cons.})^{[000]}_{\ell+2,\ell+1}\,,\\
}
where $2\times$ denotes that the multiplet appears twice.

The $\mathcal{N}=8$ stress tensor correlator was written in \cite{Chester:2014fya} in the basis
 \es{SSSScorN8}{
   \langle \overline{S}(\vec x_1,Y_1)   \overline{S}(\vec x_2,Y_2)  \overline{S}(\vec x_3,Y_3)  \overline{S}(\vec x_4,Y_4)\rangle &= \frac{1}{x_{12}^2 x_{34}^2} \biggl[ \overline{\cS^{1}}(U, V) Y_{12}^2 Y_{34}^2 
   + \overline{\cS^{2}}(U, V) Y_{13}^2 Y_{24}^2  + \overline{\cS^{3}}(U, V) Y_{14}^2 Y_{23}^2 \\
    &\hspace{-1in}{}+\overline{\cS^{4}}(U, V) Y_{13} Y_{14} Y_{23} Y_{24} +  \overline{\cS^{5}}(U, V) Y_{12} Y_{14} Y_{23} Y_{34} 
     + \overline{\cS^{6}}(U, V) Y_{12} Y_{13} Y_{24} Y_{34}  
   \biggr] \,, \\
 }
 where $Y$ are $\mathfrak{so}(8)$ null vectors. As shown in Section 2.7.1 of \cite{Binder:2019mpb},\footnote{This was for a basis of $SO(8)$ matrices $\overline X$, but as noted there its the exact same decomposition for the basis of $Y$'s.} this decomposes to the $\mathcal{N}=6$ basis in \eqref{SSSScor} as 
 \es{b8to6}{
 \{\overline{\mathcal{S}^1}\,,\overline{\mathcal{S}^2}\,,\overline{\mathcal{S}^3}\,,\overline{\mathcal{S}^4}\,,\overline{\mathcal{S}^5}\,,\overline{\mathcal{S}^6}\} \to \{\mathcal{S}^1\,,\mathcal{S}^2\,,\mathcal{S}^3\,,4\mathcal{S}^4\,,4\mathcal{S}^5\,,4\mathcal{S}^6\} \,.
 }
Finally, we can decompose the explicit $\mathcal{N}=8$ superblocks $\mathfrak{G}_{\mathcal{M}_{\Delta,\ell}^{[d_1d_2d_3d_4]}}(U,V)$ given in \cite{Chester:2014fya} into the $\mathcal{N}=6$ superblocks $\mathfrak{G}_{\mathcal{M}_{\Delta,\ell}^{[d_1d_2d_3]}}(U,V)$ given in the attached \texttt{Mathematica} file to get 
\es{N8Bpstress}{
\mathfrak{G}_{(B,+)_{1,0}^{[0020]}}\qquad &\to\qquad\frac14\mathfrak{G}_{(B,2)_{1,0}^{[011]}}\,,\\\qquad \mathfrak{G}_{(B,+)_{2,0}^{[0040]}}\qquad &\to\qquad\frac14\mathfrak{G}_{(B,2)_{2,0}^{[022]}}\,, \\
 \mathfrak{G}_{(B,2)_{2,0}^{[0200]}}\qquad &\to\qquad\frac{1}{16}\mathfrak{G}_{(B,2)_{2,0}^{[022]}}+\frac{1}{3}\mathfrak{G}_{(A,2)_{2,0}^{[011]}}+\frac{4}{35}\mathfrak{G}_{\text{Long}_{2,0}^{[000], 1}}+\frac{1}{8}\mathfrak{G}_{(B,1)_{2,0}^{[200]}}\,,\\
  \mathfrak{G}_{(A,+)_{\ell+2,\ell}^{[0020]}}\qquad &\to\qquad\frac{1}{4}\mathfrak{G}_{(A,2)_{\ell+2,\ell}^{[011]}}+\frac{(4+\ell)^2}{(5+2\ell)(7+2\ell)}\mathfrak{G}_{(A,2)_{\ell+3,\ell+1}^{[011]}}+\frac{1+\ell}{4+2\ell}\mathfrak{G}_{(A,+)_{\ell+5/2,\ell+1/2}^{[020]}}\,,\\
   \mathfrak{G}_{(A,2)_{\ell+2,\ell}^{[0020]}}\qquad &\to\qquad-\frac{1}{4}\mathfrak{G}_{(A,2)_{\ell+2,\ell}^{[011]}}-\frac{(3+\ell)^2}{(3+2\ell)(5+2\ell)}\mathfrak{G}_{(A,2)_{\ell+3,\ell+1}^{[011]}}-\frac{\ell^2}{3+4\ell(2+\ell)}\mathfrak{G}_{\text{Long}_{\ell+2,\ell}^{[000]}}\\
   &-\frac{(5+\ell)^2}{(7+2\ell)(9+2\ell)}\mathfrak{G}_{\text{Long}_{\ell+3,\ell+1}^{[000],1}}
   -\frac{1+\ell}{9+3\ell}\mathfrak{G}_{(A,1)_{\ell+5/2,\ell+1/2}^{[000],1}}-\mathfrak{G}_{(A,1)_{\ell+5/2,\ell+1/2}^{[000],2}}\,,\\
    \mathfrak{G}_{\text{Long}_{\Delta,\ell}^{[0000]}}\qquad &\to\qquad\mathfrak{G}_{\text{Long}_{\Delta,\ell}^{[000], 1}}+\frac{4 (\ell-1)^2 (-\Delta +\ell+1) (\Delta +\ell)}{(2 \ell-1) (2 \ell+1) (\ell-\Delta ) (\Delta +\ell+1)}\mathfrak{G}_{\text{Long}_{\Delta+1,\ell-1}^{[000]}}\\
&   +\frac{4 \Delta  (-\Delta +\ell+1) (\Delta +\ell)}{3 (\Delta +2) (\ell-\Delta ) (\Delta +\ell+1)}\mathfrak{G}_{\text{Long}_{\Delta+1,\ell}^{[000],2}}
     +\frac{4 (-\Delta +\ell+1) (\Delta +\ell)}{(\ell-\Delta ) (\Delta +\ell+1)}\mathfrak{G}_{\text{Long}_{\Delta+1,\ell}^{[000],3}}\\
&      +\frac{4 (\ell+1) (\ell+2) (\Delta +\ell) (\Delta +\ell+2)}{(2 \ell+1) (2 \ell+3) (\Delta +\ell+1) (\Delta +\ell+3)}\mathfrak{G}_{\text{Long}_{\Delta+1,\ell+1}^{[000]}}\\
&    +\frac{4 (\Delta +4)^2 (-\Delta +\ell+1) (\Delta +\ell)}{(2 \Delta +5) (2 \Delta +7) (\ell-\Delta ) (\Delta +\ell+1)}\mathfrak{G}_{\text{Long}_{\Delta+2,\ell}^{[000],1}}
   \,,\\
     \mathfrak{G}_{(A,\text{cons}.)_{\ell+1,\ell}^{[0000]}}\qquad &\to\qquad   \mathfrak{G}_{(A,\text{cons}.)_{\ell+1,\ell}^{[000]}}+\mathfrak{G}_{(A,\text{cons}.)_{\ell+2,\ell+1}^{[000]}}\,,
}
where for $ \mathfrak{G}_{\text{Long}_{\Delta,0}^{[0000]}}$ we should ignore the $ \mathfrak{G}_{\text{Long}_{\Delta+1,-1}^{[000]}}$ and $ \mathfrak{G}_{\text{Long}_{\Delta+1,0}^{[000],3}}$ terms on the RHS, and rescale $ \mathfrak{G}_{\text{Long}_{\Delta+1,0}^{[000],2}}$ by $\frac{\Delta-1}{\Delta+2}$.

\section{Simplifying the $S^3$ partition function}
\label{simpS3}

In this appendix we describe how to simplify the $S^3$ partition function for various $\cN = 6$ theories. In Section~\ref{simpABJ} we consider the $U(N)_k\times U(N+M)_{-k}$, and in Section~\ref{simpso2} we consider the $SO(2)_{2k}\times USp(2+2M)_{-2k}$ theory. We close by demonstrating that additional $U(1)$ factors do not change the $S^3$ partition function

\subsection{ABJ theory}
\label{simpABJ}

We will begin with the generalization of \eqref{ABJv1} to the case of two mass deformations $m_\pm$:
\begin{equation}\begin{split}\label{ABJ2m}
Z_{M,N,k}&(m_+,m_-)  \\
&=\int d^{M+N}\mu\, d^N\nu\  \frac{e^{-i\pi k(\sum_i \mu_i^2 - \sum_a \nu_a^2)}\prod_{i<j}4\sinh^2\left[\pi(\mu_i-\mu_j)\right]\prod_{a<b}4\sinh^2\left[\pi(\nu_a-\nu_b)\right]}{\prod_{i,a}4\cosh\left[\pi(\mu_i-\nu_a)+\frac{\pi m_+}2\right]\cosh[\pi(\mu_i-\nu_a)+\frac{\pi m_-}2]}\,,
\end{split}\end{equation}
which we aim to simplify from $2N+M$ integrals to an expression with merely $N$ integrals. To achieve this we will follow the methods of \cite{Honda:2013pea} which considered the special case $m_+ = m_- = 0$. We are ultimately only interested in computing $Z$ up to an overall normalization constant $Z_0$ which is independent of $m_\pm$.

Our first step is to use the determinant formula:
\begin{equation}\begin{split}\label{cauchDet}
&\frac{\prod_{i<j}2\sinh \frac{x_i-x_j}2 \prod_{a<b}2\sinh \frac{y_a-y_b}2 }{\prod_{i,a}2\cosh \frac{x_i-y_a}2 } = \prod_{i = 1}^{N+M} e^{-\frac12Mx_i}\prod_{a = 1}^Ne^{\frac12My_a} \det\left(A(x,y)\right)
\end{split}\end{equation}
where $A(x,y)$ is the matrix
\begin{equation}
A_{ij}(x,y) = \frac{\theta_{N,i}}{2\cosh \frac{x_i-y_j}2} + e^{(N+M+1/2-j)y_i}\theta_{j,N+1}\,, \quad \text{where}\quad \theta_{i,j} = \begin{cases} 1 & i\geq j\\ 0 & \text{otherwise}\end{cases}\,,
\end{equation}
which is proven in \cite{Honda:2013pea} by using a generalization of the Cauchy determinant formula. Applying this formula with $x_i = 2\pi \mu_i$ and $y_i = 2\pi\nu_i + \pi m_-$, we can rewrite \eqref{ABJ2m} as
\begin{equation}\begin{split}
&Z_{M,N,k}(m_+,m_-) = (N+M)!\,e^{-\frac \pi2MN(m_++m_-)}\int d^{M+N}\mu\, d^N\nu\ 
\prod_{j = 1}^{N+M}
e^{-\pi(ik\mu_j^2 + 2M\mu_j)} \\
\qquad&\times\prod_{a = 1}^{N}
\frac{e^{\pi(ik\nu_a^2 +2M\nu_a)}}{2\cosh\left[\pi(\mu_a-\nu_a)+\frac{\pi m_+}2\right]}\prod_{j = N+1}^{N+M}e^{(2(N+M-j)+1)\pi\mu_j} \\
\qquad&\times\sum_{\text{perms }\sk} (-1)^{\text{sgn}(\sk)}\prod_{i=1}^{N+M}\left(\frac{\theta_{N,\sk(i)}}{2\cosh\left[\pi(\mu_i-\nu_{\sk(i)})+\frac{\pi m_-}2\right]} + e^{(2(N+M-{\sk(i)})+1)\pi\mu_i}\theta_{{\sk(i)},N+1}\right)\,, \\ 
\end{split}\end{equation}
where the sum over $\sigma$ is a sum over all permutations $\sk(i)$ of $N+M$ elements.

We next take the Fourier transform of the coshines
\begin{equation}
\frac 1{2\cosh(\pi p)} = \frac1\pi\int_{-\infty}^\infty dx\,\frac{e^{2ipx}}{2\cosh(x)}\,.
\end{equation}
The $\mu$ and $\nu$ integrals then become Gaussian and can be easily performed.  We thus find that
\begin{equation}\begin{split}
Z_{M,N,k}&(m_+,m_-) \propto e^{-\frac\pi2MN(m_++m_-)}\\
&\times\sum_\sk (-1)^\sk\Bigg(\int d^Nx\,d^{N+M}y \prod_{a=1}^N\frac{e^{-\frac{2i}{k\pi}x_a(y_a-y_{\sk(a)}) + \frac2k M(y_a-y_{\sk(a)}) + i(x_am_++y_am_-)}}{4\cosh(x_a)\cosh(y_a)} \\
&\times \prod_{l = N+1}^{N+M}\left[\pi e^{-\frac{i\pi}k(N+\frac12-l)^2}\dk\left(y_l+i\pi(N+M+1/2-l)\right)e^{\frac{i}{k\pi}y_l^2+\frac2k (N+1/2-l)y_{\sk(l)}}\right]\Bigg)\,.
\end{split}\end{equation}
So long as $2M<|k|+1$, we can integrate over $x_i$, leaving
\begin{equation}\begin{split}
Z_{M,N,k}(m_+,m_-) &\propto e^{-\frac \pi 2MN(m_++m_-)}\sum_\sk (-1)^\sk\int \,d^{N+M}y \prod_{a=1}^N\frac{e^{\frac2k M(y_a-y_{\sk(a)}) + i y_am_-}}{4\cosh\left[\frac{y_a-y_{\sk(a)}}{k}-\frac {\pi m_+}2\right]\cosh(y_a)} \\
&\times \prod_{l = N+1}^{N+M}\left[\pi e^{-\frac{i\pi}k(N+\frac12-l)^2}\dk\left(y_l+i\pi(N+M+1/2-l)\right)e^{\frac{i}{k\pi}y_l^2+\frac2k (N+1/2-l)y_{\sk(l)}}\right]\,.
\end{split}\end{equation}
After a change of variable $y_a\rightarrow y_a/2$ and judicious use of the equation $\sum_a y_a = \sum_a y_{\sk(a)}$ we find that
\begin{equation}\begin{split}
Z_{M,N,k}(m_+,m_-)&\propto e^{-\frac\pi2MN(m_++m_-)}\sum_\sk (-1)^\sk\int \,d^{N+M}y \prod_{a=1}^N\frac{e^{\frac i{2} y_am_-}}{2\cosh\left[\frac {y_a}2\right]} \\
&\times \prod_{l = N+1}^{N+M}\left[\pi e^{-\frac{i\pi}k(N+\frac12-l)^2}\dk\left(y_l+i\pi(N+M+1/2-l)\right)e^{\frac{i}{4k\pi}y_l^2-\frac Mk y_{l}}\right]\\
&\times\det\left(\frac{\theta_{N,l}}{2\cosh \frac{y_j-y_l + k\pi m_+}{2k}} + e^{\frac1k(N+M+1/2-l)y_j}\theta_{l,N+1}\right) \,.
\end{split}\end{equation}
Applying \eqref{cauchDet} again, integrating over $y_{N+1}\,,\dots\,, y_{N+M}$, and then performing a final change of variables $y_a \rightarrow 2\pi y_a$, we arrive at our final expression
\begin{equation}\label{mixdU1d}\begin{split}
Z_{M,N,k}(m_+,m_-)\qquad\qquad& \\
= \frac{e^{-\frac \pi2MN m_-}Z_0}{\cosh^N \frac{\pi m_+}2}&\int d^Ny\prod_{a<b} \frac{\sinh^2\frac{\pi(y_a-y_b)}{k}}{\cosh\left[\frac{\pi (y_a-y_b)}k +\frac{\pi m_+}{2}\right]\cosh\left[\frac{\pi (y_a-y_b)}k -\frac{\pi m_+}{2}\right]}\\
&\times\prod_{a=1}^N\left(\frac{e^{i\pi y_am_-}}{2\cosh\left(\pi y_a\right)}\prod_{l=0}^{M-1}\frac{\sinh\left[\frac{\pi\big(y_a+i(l+1/2)\big)}{k}\right]}{\cosh\left[\frac{\pi\big(y_a+i(l+1/2)\big)}k-\frac{\pi m_+}2\right]}\right)\,,
\end{split}\end{equation}
Equation \eqref{ABJv2} in the main text is simply the special case of this where we set $m_- = m$ and $m_+ = 0$.

\subsection{$SO(2)\times USp(2+2M)$ theory}
\label{simpso2}

In this section we shall reduce the $SO(2)_{2k}\times USp(2+2M)_{-k}$ sphere partition function \eqref{so2v1} down to a single integral. As in the previous section, we will work with a slightly more general partition function
\begin{equation}\begin{split}\label{so2M}
Z_{M,k}&(m_+,m_-) \propto \int d\mu\,d^M\nu\,e^{2\pi i k(\mu^2 - \sum_a \nu_a^2)}\\
&\times \frac{\prod_a\sinh^2\left (2\pi\nu_a\right) \prod_{a<b} \sinh^2\left[\pi(\nu_a+\nu_b)\right]\sinh^2\left[\pi(\nu_a-\nu_b)\right]}{\prod_{b}\cosh \frac{2\pi(\mu-\nu_b)+\pi m_+}2 \cosh \frac{2\pi(\mu+\nu_b)+\pi m_+}2 \cosh \frac{2\pi (\mu-\nu_b)+\pi m_-}2 \cosh \frac{2\pi(\mu+\nu_b)+\pi m_-}2 }\,,
\end{split}\end{equation}
where we have two mass deformations $m_\pm$. We follow the derivation in \cite{Moriyama:2016kqi}, which considered the special case $m_+ = m_- = 0$. They however consider the general $SO(2N)\times USp(2N+2M)$ theory, while here we only focus on the $N = 1$ case, for which the manifest $\cN = 5$ SUSY is enhanced to $\cN = 6$.

To ease comparison with \cite{Moriyama:2016kqi}, we rewrite \eqref{so2M}:
\begin{equation}\begin{split}
Z_{M,k}&(m_+,m_-) \propto \int d\mu\,d^{M+1}\nu\,e^{\frac{i}{4\tilde k\pi}(\mu^2 - \sum_a \nu_a^2)}\\
&\times \frac{\prod_a\sinh^2 \frac{\nu_a}{\tilde k}  \prod_{a<b} \sinh^2 \frac{\nu_a+\nu_b}{2\tilde k} \sinh^2 \frac{\nu_a-\nu_b}{2\tilde k} }{\prod_{b}\cosh\left(\frac{\mu-\nu_b}{2\tilde k}+\frac{\pi m_+}2\right)\cosh\left(\frac{\mu+\nu_b}{2\tilde k}+\frac{\pi m_+}2\right)\cosh\left(\frac{\mu-\nu_b}{2\tilde k}+\frac{\pi m_-}2\right)\cosh\left(\frac{\mu+\nu_b}{2\tilde k}+\frac{\pi m_-}2\right)} \\
\end{split}\end{equation}
with $\tilde k = 2k$. Next we use the Cauchy-Vandermonde determinant, given in (2.6) of \cite{Moriyama:2016kqi}:
\begin{equation}\label{detForm20}\begin{split}
&\frac{\prod_a\sinh \frac{\nu_a}{\tilde k}  \prod_{a<b} \sinh \frac{\nu_a+\nu_b}{2{\tilde k}} \sinh \frac{\nu_a-\nu_b}{2{\tilde k}} }{\prod_{b}\cosh \frac{\mu-\nu_b+{\tilde k}\pi m_+}{2{\tilde k}} \cosh \frac{\mu+\nu_b+{\tilde k}\pi m_+}{2{\tilde k}} } \\
&=\det\begin{pmatrix}\left[\frac{\sinh\frac{\nu_a}{{\tilde k}}}{\cosh\frac{\mu+{\tilde k}\pi m_+-\nu_a}{2{\tilde k}}\cosh\frac{\mu+{\tilde k}\pi m_++\nu_a}{2{\tilde k}}}\right] \\ 
\left[\sinh\frac{b\nu_a}{{\tilde k}}\right]_{b = 1,\dots,M}\end{pmatrix}\,.
\end{split}\end{equation}
We can simplify this expression by introducing canonical position and momentum operators $\hat q$ and $\hat p$ which satisfy $[\hat q,\hat p] = 2\pi i\tilde k$. We denote the $\hat q$ eigenstates by $\ket{\nu}$, and introduce states $|b]]$ such that
\begin{equation}
2\sinh \frac{b\nu_{a}}{\tilde k} = [[b\ket{\nu_{a}}\,,
\end{equation}
allowing us to simplify the lower block of \eqref{detForm20}. The upper block can be simplified using the Fourier transform, giving us (2.9) of \cite{Moriyama:2016kqi}:
\begin{equation}
\frac{\sinh \frac{\nu_a}{{\tilde k}} }{\cosh \frac{\mu+{\tilde k}\pi m_+-\nu_a}{2{\tilde k}} \cosh \frac{\mu+{\tilde k}\pi m_++\nu_a}{2{\tilde k}} } \propto \bra{\mu}e^{\frac{i \hat p m_+}{4}}\frac1{\sinh\frac{\hat p}2}\hat\Pi_-\ket{\nu_a} \,.
\end{equation}
We then follow \cite{Moriyama:2016kqi} in performing similarity transforms
\begin{equation}
\ket\mu\rightarrow e^{-\frac{i}{4\pi\tilde k}\hat p^2} \ket\mu \,,\qquad \ket{\nu_a}\rightarrow e^{-\frac{i}{4\pi\tilde k}\hat p^2} \ket{\nu_a} \,,
\end{equation}
so that \eqref{so2M} becomes
\begin{equation}\begin{split}
Z_{M,k}&(m_+,m_-) \propto \int d\mu\,d^{M+1}\nu \det\begin{pmatrix}\left[\bra{\mu}e^{\frac{i \hat p m_+}{4}}\frac1{\sinh\frac{\hat p}2}\hat\Pi_-\ket{\nu_a}\right] \\ 
\Big[[[b|e^{\frac{i}{4\pi\tilde k}\hat p^2}\ket{\nu_{a}}\Big]_{b = 1,\dots,M}\end{pmatrix}\\
&\times\langle\mu|e^{\frac{i}{4\pi\tilde k}\hat p^2}e^{\frac{i}{4\pi\tilde k}\hat q^2}e^{\frac{i \hat p m_-}{4}}\frac1 {2\sinh \frac{\hat p}2}\hat \Pi_-e^{-\frac{i}{4\pi\tilde k}\hat q^2}e^{-\frac{i}{4\pi\tilde k}\hat p^2}|\nu_1\rangle\prod_{a=1}^M[[a|e^{-\frac{i}{4\pi\tilde k}\hat q^2}e^{-\frac{i}{4\pi\tilde k}}|\nu_{a+1}\rangle
\end{split}\end{equation}
We can now apply (2.12-5) of \cite{Moriyama:2016kqi} to simplify the matrix elements, but with the modification
\begin{equation}\begin{split}
\langle\mu|e^{\frac{i}{4\pi\tilde k}\hat p^2}e^{\frac{i}{4\pi\tilde k}\hat q^2}&e^{\frac{i \hat p m_-}{4}}\frac1 {2\sinh \frac{\hat p}2}\hat \Pi_-e^{-\frac{i}{4\pi\tilde k}\hat q^2}e^{-\frac{i}{4\pi\tilde k}\hat p^2}|\nu_1\rangle \\
&= \frac{\pi\tilde k e^{\frac{i m_-\mu}2}}{i\sinh\frac{\mu}2}\left(\dk(\mu-\nu_1)-\dk(\mu+\nu_1)\right)\,,
\end{split}\end{equation}
and so find that
\begin{equation}\begin{split}
Z(m_+,m_-) &\propto \int d\mu\,d^{M+1}\nu\  \frac{\prod_a\sinh \frac{\nu_a}{\tilde k}  \prod_{a<b} \sinh \frac{\nu_a+\nu_b}{2{\tilde k}} \sinh \frac{\nu_a-\nu_b}{2{\tilde k}} }{\prod_{b}\cosh \frac{\mu-\nu_b+{\tilde k}\pi m_+}{2{\tilde k}} \cosh \frac{\mu+\nu_b+{\tilde k}\pi m_+}{2{\tilde k}} } \\
&\times \frac{e^{\frac{i m_-\mu}2}\dk(\mu-\nu_1)}{\sinh\frac{\mu}2}\prod_{m=1}^M \dk(\nu_{m+1}-2\pi i m)\\
&\propto \frac 1 {\cosh \frac{\pi m_+}{2} }\int d\mu \frac{e^{\frac{i m_-\mu}2} \sinh\left[\frac{\mu}{2k}\right]}{\sinh\left[\frac{\mu}2\right]\cosh\left[\frac{\mu}{2k}+\frac{\pi m_+}{2}\right]} \\
&\times\frac{\prod_{l = 1}^M\sinh\left[\frac{\mu+2\pi i l}{4k}\right]\sinh\left[\frac{\mu-2\pi i l}{4k}\right]}{\prod_{l = 1}^M\cosh\left[\frac{\mu+2\pi i l}{4k}+\frac{\pi m_+}2\right]\cosh\left[\frac{\mu-2\pi i l}{4k}+\frac{\pi m_+}2\right]} \\
&\propto \frac 1 {\cosh \frac{\pi m_+}{2} }\int d\mu\  \frac{e^{i \pi m_-\mu}\cosh\left[\frac{\pi\mu}{2k}\right]\cosh\left[\frac{\pi\mu}{2k}+\frac{\pi m_+}2\right]}{\sinh\left[\pi\mu\right]\cosh\left[\frac{\pi\mu}{k}+\frac{\pi m_+}{2}\right]}\\
&\times\prod_{l = -M}^M\frac{\sinh\left[\frac{\pi(\mu+i l)}{2k}\right]}{\cosh\left[\frac{\pi(\mu+ i l)}{2k}+\frac{\pi m_+}2\right]}\,,
\end{split}\end{equation}
and this completes the derivation.

\subsection{Additional $U(1)$ factors}
\label{AddU1s}

In this section we shall show that given an $\cN=6$ Chern-Simon gauge theory with gauge group $G = SU(N)\times U(N+M)$ or $G = USp(2+2M)$, the $S^3$ partition function for the theory $G\times U(1)^L$ is equivalent to that of the theory $G\times U(1)$, up to an overall constant.  To this end we note that the mass-deformed $S^3$ partition function for the $G\times U(1)^L$ can be generically written as
\begin{equation}\label{manyU1}
Z_{G\times U(1)^L}(m_+,m_-) = \int d\chi_1\dots d\chi_N\,e^{i\pi\sum_{ab}K_{ab}\chi_a\chi_b} Z_G(m_++2q\cdot\chi,m_-+2q\cdot\chi)\,,
\end{equation}
where $K_{ab}$ is the matrix of Chern-Simons levels for the $U(1)$s $q = (q_1,\dots,q_L)$ are the charges of the (bi)fundamentals under each $U(1)$, and $Z_G(m_+,m_-)$ is the $S^3$ partition function for the theory without any $U(1)$ factors. In order for $G\times U(1)^N$ to have $\cN = 6$ supersymmetry, $K_{ab}$ and $q_a$ must satisfy the condition
\begin{equation}\label{tachCond}
\sum_{a,b} K^{ab}q_aq_b = \frac 1k_G
\end{equation}
for some $G$ dependent constant $k_G$, where $K^{ab}$ is the inverse of $K_{ab}$ \cite{Schnabl:2008wj}.

To simplify \eqref{manyU1}, we first perform a change of basis of $\chi_a$ such that $q_a = (1,0,\dots,0)$. Because $K_{ab}$ is symmetric, we can then perform a second change of basis to $\chi_2\,,\dots\,,\chi_L$, so that $K_{ab}$ take the form
\begin{equation}
K_{ab} = \begin{pmatrix} K_{11} & K_{12} & 0 & \hdots \\ K_{12} & 1 & 0 & \hdots \\ 0 & 0 & 1 & \hdots \\
\vdots & \vdots & \vdots & \ddots \end{pmatrix}\,.
\end{equation}
We can now integrate over $\chi_2\,,\chi_3\,,\dots$ leaving us with
\begin{equation}
Z_{G\times U(1)^L}(m_+,m_-) \propto \int d\chi_1\,e^{i\pi(K_{11}-K_{12}^2)\chi^2} Z_G(m_++2\chi_1,m_-+2\chi_1)\,,
\end{equation}
We then note that, in this basis, the condition \eqref{tachCond} becomes:
\begin{equation}
K_{11}-K_{12}^2 = k_G\,,
\end{equation}
and so 
\begin{equation}
Z_{G\times U(1)^L}(m_+,m_-) \propto \int d\chi_1\,e^{i\pi k_G\chi^2} Z_G(m_++2\chi_1,m_-+2\chi_1)\,.
\end{equation}
We now simply recognize the right-hand side of this equation is the partition function for the $G\times U(1)$ theory, and have hence shown what we set out to prove.

\bibliographystyle{ssg}
\bibliography{N6bootdraft}

\end{document}